\documentclass[twocolumn,aps,prb,showpacs,groupedaddress]{revtex4-1}
\usepackage{amsmath}
\usepackage{graphicx}

\usepackage{pifont}
\usepackage{amsfonts}
\usepackage{subfigure}
\usepackage{setspace}
\usepackage{threeparttable}
\usepackage{enumerate}
\usepackage{xspace}
\usepackage{textcomp}
\usepackage{bm}
\usepackage{graphicx,color,epsfig}
\usepackage{soul}
\usepackage{multirow}

\begin{document}

\title{Variational study of the Kitaev-Heisenberg-Gamma model}

\author{Shang-Shun Zhang$^{1}$, G\'abor B. Hal\'asz$^{2,3}$, Wei Zhu$^{4}$, and Cristian D. Batista$^{1,5}$}

\address{$^1$Department of Physics and Astronomy, University of Tennessee, Knoxville,
Tennessee 37996-1200, USA}
\address{$^2$Materials Science and Technology Division, Oak Ridge
National Laboratory, Oak Ridge, Tennessee 37831, USA}
\address{$^3$Quantum
Science Center, Oak Ridge, Tennessee 37831, USA}
\address{$^4$ Key Laboratory for Quantum Materials of Zhejiang Province, School of Science, Westlake University, 18 Shilongshan Road, Hangzhou, Zhejiang 310024, China}
\address{$^5$Quantum Condensed Matter Division and Shull-Wollan Center, Oak Ridge National Laboratory, Oak Ridge, Tennessee 37831, USA}

\begin{abstract}
We compute the low-energy excitation spectrum and the dynamical spin
structure factor of the Kitaev-Heisenberg-Gamma model through a
variational approach based on the exact fractionalized excitations
of the pure Kitaev honeycomb model. This novel approach reveals the
physical reason for the asymmetric stability of the Kitaev spin
liquid phases around the ferromagnetic and antiferromagnetic Kitaev
limits. Moreover, we demonstrate that the fractionalized excitations
form bound states in specific regions of each Kitaev spin liquid
phase and that certain phase transitions induced by Heisenberg and
Gamma interactions are driven by the condensation of such a bound
state. Remarkably, this bound state appears as a sharp mode in the
dynamical spin structure factor, while its condensation patterns at
the appropriate phase transitions provide a simple explanation for
the magnetically ordered phases surrounding each Kitaev spin liquid
phase.
\end{abstract}
\maketitle

\section{Introduction} \label{sec-int}

Quantum spin liquids are highly unconventional, fundamentally
quantum phases of magnetic systems that do not order by spontaneous
symmetry breaking and are instead characterized by long-range
quantum entanglement,~\cite{balents2010spin, savary2016quantum}
along with a rich variety of exotic phenomena, including topological
order, emergent gauge theories, and quasiparticle fractionalization.
Indeed, the spin degrees of freedom in quantum spin liquids appear
to be fractionalized into nonlocal quasiparticle excitations that
carry internal gauge charges and exhibit nontrivial anyonic particle
statistics. In addition to their fundamental appeal, these anyonic
quasiparticles facilitate topological braiding processes and thus
provide manifestly fault-tolerant schemes of quantum
computation.\cite{kitaev2003fault, nayak2008nonabelian}

The Kitaev spin liquid phase on the honeycomb lattice is
particularly amenable to theoretical studies due to its exactly
solvable limit described by the famous Kitaev honeycomb
model.\cite{kitaev2006anyons} Remarkably, it has been demonstrated
that the bond-directional spin interactions of the Kitaev model are
naturally realized between effective spin-one-half magnetic moments
in strongly spin-orbit-coupled $4d$ and $5d$
systems,\cite{jackeli2009mott} and this realization has led to the
discovery of several candidate materials in which the microscopic
spin Hamiltonian is believed to be well approximated by the Kitaev
model.\cite{rau2016spin, trebst2017kitaev, hermanns2018physics,
takagi2019concept} These Kitaev materials include the honeycomb
iridates Na$_2$IrO$_3$,\cite{singh2010antiferromagnetic,
liu2011longrange, choi2012spin, ye2012direct, comin2012novel,
chun2015direct} $\alpha$-Li$_2$IrO$_3$,\cite{singh2012relevance,
williams2016incommensurate}
H$_3$LiIr$_2$O$_6$,\cite{kitagawa2018spin}, and
Ag$_3$LiIr$_2$O$_6$,\cite{bahrami2019thermodynamic} as well as the
analogous ruthenium system $\alpha$-RuCl$_3$.\cite{plumb2014spin,
sandilands2015scattering, sears2015magnetic,
majumder2015anisotropic, johnson2015monoclinic, sandilands2016spin,
banerjee2016proximate, banerjee2017neutron, do2017majorana} While
the precise microscopic Hamiltonians of these materials are still
subject to intense debate,\cite{yamaji2014first,
sizyuk2014importance, kim2015kitaev, chaloupka2015hidden,
rousochatzakis2015phase, nishimoto2016strongly, kim2016crystal,
winter2016challenges, chaloupka2016magnetic, winter2017breakdown,
winter2017models, eichstaedt2019deriving} the simplest
nearest-neighbor model one may consider is the
Kitaev-Heisenberg-Gamma model,\cite{chaloupka2010kitaev,
chaloupka2013zigzag, rau2014generic} where isotropic (Heisenberg)
interactions and symmetric off-diagonal anisotropic (Gamma)
interactions are superimposed on the diagonal anisotropic (Kitaev)
interactions.

Even though none of the candidate materials seem to realize the
Kitaev spin liquid, and most of them are found to be magnetically
ordered at the lowest temperatures, there are nevertheless direct
experimental indications of dominant Kitaev
interactions.~\cite{chun2015direct} The Kitaev materials are thus
expected to be proximate to the Kitaev spin
liquid,~\cite{banerjee2016proximate} making it important to
understand the magnetically ordered phases surrounding the Kitaev
spin liquid, along with the corresponding quantum phase transitions
giving rise to magnetic order. To this end, the phase diagram of the
Kitaev-Heisenberg-Gamma model has been studied extensively using a
wide range of techniques, including exact
diagonalization,\cite{chaloupka2010kitaev, chaloupka2013zigzag,
rau2014generic, rau2014trigonal, gotfryd2017phase} density-matrix
renormalization group,\cite{gohlke2017dynamics, gohlke2017quantum,
gordon2019theory} tensor-network methods,\cite{iregui2014probing,
lee2020magnetic} slave-particle mean-field
theories,\cite{schaffer2012probing} and variational Monte Carlo
approaches.\cite{wang2019one} These works yield consistent phase
diagrams, predicting a rich variety of magnetically ordered phases
around the ferromagnetic (FM) and antiferromagnetic (AFM) Kitaev
spin liquid phases. However, while they seem to indicate that some
of the phase transitions are close to continuous (i.e., weakly first
order), the underlying mechanisms driving these phase transitions
are still not properly understood.

Since quantum spin liquids do not break any symmetries and possess
no local order parameters or any other ``smoking-gun'' signatures,
they are notoriously difficult to identify in an experimental setup.
It is conventional wisdom that, as a result of fractionalization,
quantum spin liquids are characterized by diffuse continuum features
in spectroscopic probes because any local magnetic excitation
(magnon) created by such a probe immediately decomposes into several
nonlocal quasiparticle excitations (spinons). Indeed, for the Kitaev
spin liquid, such continuum features have been proposed to be
observable in inelastic neutron scattering,\cite{baskaran2007exact,
knolle2014dynamics, knolle2015dynamics, song2016low} Raman
scattering,\cite{knolle2014raman} and resonant inelastic x-ray
scattering.\cite{halasz2016resonant, halasz2019observing} However,
it is important to emphasize that the presence of such continuum
features does not necessarily indicate an underlying quantum spin
liquid, as it can also originate from disorder and/or thermal
fluctuations in classical phases with no long-range
entanglement.\cite{samarakoon2017comprehensive, zhu2017disorder,
samarakoon2018classical}

The main difference between the two scenarios is that the continuum
features of quantum spin liquids are produced by coherent (albeit
nonlocal) quasiparticles while those appearing in classical phases
have completely incoherent (diffusive) origins. To confirm the
presence of an underlying quantum spin liquid, one must therefore
demonstrate the coherent nature of its nonlocal quasiparticle
excitations. For example, these nonlocal excitations, which
typically correspond to continuum features in dynamical responses,
may also form local bound states, thereby producing sharp modes and
providing evidence for coherent behavior. Surprisingly, the
distinctive signature of the quantum spin liquid in this scenario is
then a sharp mode, in conjunction with the continuum features and
the lack of magnetic order.

In this paper, we use a variational approach based on the exactly
solvable Kitaev honeycomb model to demonstrate that such local bound
states of nonlocal excitations are prominent in both Kitaev spin
liquid phases of the Kitaev-Heisenberg-Gamma model. Specifically, we
focus on the nonlocal flux pair and Majorana fermion excitations of
the pure Kitaev model and investigate their dynamics and
interactions resulting from the Heisenberg and Gamma perturbations.
We find that the flux pairs, which are static in the pure Kitaev
model, acquire an emergent dynamics and may bind Majorana fermions
to form local magnon-like excitations. In turn, such a local
excitation manifests in the dynamical spin structure factor as a
sharp mode descending from a diffuse continuum feature. Moreover,
when such a bosonic magnon-like excitation becomes gapless, its
condensation directly gives rise to magnetic order. We argue that
this condensation is the driving force behind the weakly first-order
transitions out of the Kitaev spin liquid phases and, by considering
the precise condensation patterns, we account for the resulting
magnetically ordered phases found in previous works. Finally, by
studying the dynamics of both nonlocal and local excitations, we
explain the asymmetric stability of the Kitaev spin liquid phases
around the FM and AFM Kitaev limits.

The structure of this paper is as follows. In Sec.~II, we introduce
the model Hamiltonian and the variational approach for computing its
low-energy excitation spectrum. In Sec.~III, we study the dynamics
of flux-pair excitations while, in Sec.~IV, we consider magnon-like
bound states formed by a flux pair and a Majorana fermion. In
Sec.~V, we determine the magnetic orders obtained by condensing
these magnon-like excitations while, in Sec.~VI, we calculate their
direct contributions to the dynamical spin structure factor.
Finally, we conclude the paper with a brief summary and outlook in
Sec.~VII.

\section{General formulation} \label{sec-gen}

\subsection{Kitaev-Heisenberg-Gamma model} \label{sec-gen-KH}

\begin{figure}[!t]
\includegraphics[scale=1.5]{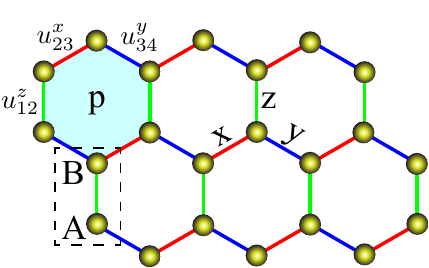}
\caption{Kitaev honeycomb model with bond-dependent Ising
interactions on the honeycomb lattice. The three bond types ($x$, $y$, $z$) are marked by three different colors. The bond variables
$u_{\langle ij \rangle}$ are $\mathbb{Z}_2$ gauge fields whose fluxes around hexagonal plaquettes $p$
coincide with the plaquette operators $W_p$.} \label{fig:model}
\end{figure}

The Kitaev-Heisenberg-Gamma model\cite{chaloupka2010kitaev,
chaloupka2013zigzag, rau2014generic} is obtained by including
Heisenberg and Gamma interactions on top of the exactly solvable
Kitaev honeycomb model~\cite{kitaev2006anyons}. The corresponding
Hamiltonian,
\begin{equation}\label{eq0}
{\cal H} = {\cal H}_K + {\cal H}_H + {\cal H}_{\Gamma},
\end{equation}
includes three contributions,
\begin{eqnarray}
\label{eq:model} {\cal H}_K &=& K \sum_{\gamma =\{ x, y, z\}}
\sum_{\langle jk \rangle \parallel \gamma } \sigma_j^{\gamma}
\sigma_k^{\gamma},
\nonumber  \\
{\cal H}_H &=& J \sum_{\langle jk \rangle} {\bm \sigma}_j \cdot {\bm
\sigma}_k,
\nonumber \\
{\cal H}_{\Gamma} &=&  \Gamma \sum_{\alpha\neq \beta \neq \gamma} \sum_{\langle jk \rangle \parallel
\gamma} {\sigma}_j^{\alpha} {\sigma}_k^{\beta},
\end{eqnarray}
where $J$ is the isotropic Heisenberg interaction, $\Gamma$ is a
symmetric off-diagonal exchange anisotropy, and $K$ is the Kitaev
interaction, corresponding to diagonal exchange anisotropy. As it is
illustrated in Fig.~\ref{fig:model}, the Kitaev terms are
bond-dependent Ising interactions between spin components that
depend on the bond orientation.

\subsection{Exactly solvable Kitaev limit} \label{sec-gen-ex}

The pure Kitaev model ($J=\Gamma=0$) is exactly solvable by means of
a simple mapping in which each spin operator is expressed as a
product of two Majorana fermions: $\sigma_j^{\gamma} = i
b_j^{\gamma} c_j^{\phantom{\gamma}}$.\cite{kitaev2006anyons} The
``matter'' Majorana fermion $c_j$ does not carry any flavor and is
associated with the site $j$. In contrast, the ``bond'' Majorana
fermions $b_j^{\gamma}$ have three flavors, $\gamma = x,y,z$,
associated with the three orientations of the bonds connected to the
site $j$. In terms of these Majorana fermions, the Kitaev
Hamiltonian becomes
\begin{eqnarray} \label{eq:kitaev}
{\cal H}_K = -i K \sum_{\gamma} \sum_{\langle jk \rangle \parallel
\gamma} u_{\langle jk \rangle \parallel \gamma} c_j c_k,
\end{eqnarray}
where $u_{\langle jk \rangle \parallel \gamma} \equiv i b_j^{\gamma}
b_k^{\gamma}$ are conserved bond variables that can be identified as
static $\mathbb{Z}_2$ gauge fields coupled to the matter fermions.
In each bond-fermion sector characterized by the $\mathbb{Z}_2$
gauge fields, $u_{\langle jk \rangle} = \pm 1$, the quadratic
matter-fermion Hamiltonian can then be diagonalized:
\begin{eqnarray} \label{eq:fermions}
{\cal H}_K^u = \sum_{\alpha} \varepsilon_{\alpha}^u
(f_{\alpha}^u)^{\dag} f_{\alpha}^u.
\end{eqnarray}
The superscript $u$ of the fermion operators, $f_{\alpha}^u$, and
the corresponding eigenenergies, $\varepsilon_{\alpha}^u \geq 0$,
indicates their dependence on the $\mathbb{Z}_2$ gauge-field
configuration. Note that, in the following, we use the terms
``fermions'' and ``matter fermions'' interchangeably, unless we make
an explicit distinction between them.

Since the  Majorana fermion representation of the spin operators
enlarges the local Hilbert space, it gives rise to a local
$\mathbb{Z}_2$ gauge transformation, $D_j^{\phantom{z}} = b_j^x b_j
^y b_j^z c_j^{\phantom{z}}$, which commutes with any physical
observable and anticommutes with the bond variables $u_{\langle jk
\rangle}$ that include the site $j$. The physical states are then
characterized by gauge-invariant products or ``fluxes'' of the
$\mathbb{Z}_2$ gauge fields, corresponding to their circulation
(i.e., Wilson loop) around each hexagonal plaquette ($p$):
\begin{equation} \label{eq:W}
W_p = \prod_{\langle jk \rangle \in \partial p} u_{\langle jk
\rangle} = \pm 1.
\end{equation}
An eigenvalue $-1$ ($+1$) of the flux operator $W_p$ indicates the
presence (absence) of a $\mathbb{Z}_2$ gauge flux $\Phi$ or,
equivalently, a vison at the plaquette $p$.

To obtain a general eigenstate of the Kitaev model, we start from a
product state $| u \rangle \otimes | n^u \rangle$ in the enlarged
Hilbert space of the Majorana fermions, where the $\mathbb{Z}_2$
gauge field configuration, $| u \rangle$, is a simultaneous
eigenstate of all bond variables, $u_{\langle jk \rangle} = \pm 1$,
while the matter-fermion state $| n^u \rangle$ is a simultaneous
eigenstate of all fermion numbers, $n_{\alpha}^u =
(f_{\alpha}^u)^{\dag} f_{\alpha}^u = \{ 0,1 \}$, corresponding to a
Slater determinant in terms of the matter fermions $c_j$. The
physical eigenstate is then obtained by projecting this product
state into the physical Hilbert space through an appropriate
projector,
\begin{eqnarray} \label{eq:P}
P = \frac{1}{2^{N}} \prod_j (1 + D_j),
\end{eqnarray}
which enforces $D_j = +1$ for all sites $j = 1, \ldots, N$ of the
lattice. Consequently, the physical eigenstate
\begin{eqnarray} \label{eq:state}
| \Phi; f \rangle = 2^{N/2} P \, \big\{ | u \rangle \otimes | n^u
\rangle \big\},
\end{eqnarray}
labeled schematically by its flux ($\Phi$) and fermion ($f$)
configurations, is manifestly gauge invariant. Note that the same
physical eigenstate, $| \Phi; f \rangle$, can be obtained by
projecting distinct Majorana-fermion states $| u_1 \rangle \otimes |
n^{u_1} \rangle$ and $| u_2 \rangle \otimes | n^{u_2} \rangle$
related by gauge transformations. While these states belong to
distinct bond-fermion sectors, $u_1$ and $u_2$, they correspond to
the same flux sector $\Phi$.

Due to the extensive projection described by Eq.~\eqref{eq:P}, it is
not surprising that the ground state of the Kitaev model,
characterized by flux operators $W_p = +1$ for all $p$ and fermion
numbers $n_{\alpha}^u = 0$ for all $\alpha$, is a strongly entangled
quantum spin liquid~\cite{kitaev2006anyons}. While the flux
excitations, characterized by $W_p = -1$ for some $p$, are gapped,
the fermion excitations, characterized by $n_{\alpha}^u = 1$ for
some $\alpha$, have a single gapless Dirac point.

\subsection{Variational approach} \label{sec-gen-var}

\begin{figure*}[!t]
\includegraphics[width=0.8\textwidth]{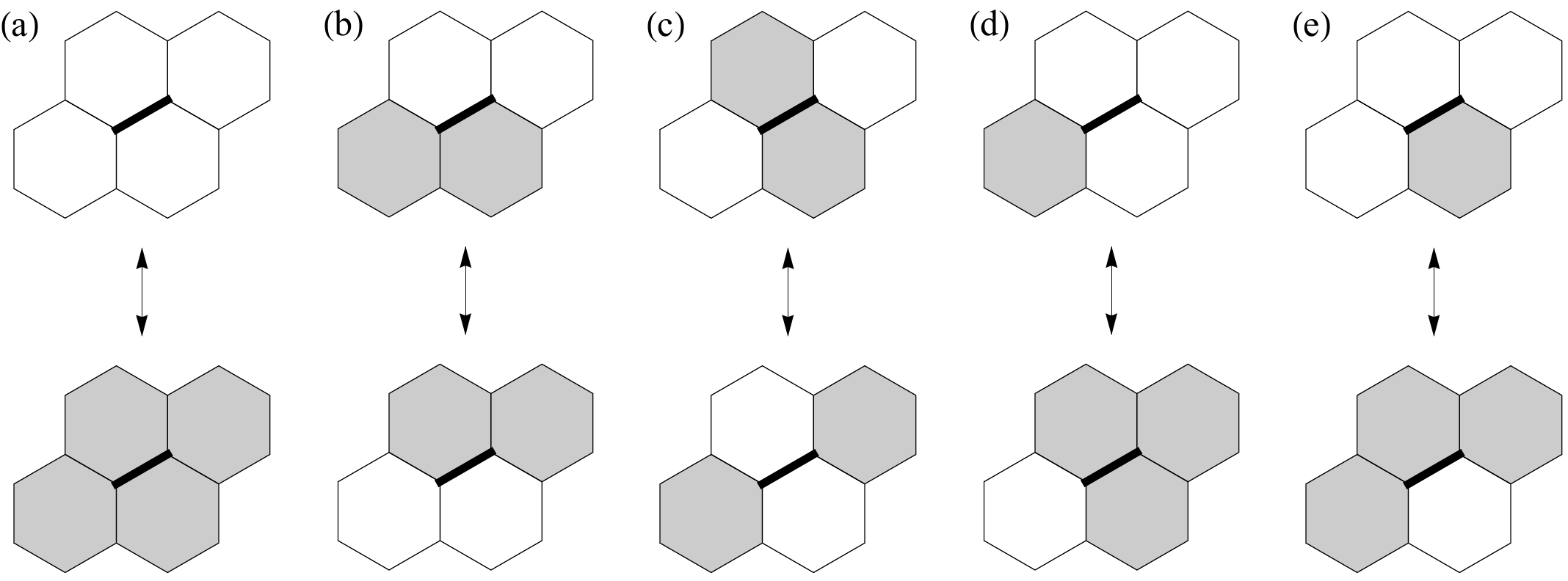}
\caption{Distinct matrix elements of the Heisenberg interactions in
terms of the $J = \Gamma= 0$ flux sectors connected.}
\label{fig:matrix}
\end{figure*}

\begin{figure*}[!t]
\includegraphics[width=0.95\textwidth]{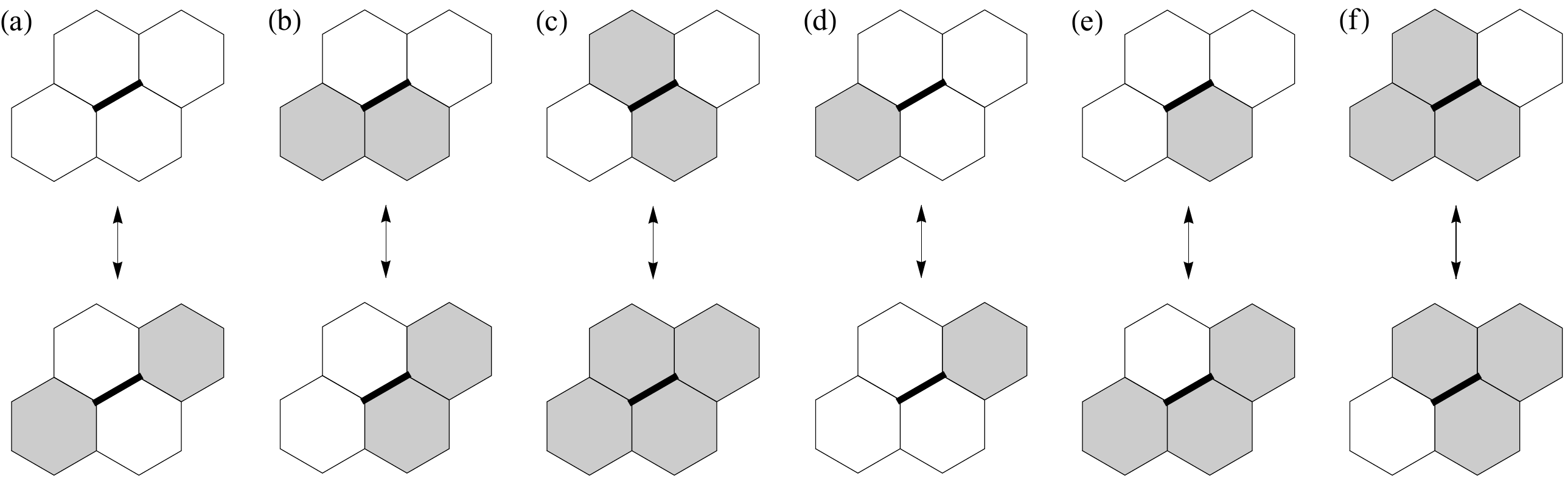}
\caption{Distinct matrix elements of the Gamma interactions in terms
of the $J = \Gamma= 0$ flux sectors connected.}
\label{fig:matrix_gamma}
\end{figure*}

Beyond the exactly solvable Kitaev limit, one can treat small enough
Heisenberg and Gamma interactions as perturbations. In general,
perturbation theory would require the evaluation of all matrix
elements of ${\cal H}_H$ and ${\cal H}_{\Gamma}$ between the exact
Kitaev eigenstates. To simplify the problem, we instead take a
variational approach and only calculate these matrix elements within
an appropriate variational subspace. Importantly, this variational
subspace is naturally constrained by the various selection rules on
the flux and fermion quantum numbers which make most of the above
matrix elements vanish.

Let us first concentrate on the fluxes. For a Heisenberg term along
an $x$ bond, $\langle jk \rangle \parallel x$, the component
$\sigma_j^x \sigma_k^x$ simply renormalizes the corresponding Kitaev
term, while the components $\sigma_j^y \sigma_k^y$ and $\sigma_j^z
\sigma_k^z$ each invert the eigenvalues, $\Phi_p = \pm 1$, of all
four fluxes at the plaquettes $p$ surrounding the bond $\langle jk
\rangle$. In contrast, for a Gamma term along the same $x$ bond,
$\langle jk \rangle \parallel x$, the components $\sigma_j^y
\sigma_k^z$ and $\sigma_j^z \sigma_k^y$ each invert the eigenvalues
of only two fluxes at the plaquettes $p$ that are separated by the
bond $\langle jk \rangle$. The symmetry-inequivalent flux sectors
connected by these Heisenberg and Gamma terms are depicted in
Figs.~\ref{fig:matrix} and \ref{fig:matrix_gamma}, respectively.

From the flux selection rules of Figs.~\ref{fig:matrix} and
\ref{fig:matrix_gamma}, it is clear that both Heisenberg and Gamma
interactions naturally result in flux-pair hopping [see
Figs.~\ref{fig:matrix}(b) and \ref{fig:matrix_gamma}(b)].
Remarkably, a flux-pair excitation, consisting of two flux
excitations at neighboring plaquettes, can propagate coherently and,
thus, can be treated as a well-defined quasiparticle even in the
presence of the gapless fermions. Indeed, a flux-pair excitation
does not lead to an orthogonality catastrophe because it corresponds
to a local perturbation (the flip of a single hopping amplitude
$u_{\langle jk \rangle} = \pm 1$) from the perspective of the
fermions which are, in turn, gapless at a Dirac point
only~\cite{hentschel2007orthogonality}. In other words, the matrix
elements of ${\cal H}_H$ and ${\cal H}_{\Gamma}$ connecting
different flux sectors in Figs.~\ref{fig:matrix}(b) and
\ref{fig:matrix_gamma}(b) are ${\cal O}(1)$ between the respective
fermion vacua and exponentially small in the number of fermions
created or annihilated.

Focusing on flux-pair excitations, the variational subspace is then
restricted to flux sectors containing a single flux pair at various
positions [see Figs.~\ref{fig:matrix}(b) and
\ref{fig:matrix_gamma}(b)]. As shown in
Refs.~\onlinecite{knolle2014dynamics} and
\onlinecite{knolle2015dynamics}, the fermion vacuum of such a flux
sector has odd fermion parity.\footnote{Note that the fermion parity
is defined with respect to the fermion vacuum of the flux-free
sector and accounts for both bond and matter fermions.} Therefore,
the flux-pair excitation has fermionic statistics and is
topologically equivalent to a bond fermion. The hopping problem of
such a bare flux pair is obtained by calculating the matrix elements
of ${\cal H}_H$ and ${\cal H}_{\Gamma}$ between the fermion vacua of
the appropriate flux sectors. Another excitation of interest is the
bound state of a flux pair (i.e., bond fermion) and a matter fermion
which has bosonic statistics and is topologically equivalent to a
spin flip or, in other words, a magnon (i.e., topologically
trivial). This composite flux pair is important for us because it
directly manifests in the dynamical spin structure factor and its
condensation immediately leads to magnetic order. To demonstrate its
formation and solve the hopping problem, we calculate the matrix
elements of ${\cal H}_H$ and ${\cal H}_{\Gamma}$ between one-fermion
states of the appropriate flux sectors.

As shown in Figs.~\ref{fig:matrix} and \ref{fig:matrix_gamma}, the
Heisenberg and Gamma interactions also give rise to many other
processes beyond flux-pair hopping. First, the energy of the ground
state is renormalized because of matrix elements connecting the
ground-state flux sector (i.e., the flux-free sector) with excited
flux sectors [see Figs.~\ref{fig:matrix}(a) and
\ref{fig:matrix_gamma}(a)]. Next, the ``closed'' flux-pair
excitations of Figs.~\ref{fig:matrix}(b) and
\ref{fig:matrix_gamma}(b) can hybridize with ``open'' flux-pair
excitations [see Fig.~\ref{fig:matrix}(c)] and four-flux excitations
[see Fig.~\ref{fig:matrix_gamma}(c)]. Finally and most
interestingly, the remaining processes describe the hopping of
single-flux excitations [see Fig.~\ref{fig:matrix_gamma}(d)] and
their hybridization with three-flux excitations [see
Figs.~\ref{fig:matrix}(d), \ref{fig:matrix}(e), \ref{fig:matrix_gamma}(e), and \ref{fig:matrix_gamma}(f)]. While
such single-flux excitations seem natural to consider in conjunction
with flux-pair excitations, we do not expect a single-flux
excitation to be a coherent quasiparticle as it corresponds to a
vortex (rather than a local perturbation) from the perspective of
the gapless fermions. Therefore, we concentrate on flux-pair hopping
[see Figs.~\ref{fig:matrix}(b) and \ref{fig:matrix_gamma}(b)] in the
following and only consider the hybridization between ``open'' and
``closed'' flux pairs [see Fig.~\ref{fig:matrix}(c)] in
Sec.~\ref{sec-open}.

\begin{figure}[!h]
\includegraphics[width=0.46\textwidth]{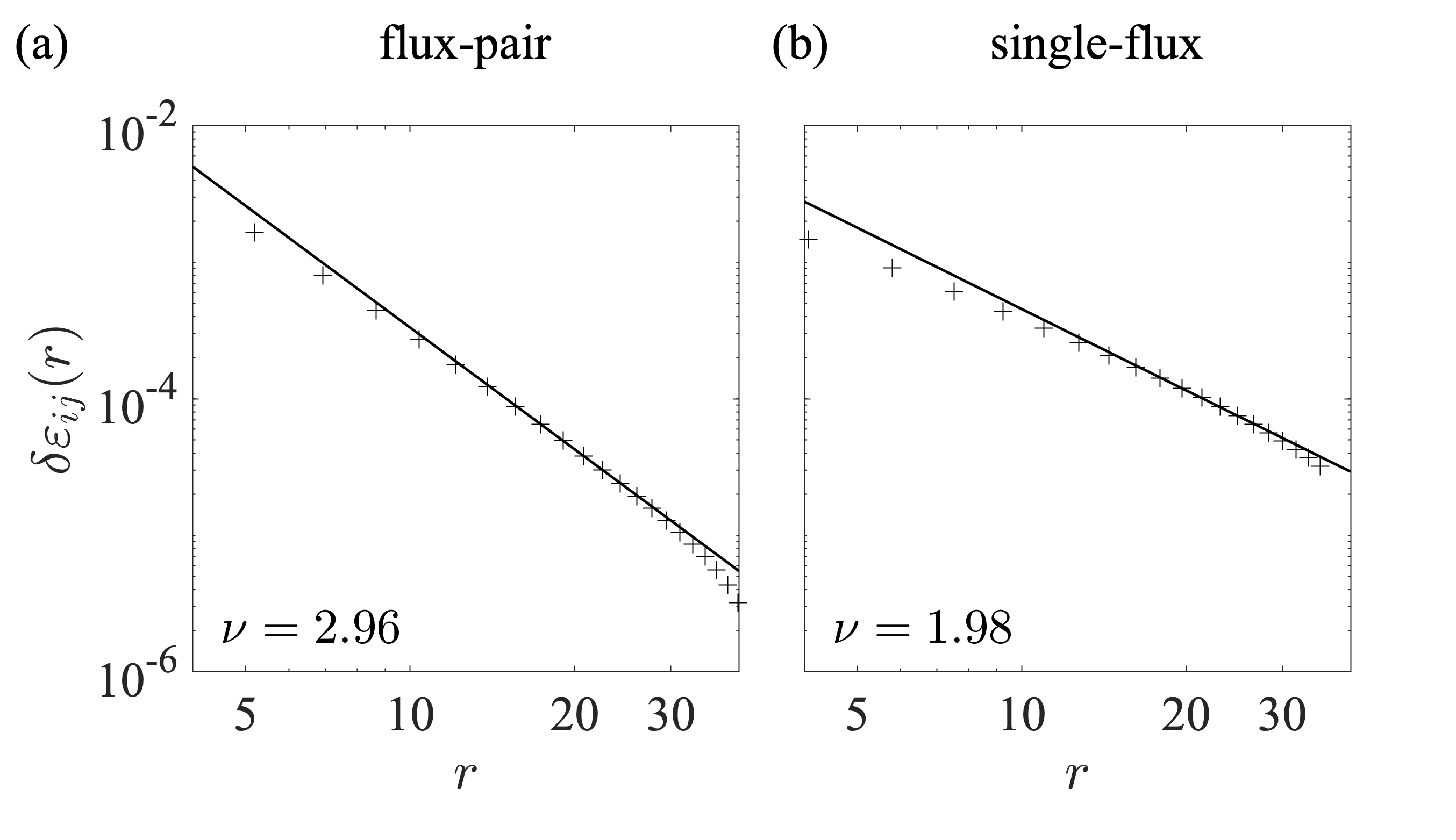}
\caption{Radial dependence of the bond-energy distortion $\delta
\varepsilon_{ij}$ induced by: (a) a flux pair and (b) a single flux. In each case, the asymptotic behavior obeys a power law  $\delta
\varepsilon_{ij}\sim r^{-\nu}$. The exponent $\nu$ is obtained by  fitting the bond-energy distortion
computed on a honeycomb lattice of $80\times 80$ unit cells. The fitting range  $17 < r < 25$ is chosen to capture the asymptotic behavior while minimizing the boundary effects.} \label{fig:locality}
\end{figure}

To verify that a flux pair is a well-defined quasiparticle while a single flux is not, we investigate how the presence of each distorts the matter-fermion vacuum. We first consider the effect of a flux pair. In the pure Kitaev limit, the distortion of the matter-fermion vacuum can be
studied by computing the spatial variation of the bond energy,
$\varepsilon_{ij} = \langle G \rvert K
\sigma_i^{\gamma} \sigma_j^{\gamma} \rvert G\rangle $, where $\rvert
G\rangle$ is the ground state of the corresponding flux sector. For
the flux-free sector, this bond energy is the same for all
bonds by symmetry and is numerically found to be $\varepsilon^0
\simeq -0.5249 \rvert K\rvert$. The gapless nature of the matter
fermions then implies that the distortion of the bond energy,
$\delta \varepsilon_{ij} \equiv \varepsilon_{ij} -\varepsilon^{0}$, induced by a flux
pair, falls off as a power law $r^{-\nu}$ for long distances $r$
between the bond $\langle ij\rangle$ and the flux pair. The
situation is similar to Friedel oscillations
induced by an impurity in half-filled graphene, where the decay exponent is known to be $\nu=3$ in both
sublattices.\cite{Basci10} This power-law behavior is confirmed by
our numerical results [see Fig.~\ref{fig:locality}(a)], 
for the distortion around a flux pair in a specific direction with no oscillatory behavior. The value of the exponent,
$\nu=2.96$, is very close to the expected value, $\nu=3$, for the
true asymptotic behavior, and its small deviation can be attributed  to finite-size effects. In turn, since the 2D integral $\int
d^2{\bm r} \, r^{-\nu}$ is convergent for $\nu > 2$, a
distortion decaying as $1/r^3$ can be regarded as localized.
Therefore, the two matter-fermion vacua connected by the matrix
element in Fig.~\ref{fig:matrix}(b) must have an $\mathcal{O}(1)$ overlap, which is numerically found to be $\simeq 0.78$. From this result, we also expect that matrix elements involving multimatter-fermion states can be safely neglected.

Next, we investigate the effect of a single flux. We start by noting that, for a closed system with periodic boundary conditions, fluxes must always be inserted in pairs because of the global constraint $\prod_{p}W_p = 1$. Therefore, we introduce two fluxes separated by the largest possible distance on a finite lattice, and we only concentrate on one of these fluxes. Similarly to the case of a flux pair, the spatial variation of the bond energy, $\delta \varepsilon_{ij}$, as defined above,
falls off as a power law for long enough distances $r$ between the bond $\langle ij\rangle$ and the single flux. However, in this case, we find an exponent $\nu \simeq 1.98$ [see Fig.~\ref{fig:locality}(b)], which is close to the critical value $\nu = 2$ separating the localized and the delocalized scenarios. 
In other words, the distortion of the matter-fermion vacuum due to a single flux is found to be marginally delocalized.

In view of these considerations, we exclude single-flux excitations from our variational analysis. By doing so, we implicitly assume that single-flux excitations do not play a relevant role in the instabilities induced by the Heisenberg and Gamma terms. Note also that these excitations are not expected to give a substantial contribution to dynamical correlation functions of local observables, such as the dynamical spin structure factor, because local operators necessarily create fluxes in pairs.

\subsection{Computation of matrix elements}
\label{calculation_general}

In this subsection, we explain how to compute the matrix element of
an arbitrary Hamiltonian ${\cal H}^{\prime}$ between two eigenstates
of the pure Kitaev model specified in the form of
Eq.~(\ref{eq:state}). For the purposes of this work, we consider two
different classes of eigenstates. First, there are ``conventional
eigenstates'' containing a single topologically trivial excitation,
for example, a bosonic bound state of a flux pair and a matter
fermion. Second, there are ``fractionalized eigenstates'' containing
a topologically nontrivial (i.e., fractionalized) excitation, for
example, a fermionic bare flux pair. Importantly, for a closed
system, an eigenstate containing a single fractionalized excitation
is unphysical and annihilated by the projector in Eq.~(\ref{eq:P}).
This problem can be circumvented by considering an eigenstate with
two fractionalized excitations: the original excitation of interest
and another excitation as far away as possible. In this work,
however, instead of including another excitation, we modify the
projector in Eq.~(\ref{eq:P}) such that it enforces $D_j = -1$ at a
far-away reference site $j$ and $D_{j'} = +1$ at all other sites
$j'$:
\begin{eqnarray} \label{eq:Pbar}
\bar{P}_j = \frac{1}{2^{N}} (1 - D_j) \prod_{j'} (1 + D_{j'}).
\end{eqnarray}
In contrast to the original projector $P$, this modified projector
$\bar{P}_j$ annihilates states with even fermion parity but not
those with odd fermion parity. Therefore, replacing $P$ with
$\bar{P}_j$ is physically equivalent to including a fermion
excitation around site $j$.

Exploiting $[{\cal H}^{\prime},P]=0$ and $P^{2}=P$, the general
matrix element between two conventional eigenstates reads:
\begin{equation}
\langle\Phi_{2};f_{2}\rvert{\cal
H}^{\prime}\rvert\Phi_{1};f_{1}\rangle = 2^{N}\langle
n^{u_2}\rvert\otimes\langle u_2\rvert P{\cal H}^{\prime}\rvert
u_1\rangle\otimes\rvert n^{u_1}\rangle.\label{eq:recipe1}
\end{equation}
Similarly, due to $[{\cal H}^{\prime},\bar{P}_j]=0$ and
$\bar{P}_j^{2}=\bar{P}_j$, the general matrix element between two
fractionalized eigenstates takes an analogous form:
\begin{equation}
\langle\Phi_{2};f_{2}\rvert{\cal
H}^{\prime}\rvert\Phi_{1};f_{1}\rangle = 2^{N}\langle
n^{u_2}\rvert\otimes\langle u_{2}\rvert\bar{P}_j{\cal
H}^{\prime}\rvert u_{1}\rangle\otimes\rvert
n^{u_1}\rangle.\label{eq:recipe2}
\end{equation}
In each case, the two states connected by ${\cal H}^{\prime}$ may
have different gauge choices $u_1$ and $u_2$ as the appropriate
projector ($P$ or $\bar{P}_j$) automatically takes care of the gauge
transformation between these two gauge choices.

Since the bond fermions are static, the bond-fermion matrix elements
in Eqs.~(\ref{eq:recipe1}) and (\ref{eq:recipe2}) can be calculated
straightforwardly. Thus, Eqs.~(\ref{eq:recipe1}) and
(\ref{eq:recipe2}) are reduced to the general form
$\langle0^{u_2}\rvert \hat{\psi}_{1}
\hat{\psi}_{2}...\hat{\psi}_{2 N}\rvert0^{u_1}\rangle$ in
matter-fermion space, where $\rvert0^{u_{1,2}}\rangle$ is the
matter-fermion vacuum of the gauge sector $\rvert u_{1,2}\rangle$,
and each $\hat{\psi}_{i} = \sum_k (a_k^{i} f_{k}^{(u_1)} + {\bar
a}_k^{i} f_{k}^{(u_1)\dagger} )$ is expressed as a linear
combination of the normal matter fermion modes $f_{k}^{(u_1)}$
and $f_{k}^{(u_1)\dagger}$. The matter-fermion matrix element is
then computed by introducing a modified form of Wick's theorem,
which is necessary because the two states
$\rvert0^{u_1}\rangle$ and $\rvert0^{u_2}\rangle$
correspond to different fermion vacua. In general, these fermion
vacua are related by
\begin{eqnarray}
\rvert0^{u_2}\rangle & = & \rvert C \rvert e^{-\frac{1}{2} \sum_{kl}f^{(u_1)\dagger}_k M^{*}_{kl}f^{(u_1)\dagger}_l }\rvert 0^{u_1} \rangle,\label{eq:gs_trans}
\end{eqnarray}
where the anti-symmetric matrix $M=X^{-1}Y$ and the normalization
constant $C=\det[X]^{1/2}$ are both derived from the unitary
transformation
\begin{eqnarray}
\left(\begin{array}{c}
f_{n}^{(u_2)}\\
f_{n}^{(u_2)\dagger}
\end{array}\right) & = & \left(\begin{array}{cc}
X_{nm}^{*} & Y_{nm}^{*}\\
Y_{nm} & X_{nm}
\end{array}\right)\left(\begin{array}{c}
f_{m}^{(u_1)}\\
f_{m}^{(u_1)\dagger}
\end{array}\right).\label{eq:trans}
\end{eqnarray}
By expanding the exponential in Eq.~(\ref{eq:gs_trans}), the
matter-fermion matrix element becomes
\begin{eqnarray}
&& \langle0^{u_2}\rvert\hat{\psi}_{1}\hat{\psi}_{2}...\hat{\psi}_{2N}\rvert0^{u_1}\rangle \label{eq:expansion} \\
&& = \rvert C \rvert \langle0^{u_1} \rvert \sum_{n=0}^{\infty}
\frac{1}{n!}\left(\frac{1}{2} \sum_{kl} f^{(u_1)}_k M_{kl}
f^{(u_1)}_l
\right)^{n}\hat{\psi}_{1}...\hat{\psi}_{2N}\rvert0^{u_1}\rangle,
\nonumber
\end{eqnarray}
and can be evaluated through Wick's theorem. If $q$ pairs of ${\hat
\psi}$ operators are contracted, the remaining $2N-2q$ ${\hat \psi}$
operators must be paired with $f$ operators. In this case, only
the term of order $n=N-q$ gives a finite contribution because
$\langle0^{u_1} \rvert f^{(u_1)}_k f^{(u_1)}_l
\rvert0^{u_1}\rangle=0$. If we consider the partitions of
$2N-2q$ ${\hat \psi}$ operators into $N-q$ pairs, for each partition
there are $n!=(N-q)!$ equivalent ways of contracting each pair with
$\frac{1}{2}\sum_{kl} f^{(u_1)}_k M_{kl} f^{(u_1)}_l$. This
factor cancels  the prefactor $1/n!= 1/(N-q)!$. Summing over all
possible cases, $ 0 \leq q \leq N$, we then obtain
\begin{widetext}
\begin{eqnarray}
\langle0^{u_2}\rvert\hat{\psi}_{1}\hat{\psi}_{2}...\hat{\psi}_{2N}\rvert0^{u_1}\rangle & = & \rvert C \rvert\sum_{P}\eta_{P}\prod_{q=1}^{N}
\left[\langle0^{u_1}\rvert\hat{\psi}_{P_{q}(1)}\hat{\psi}_{P_{q}(2)}+ \left( \frac{1}{2} \sum_{kl} f^{(u_1)}_k M_{kl} f^{(u_1)}_l \right) \hat{\psi}_{P_{q}(1)}\hat{\psi}_{P_{q}(2)}\rvert0^{u_1}\rangle\right]\nonumber \\
 & = & \rvert C\rvert\sum_{P}\eta_{P}\prod_{q=1}^{N}\langle0^{u_2} \rvert\hat{\psi}_{P_{q}(1)}\hat{\psi}_{P_{q}(2)}\rvert0^{u_1}\rangle,
\end{eqnarray}
\end{widetext}
where $P$ sums over all the possible partitions  of the $2N$ ${\hat
\psi}$ operators into $N$ pairs and $(P_{q}(1),P_{q}(2))$ is the
index of the $q$-th pair. The coefficient $\eta_{P}= {\rm
sgn}{(P)}=\pm1$ arises from the fermionic statistics of the ${\hat
\psi}$ operators. By introducing a skew-symmetric matrix $A$ of
dimension $2N$, whose matrix elements are given by the generalized
contraction $  A_{ij} \equiv \langle0^{u_2}\rvert
\hat{\psi}_{i}\hat{\psi}_{j} \rvert0^{u_1}\rangle \equiv
\sum_{kl} \left(a_{k}^{i} \delta_{kl} -\bar{a}_{k}^{i}M_{kl}
\right)\bar{a}_{l}^{j}$ for $i < j$, the above result takes the
concise form
\begin{eqnarray}
\langle0^{u_2}\rvert\hat{\psi}_{1}\hat{\psi}_{2}...\hat{\psi}_{2N}\rvert0^{u_1}\rangle & = & \rvert C \rvert ~ \text{Pf}\left[A\right],\label{eq:wick}
\end{eqnarray}
where $\text{Pf}\left[A\right]$ is the Pfaffian of the matrix $A$.

\section{Dynamics of bare flux pairs}

\subsection{General formulation}

In this section, we focus on bare flux pairs, which are coherently
propagating fermionic quasiparticles and are topologically
equivalent to bond fermions. The relevant variational subspace
contains all eigenstates of the pure Kitaev model that have two flux
excitations at neighboring plaquettes and no matter-fermion
excitations. Each state from the variational subspace then
corresponds to a bond on the honeycomb lattice, and the state
corresponding to bond $\langle jk \rangle \parallel \gamma$ (with $j \in
A$) can be written as
\begin{eqnarray}
| j, \gamma \rangle & = & 2^{N/2} P \, \big\{ b_{j}^{\gamma} |
u_0 \rangle \otimes | 0^{(j,\gamma)} \rangle
\big\},\label{eq:Wannier-1}
\end{eqnarray}
where $| u_0 \rangle$ is the trivial gauge sector with $u_{\langle
lm \rangle} = +1$ for all bonds $\langle lm \rangle$, while $|
0^{(j,\gamma)} \rangle$ is the matter-fermion vacuum of the
gauge sector $b_{j}^{\gamma} | u_0 \rangle$. The labels $\gamma$ and $j$ specify the orientation (also denoted as ``flavor'') and the position of the flux pair, respectively.

\begin{figure}[t!]
\centering
\includegraphics[scale=0.35]{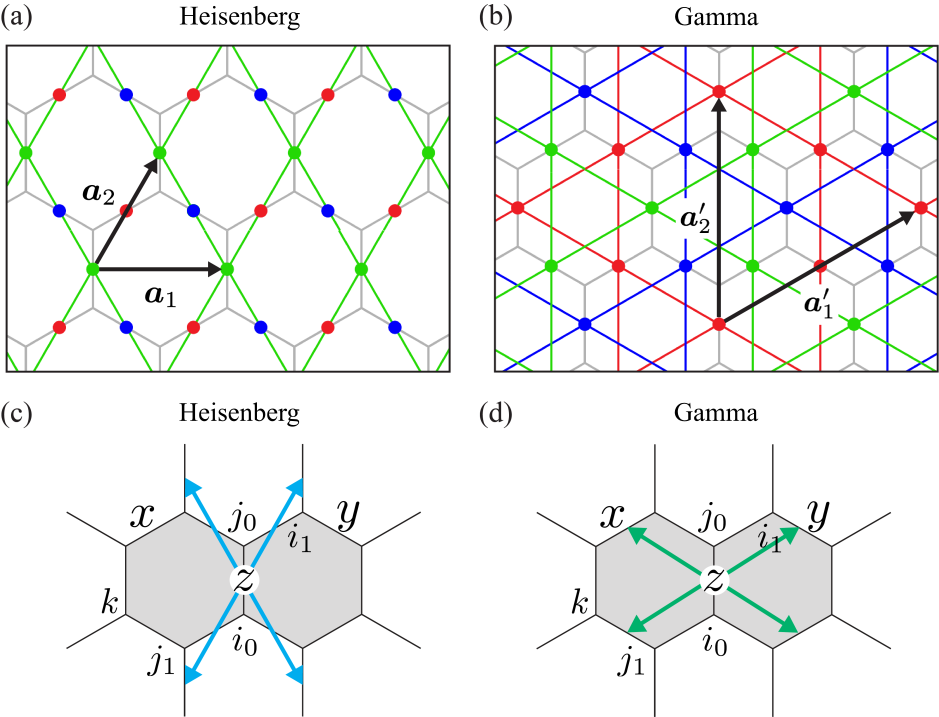}
\caption{Sublattices  of the dual Kagome lattice on which the
flux pair propagates through (a) Heisenberg and (b) Gamma
interactions. The primitive lattice vectors of the triangular
Bravais lattice of the dual Kagome lattice are indicated in each
figure. (c) The blue arrows indicate the non-zero hopping amplitudes
of a flux pair via Heisenberg interaction. (d) The green arrows
indicate the non-zero hopping amplitudes of a flux pair via Gamma
interaction.} \label{dual_lattice}
\end{figure}

The Hamiltonian restricted to the above variational subspace describes a tight-binding problem on the dual Kagome lattice formed by the center of bonds of the honeycomb lattice (see Fig.~\ref{dual_lattice} (a)).
The dual Kagome lattice has a primitive unit cell of three lattice sites corresponding to the bonds with different orientations, which is shown by the three colors (red, blue, green) in Fig.~\ref{dual_lattice} (a).
The primitive vectors ${\bm a}_{1,2}$ coincide with the ones of the original honeycomb lattice. 
The Heisenberg interaction ($J$) generates effective hopping amplitudes within one of the three sublattices of the Kagome lattice that are shown by one particular color in Fig.~\ref{dual_lattice}~(a), which is diagonal in the flavor index of the flux pair.
In contrast, the Gamma interaction ($\Gamma$) propagates the flux pair within one of the three \textit{Kagome} sublattices that are shown by one particular color in Fig.~\ref{dual_lattice}~(b), which is off-diagonal in the flavor index of the flux pair. 
Each sublattice forms an enlarged Kagome lattice spanned by the primitive vectors ${\bm a}_{1,2}^{\prime}$. 
For either $J=0$ or $\Gamma=0$, the respective ``sublattice'' index is a good quantum number, which leads to a triple degeneracy of the flux pair spectrum.


\begin{figure}[t!]
\centering
\includegraphics[scale=0.25]{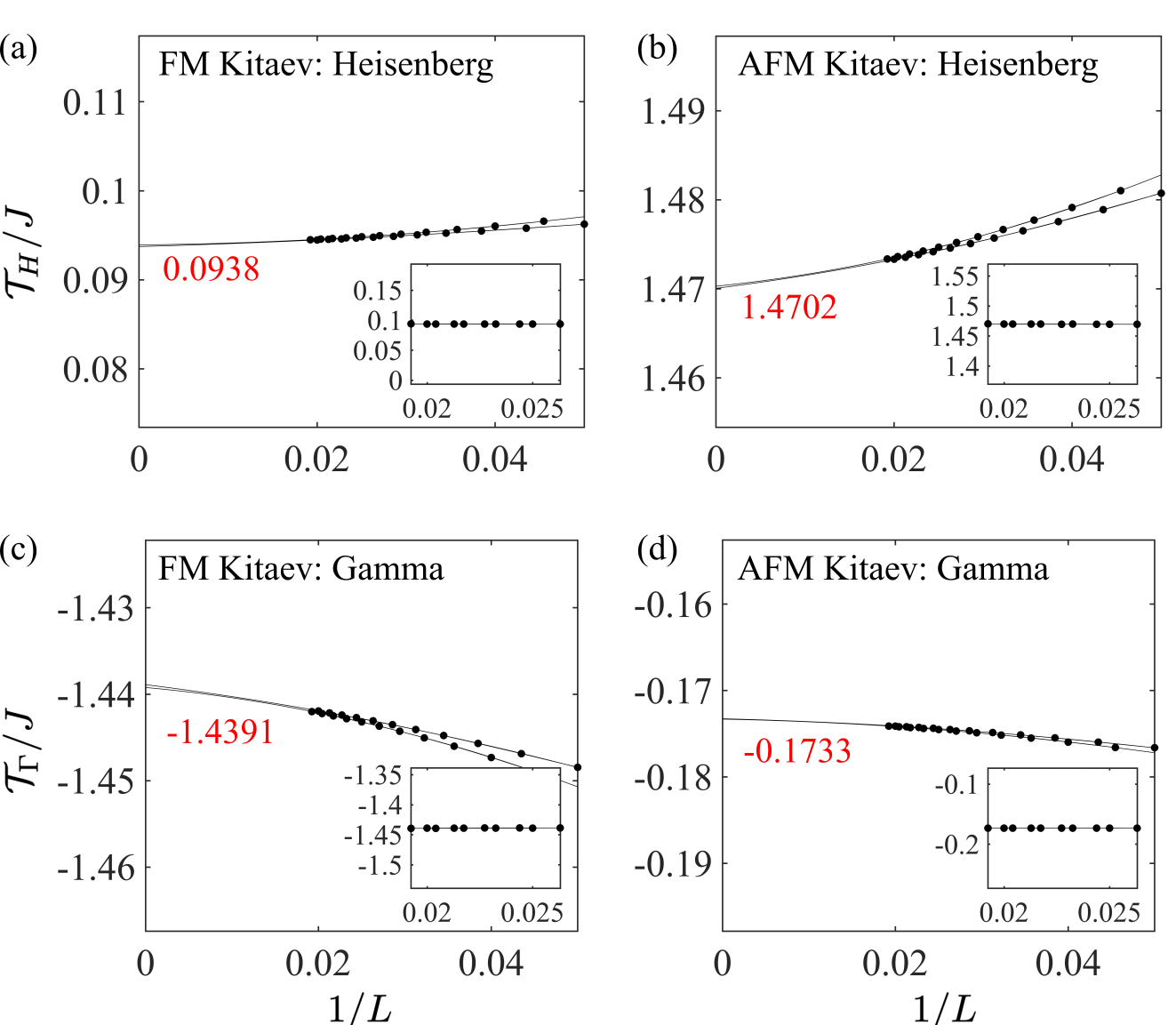}
\caption{Finite-size scaling of the flux-pair hopping amplitudes
under Heisenberg and Gamma interactions on top of the FM and AFM Kitaev models. Insets: Extrapolated hopping amplitudes as a function of $1/L_{\text{max}}$, where $L_{\text{max}}$ refers to the maximal system size used in the finite-size fitting.} \label{hopping_fp_odd}
\end{figure}

To compute each hopping amplitude, we consider a flux pair $\rvert i_0,z\rangle$ on the  bond $\langle i_0j_0 \rangle \parallel z$ (see Fig.~\ref{dual_lattice}~(c) or (d)). 
The hopping amplitude of a flux pair on bonds $x$ or $y$ is related to this case by the three-fold rotation symmetry of the model.
The Heisenberg interaction moves this flux pair to the neighboring sites of the dual (Kagome) lattice with {\it the same} bond orientation. 
The hopping amplitude that takes the flux pair $\rvert i_0,z\rangle$ to $\rvert i_1,z\rangle$
(see Fig.~\ref{dual_lattice}~(c) for lattice site labels) is
\begin{eqnarray}\label{eq:hop1_odd}
T_{i_1,i_0}^{zz} &=& J u_{\langle i_0 j_0\rangle_{z}}^{0} \bigg[ \langle 0^{(i_1,z)} \rvert ic_{i_1}c_{j_0}\rvert 0^{(i_0,z)} \rangle
\nonumber \\
 &+& u_{\langle i_1 j_0\rangle_{x}}^{0}\langle 0^{(i_1,z)}\rvert 0^{(i_0,z)} \rangle \bigg],
\end{eqnarray}
where $i_{\mu}\in A$ and $j_{\mu} \in B$, $\mu=0,1$.
This formula also applies to the other hopping processes related by mirror symmetry about the $z$-bond shown by Fig.~\ref{dual_lattice} (c). The remaining two hopping amplitudes are  obtained by complex conjugation of Eq.~\eqref{eq:hop1_odd}. 
There are in total six independent hopping parameters including all the three flavors of flux pairs. For the isotropic Kitaev model, the six hopping parameters are equal and real:
\begin{eqnarray}
T_{-{\bm a}_1}^{xx} &=& T_{-{\bm a}_2}^{xx} = T_{{\bm a}_1-{\bm a}_2}^{yy} = T_{{\bm a}_1}^{yy} = T_{{\bm a}_2}^{zz}   
=  T_{{\bm a}_2-{\bm a}_1}^{zz}  \!   \equiv {\cal T}_H \!\! \in \mathbb R,
\nonumber \\
\end{eqnarray}
where the subscript denotes the relative vector between the final and the initial flux pair.
The hopping amplitude ${\cal T}_H$ has been computed for FM and AFM Kitaev models as shown in Figs.~\ref{hopping_fp_odd} (a) and (b), which has an apparent system size dependence due to the gapless nature of the matter fermion spectrum.
By fitting the numerical results with a second degree polynomial in  $1/L$, we obtain the following extrapolation of the hopping amplitudes to the thermodynamic limit: ${\cal T}_H\simeq 0.0938 J$ for the FM Kitaev model and ${\cal T}_H\simeq 1.4702 J$ for the AFM Kitaev model.
It is interesting to note that the two amplitudes (for the same value of $J$) are different by one order of magnitude between the FM and AFM Kitaev models.

Through the Gamma interaction, the flux pair $\rvert i_0,z\rangle$ propagates to its four second nearest neighbors of the dual (Kagome) lattice as shown in Fig.~\ref{dual_lattice}~(d).
The one that takes the flux pair $\rvert i_0,z\rangle$ to $\rvert i_1,y\rangle$ 
reads
\begin{eqnarray}\label{eq:hop2_odd}
T_{i_1,i_0}^{yz}  &=& \Gamma u_{\langle i_0 j_0 \rangle _{z} }^{0}  \bigg[ \langle 0^{(i_1,y)} \rvert ic_{i_1}c_{j_0}\rvert 0^{(i_0,z)} \rangle
\nonumber \\
& - &  u_{\langle i_1 j_0\rangle _{x} }^{0}\langle 0^{(i_1,y)}\rvert 0^{(i_0,z)} \rangle \bigg] .
\end{eqnarray}
The other hopping amplitude that takes the flux pair $\rvert i_0,z\rangle$ to $\rvert k,y\rangle$ reads
\begin{eqnarray}\label{eq:hop3_odd}
T_{k,i_0}^{yz}  &=& \Gamma u_{\langle k j_1 \rangle _{\gamma} }^{0}  \bigg[ \langle 0^{(k,y)} \rvert ic_{i_0}c_{j_1}\rvert 0^{(i_0,z)} \rangle
\nonumber \\
 &-&  u_{\langle i_0 j_1 \rangle_{x} }^{0}\langle 0^{(k,y)}\rvert 0^{(i_0,z)} \rangle \bigg].
\end{eqnarray}
The above two formulas also apply to the remaining two hopping processes related by mirror symmetry about the $z$-bond.
There are also six independent hopping parameters. For the isotropic Kitaev model, the six hopping parameters are equal and real:
\begin{eqnarray}
T_{{\bm a}_2}^{yz} &=& T_{-{\bm a}_1}^{yz} = T_{-{\bm a}_1+{\bm a}_2}^{xz} = T_{{\bm a}_1}^{xz} = T_{{\bm a}_2}^{xy}   
=  T_{{\bm a}_1-{\bm a}_2}^{xy} \! \equiv {\cal T}_{\Gamma} \!\! \in \mathbb R.
\nonumber \\
\end{eqnarray}
The finite size scaling of $ {\cal T}_{\Gamma}$ is shown in Figs.~\ref{hopping_fp_odd} (c) and (d). The extrapolation to the thermodynamic limit
gives  ${\cal T}_{\Gamma} \simeq -1.4391 \Gamma$ for the FM Kitaev model and ${\cal T}_{\Gamma} \simeq -0.1733 \Gamma$ for the AFM Kitaev model.
As in the previous case, the two amplitudes differ by one order of magnitude.

\subsection{Asymmetry between FM and AFM Kitaev models}

We have observed that there is an apparent difference in magnitude between the hopping amplitudes of the flux pairs for the FM and AFM Kitaev models.
As we demonstrate below, this asymmetry arises from the fact that each flux-pair hopping process has two contributions which interfere constructively for one sign of  Kitaev interaction and destructively for the other.

Without loss of generality, we consider the matrix element shown in Fig.~\ref{fig:matrix} (b) due to the Heisenberg interaction
on the $x$-bond. This matrix element has contributions from the two spin-exchange interactions, $J \sigma_i^y \sigma_j^y$ and $J\sigma_i^z \sigma_j^z$, denoted as ${\cal T}_y$ and $ {\cal T}_z$, respectively.
Note that the FM and AFM Kitaev interactions are related to each other by a four-site unitary transformation.
The two Heisenberg interactions acquire a relative ``-'' sign under this transformation, namely,
$\sigma_i^y \sigma_j^y \rightarrow -\sigma_i^y \sigma_j^y$ and $\sigma_i^z \sigma_j^z \rightarrow \sigma_i^z \sigma_j^z$.
It then follows that, if ${\cal T}^{FM} = {\cal T}_y + {\cal T}_z$ for the FM Kitaev model, we have ${\cal T}^{AFM} = {\cal T}_y - {\cal T}_z$ for the AFM Kitaev model.
Given the similar absolute values of ${\cal T}_y$ and ${\cal T}_z$, the net hopping amplitudes
${\cal T}^{FM}$ and ${\cal T}^{AFM}$ are dramatically different [see Figs.~\ref{hopping_fp_odd} (a) and (b)].
The relative sign between ${\cal T}_y$ and ${\cal T}_z$  can be determined using a simple argument.
We will denote the initial and final flux-pair configurations by $\Phi_1$ and $\Phi_2$, and their respective ground states
by $\rvert \Phi_1; \omega_1 \rangle$ and $\rvert \Phi_2; \omega_2 \rangle$.
In the spirit of our approximation scheme, we can estimate the product of the two matrix elements:
\begin{eqnarray}\label{eq:interference_Heis}
{\cal T}_y {\cal T}_z &=& J^2 \langle \Phi_1; \omega_1 \rvert \sigma_i^y \sigma_j^y \rvert \Phi_2; \omega_2 \rangle \langle \Phi_2; \omega_2 \rvert \sigma_i^z \sigma_j^z\rvert \Phi_1; \omega_1 \rangle \nonumber \\
&\simeq & J^2 \sum_{f_2} \langle \Phi_1; \omega_1 \rvert \sigma_i^y \sigma_j^y \rvert \Phi_2;f_2 \rangle \langle \Phi_2; f_2 \rvert \sigma_i^z \sigma_j^z\rvert \Phi_1;\omega_1 \rangle \nonumber \\
&=&  -J^2 \langle \Phi_1;\omega_1\rvert \sigma_i^x \sigma_j^x \rvert \Phi_1;\omega_1 \rangle \nonumber \\
&=&  -{J^2\over K} \varepsilon_{ij}^{\Phi_1} = 0.5851 {J^2} \text{sign}(K),
\end{eqnarray}
where, in the second step, the ground state projector $\rvert \Phi_2; \omega_2 \rangle \langle \Phi_2;\omega_2\rvert$ is replaced with the identity $\sum_{f_2} \rvert \Phi_2; f_2 \rangle \langle \Phi_2; f_2 \rvert = I$ in the gauge sector $\Phi_2$ ($f_2$ runs over all the eigenstates of the pure Kitaev model in the gauge sector $\Phi_2$).
We note that this approximation is basically the same as the one that we are using in our variational scheme, where we ignore the tunneling of $\rvert \Phi_1;\omega_1 \rangle $ into multi-matter-fermion state $\rvert \Phi_2;f_2 \neq \omega_2 \rangle $.
The simple outcome is that the above product is proportional to the bond energy $\varepsilon_{ij}^{\Phi_1}$ on the bond where the Heisenberg interaction applies in the ground state $\rvert \Phi_1;\omega_1 \rangle$ of the pure Kitaev model.
By following the scheme described in the previous subsection, we obtain ${\cal T}_y {\cal T}_z \simeq 0.5382 {J^2} \text{sign}(K)$, which is very close to the value obtained in Eq.~\eqref{eq:interference_Heis}. Besides providing a sanity check for the approximation scheme adopted in this work,  this simple analysis explains that ${\cal T}_y$ and ${\cal T}_z$ have \textit{opposite} sign for the FM Kitaev model and the \textit{same} sign for the AFM Kitaev model.

Now we consider the Gamma interaction acting on the $x$- bond, that also  includes two terms: $J\sigma_i^y \sigma_j^z$ and $J\sigma_i^z \sigma_j^y$.
Both terms contribute to the matrix elements shown in Fig.~\ref{fig:matrix_gamma} (b), which are denoted as ${\cal T}_{yz} $ and $ {\cal T}_{zy}$, respectively.
Under the same unitary transformation that relates the FM and AFM Kitaev model,
the above two Gamma terms transform according to: $\sigma_i^y \sigma_j^z \rightarrow -\sigma_i^y \sigma_j^z$ and $\sigma_i^z \sigma_j^y \rightarrow \sigma_i^z \sigma_j^y$. Therefore,  ${\cal T}_{yz} $ and $ {\cal T}_{zy}$ have  opposite relative sign for the FM and AFM Kitaev models.
Once again, this relative sign is determined by the sign of the Kitaev interaction:
\begin{eqnarray}\label{eq:interference_gamma}
{\cal T}_{yz} {\cal T}_{zy} \simeq  {\Gamma^2\over K} \varepsilon_{ij}^{\Phi_1} =  - 0.5851 {\Gamma^2} \text{sign}(K).
\label{apTyz}
\end{eqnarray}
Thus, ${\cal T}_{yz}$ and ${\cal T}_{zy}$ have the \textit{same} sign for a FM Kitaev model and \textit{opposite} sign for an AFM Kitaev model.
According to the calculation, ${\cal T}_{yz} {\cal T}_{zy} \simeq -0.5102 {\Gamma^2} \text{sign}(K)$ which is in a good agreement with the approximated value given in Eq.~\eqref{apTyz}.

\begin{figure*}[t!]
\centering
\includegraphics[scale=0.5]{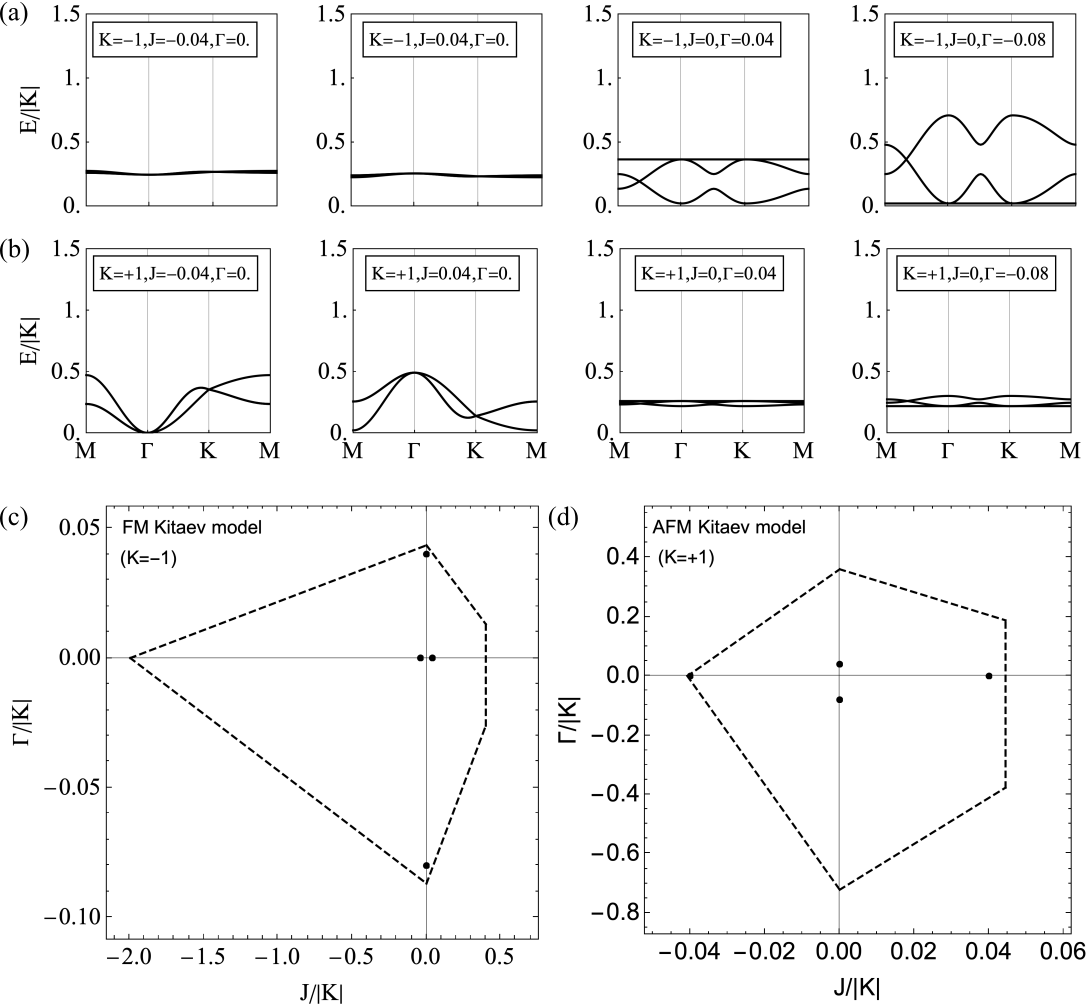}
\caption{(a)-(b) Dispersion of a bare flux pair (corresponding to the odd-fermion-parity
sector) for the FM (a) and AFM (b) Kitaev models with four representative perturbations indicated by black dots in
panels  (c) and (d). (c)-(d) Stability ranges of the FM (c) and AFM (d) Kitaev spin liquids based on a partial variational subspace restricted to bare flux pairs. The dashed lines mark the boundary of each Kitaev spin liquid where bare flux pairs become soft.} \label{pd_odd}
\end{figure*}

\subsection{Stability analysis}

The projection of the full Hamiltonian in Eq.~(\ref{eq0})
to the subspace spanned by the states \eqref{eq:Wannier-1} leads to the tight-binding Hamiltonian:
\begin{eqnarray}
{\cal H}_{odd} & = & \Delta \sum_{\alpha,{\bm r} \in A} \tilde{\chi}_{\bm{r}}^{\alpha\dagger} \tilde{\chi}^{\alpha}_{\bm{r}} \nonumber \\
&+& \left[\sum_{\alpha\beta,\bm{\delta}_{\beta \alpha}}\sum_{{\bm r} \in A} {\cal T}^{\beta\alpha} \tilde{\chi}^{\beta\dagger}_{\bm{r}+\bm{\delta}_{\beta \alpha}} \tilde{\chi}^{\alpha}_{\bm{r}}
+ h.c.\right],
\label{eq:Hodd}
\end{eqnarray}
where $\Delta\simeq 0.26 |K+J|$ is the energy gap of the flux-pair excitation of the pure Kitaev model, $\tilde{\chi}^{\alpha\dagger}_{{\bm r}_j},\tilde{\chi}^{\alpha}_{{\bm r}_j}$ are fermionic creation and annihilation operators of a flux pair state $\rvert {\bm r}_j, \alpha \rangle \equiv \tilde{\chi}^{\alpha\dagger}_{{\bm r}_j} \left( \rvert u_0 \rangle \otimes \rvert 0^{u_0} \rangle \right)$ ($u_0$ refers to the trivial gauge sector as in Eq.~(\ref{eq:Wannier-1})) which inserts a flux-pair and simultaneously polarizes the matter fermion vacuum,
${\cal T}^{\beta=\alpha}={\cal T}_H$ and ${\cal T}^{\beta\neq\alpha}={\cal T}_{\Gamma}$, 
the bond vectors $\bm{\delta}_{\beta \alpha}$ are summed over the sets:
${\bm \delta}_{xx} = \{-{\bm a}_2,-{\bm a}_1\}$,
${\bm \delta}_{yy} = \{ {\bm a}_1,{\bm a}_1-{\bm a}_2 \}$, ${\bm \delta}_{zz} = \{ -{\bm a}_1+{\bm a}_2,  {\bm a}_2 \}$, ${\bm \delta}_{yz} = \{ {\bm a}_2, -{\bm a}_1 \}$,
${\bm \delta}_{xz} = \{ -{\bm a}_1+{\bm a}_2, {\bm a}_1 \}$ and ${\bm \delta}_{xy} = \{ {\bm a}_2, {\bm a}_1-{\bm a}_2 \}$.
This tight-binding Hamiltonian can be diagonalized in  momentum space:
\begin{equation}
{\cal H}_{odd} = \sum_{\bm q} \tilde{X}_{\bm q}^{\dagger} \left( \Delta I_{3\times 3}+ M_{{\bm q}}\right) \tilde{X}_{\bm q},
\end{equation}
where $I_{3\times3}$ is the $3\times 3$ unit matrix, $\tilde{X}_{\bm q}\equiv (\tilde{X}_{{\bm q}}^x,\tilde{X}^y_{{\bm q}},\tilde{X}^z_{{\bm q}})^T$ with
\begin{equation}
\tilde{\chi}^{\alpha}_{\bm r} = \sqrt{{2\over N}} \sum_{{\bm q}} \tilde{X}^{\alpha}_{ {\bm q}} e^{i{\bm q}\cdot{\bm r}},
\end{equation}
and
\begin{widetext}
\begin{equation}
M_{{\bm q}}=\left(\begin{array}{ccc}
2{\cal T}_H [\cos(q_1)+\cos(q_2)] & {\cal T}_{\Gamma} [e^{-iq_2} + e^{-i(q_1-q_2)}]   &   {\cal T}_{\Gamma} [e^{i(q_1-q_2)} + e^{-iq_1}]  \\
 {\cal T}_{\Gamma} [e^{iq_2} + e^{i(q_1-q_2)}]   &   2{\cal T}_H [\cos(q_1)+\cos(q_1-q_2)]      &    {\cal T}_{\Gamma} [e^{-iq_2} + e^{iq_1}]  \\
 {\cal T}_{\Gamma} [e^{-i(q_1-q_2)} + e^{iq_1}]     &   {\cal T}_{\Gamma} [e^{iq_2} + e^{-iq_1}]      &    2{\cal T}_H [\cos(q_1-q_2)+\cos(q_2)]
\end{array}\right).
\end{equation}
\end{widetext}
The resulting   tight-binding spectrum of the flux-pair is shown in Figs.~\ref{pd_odd}~(a) and (b) for different sets of model parameters $(K=\pm1,J,\Gamma=0)$, $(K=\pm1,J=0,\Gamma)$.
As expected, for equal strengths of either $J$ or $\Gamma$, the bandwidth  is very different between the
FM and AFM Kitaev models. 
Interestingly, in the  $J=0$ limit, the flux pair propagates through the Gamma interaction on one of the three Kagome sublattices shown in Fig.~\ref{dual_lattice}~(b). It is well-known that the  spectrum of this tight-binding problem hosts a completely flat band, which corresponds to a localized single-particle state that circulates around the six lattice sites of a hexagon and carries momentum $\pi$. This flat band becomes the lowest energy band for $\Gamma<0$.

The tight-binding spectrum becomes gapless for a critical strength of the perturbations, indicating a potential instability of the Kitaev liquid state.
As we will see in the next section, a (bosonic) bound state between a flux pair and a matter fermion can become gapless for a lower strength of the perturbations. While more details  of the instability analysis are presented in the next section, here we aim to provide a qualitative understanding of the stability of the FM or AFM Kitaev spin liquid. 
Fig.~\ref{pd_odd} (c) and (d) show the boundaries in the $J-\Gamma$ space where the flux-pair spectrum becomes gapless. According to this analysis, the AFM Kitaev spin liquid is  more fragile against the inclusion of a Heisenberg term,
while the FM Kitaev spin liquid is more fragile against the inclusion of a Gamma term.
Both results are entirely consistent with numerical results pulished in previous works~\cite{gotfryd2017phase,gohlke2017dynamics}.
As we will see in Sec.~\ref{dynamics}, besides the stability analysis,  the flux pair dispersion induced by the Heisenberg  and Gamma terms determines the dispersion of the continuum of excited states that contribute to the dynamical spin structure factor.




\subsection{Hybridization with open flux pairs} \label{sec-open}

\begin{figure}[h!]
\centering
\includegraphics[scale=0.4]{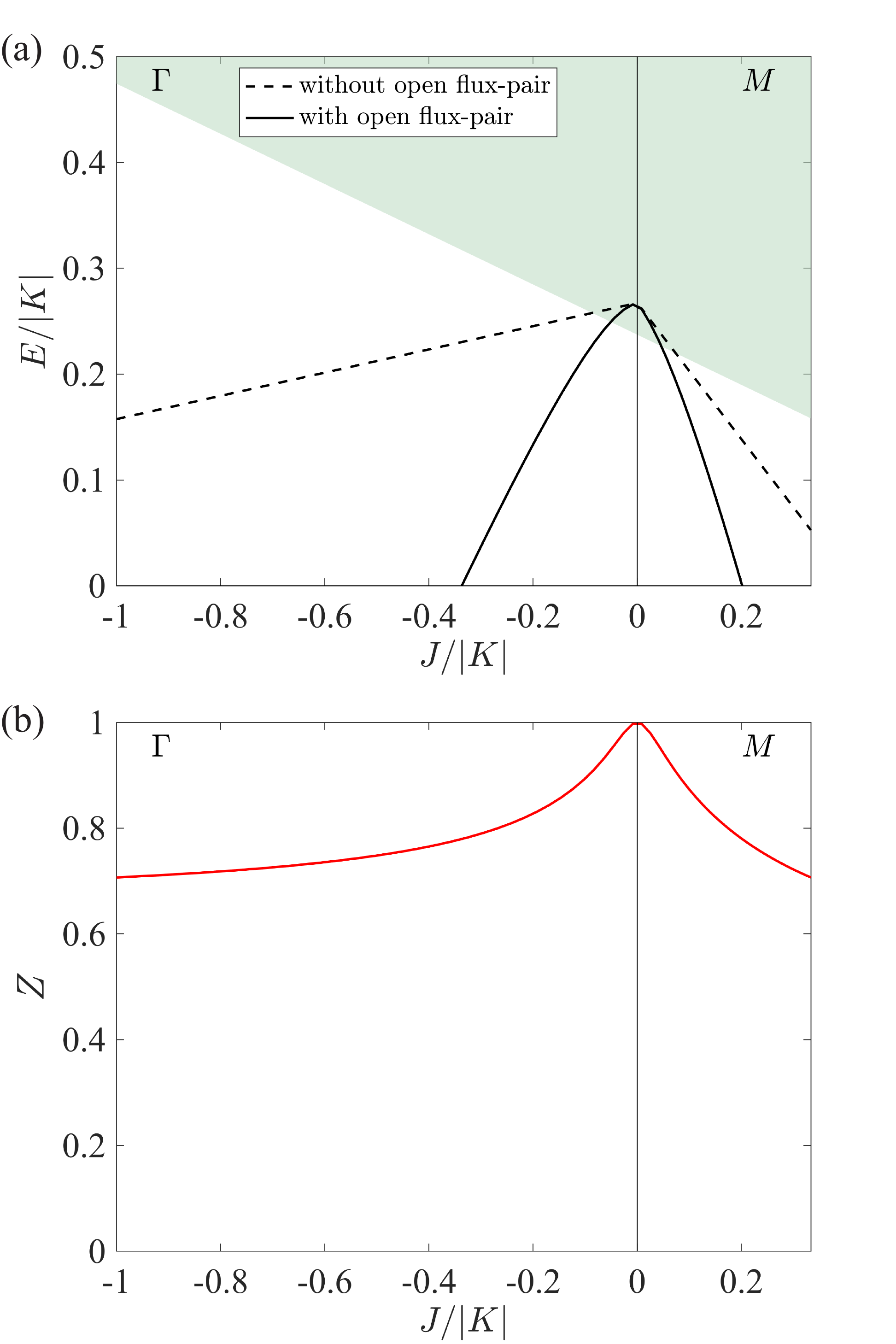}
\caption{(a) Minimal energy of a closed flux pair for
the FM Kitaev model in the presence/absence of hybridization with
the open flux pair continuum (green shaded region). (b) Quasi-particle residue of a
closed flux pair hybridized with the open flux pair continuum.} \label{open_FP}
\end{figure}

Finally, we discuss the  quantum tunneling between the closed and open flux pairs shown in Fig.~\ref{fig:matrix}~(c).
Both perturbations, $J$ and $\Gamma$, contribute to this  tunneling.
Given that the two flux sectors have ground states with opposite fermion parities,
the fermion vacuum of the closed flux sector can tunnel into a continuum of excited states with odd number of fermions in the open flux sector.
In the pure Kitaev limit,  the closed flux pair has  higher excitation energy than the open flux pair, implying that the closed flux pair is immersed in the matter fermion continuum of the open flux sector. The energy difference between the two excitations is $\delta \simeq 0.03K$.

For weak tunneling, $\vert \lambda(\varepsilon\sim \delta) \rvert  \ll \delta$,
the lifetime of the  closed flux pair is $\tau \propto 1/[\rho(\delta) \rvert \lambda(\delta)\rvert^{2}] \propto 1/[\delta \rvert \lambda(\delta)\rvert^{2}]$,
where $\rho(\varepsilon)\propto \varepsilon$ is the density of states of the matter fermions around the Dirac point.
A stronger tunneling can eventually push the energy  of a closed flux pair  below the continuum.
Fig.~\ref{open_FP}~(a) shows the matter fermion continuum of the open flux pair and the energy level of the closed flux pair as a function of $J$
for the FM Kitaev model and for two different values of the momentum of the flux pair.
The tunneling between the two sectors produces a large renormalization of the energy of the closed flux pair.
In particular, the spectrum of the flux pair becomes gapless for a smaller absolute  value of the critical Heisenberg interaction $J_c$.
For instance, $J_c$ reduces from $0.40\rvert K\rvert $ to $0.20\rvert K\rvert $ for an AFM Heisenberg interaction, and from $2.0\rvert K\rvert $ to $0.34 \rvert K\rvert $ for a FM Heisenberg interaction.
As shown in Fig.~\ref{open_FP}~(b), the quasi-particle residue of the closed flux pair remains above $70\%$ before the mode becomes soft.


\section{Magnon-like bound states}

\subsection{General formulation}

In this section, we aim to describe topologically trivial bound
states of flux pairs and matter fermions. The relevant variational
subspace contains all eigenstates of the pure Kitaev model that have
two flux excitations at neighboring plaquettes and a single
matter-fermion excitation. In the notation of
Eq.~(\ref{eq:Wannier-1}), a general state from this variational
subspace can then be written as
\begin{eqnarray}
| j, \gamma, n \rangle & = & 2^{N/2} P \, \big\{ f_n^{(j,\gamma)\dagger}  \chi_{j}^{\gamma \dagger} | u_0 \rangle \otimes | 0^{(j,\gamma)} \rangle \big\},\label{eq:Wannier-2}
\end{eqnarray}
where $j \in A$, $f_n^{(j,\gamma)}$ (labeled by $n$) are the matter fermions in the gauge sector $b_{j}^{\gamma} | u_0 \rangle$ specified by the superscript $(j,\gamma)$ and $u_0$ refers to the trivial gauge sector. Generally,
the matrix elements of the Heisenberg and Gamma terms that connect
these eigenstates give rise to independently propagating flux pairs
and matter fermions. However, due to the interactions generated by
such non-Kitaev terms, these excitations may also form bosonic
(magnon-like) bound states.

Once again, the low-energy dynamics of flux pairs is governed by an
effective low-energy Hamiltonian that results from projecting
$\mathcal{H}$ on the subspace spanned by the  basis (\ref{eq:Wannier-2}):
\begin{eqnarray}\label{H_even}
{\cal H}_{even}^{j,\beta,m; i,\alpha,n}= \langle j,\beta,m \rvert ({\cal H}_K + {\cal H}_H + {\cal H}_{\Gamma}) \rvert i,\alpha,n \rangle.
\end{eqnarray}
The Kitaev term is diagonal on this basis, $\langle j,\beta,m \rvert {\cal H}_K\rvert i,\alpha,n \rangle = \delta_{\beta \alpha}\delta_{ij}\delta_{nm}(\Delta+\varepsilon_n)$. 
In contrast, the Heisenberg and Gamma interactions generate off-diagonal elements of ${\cal H}_{even}$, namely a hopping matrix $({T}_{ij}^{\beta  \alpha})_{mn} =  \langle j,\beta,m \rvert ({\cal H}_H + {\cal H}_{\Gamma}) \rvert i,\alpha,n \rangle$ which is a generalization of the hopping amplitudes $T_{ij}^{\beta \alpha} = \langle j,\beta \rvert ({\cal H}_H + {\cal H}_{\Gamma}) \rvert i,\alpha \rangle$ introduced for a flux pair.
For concreteness, the hopping matrix associated with the flux-pair hopping processes in Fig.~\ref{dual_lattice}~(c,d) is given by (the lattice site labels are shown in the figures)
 \begin{eqnarray}\label{eq:hop1_even}
   \left(T_{i_1,i_0}^{zz}\right)_{mn} && = J u_{\langle i_0 j_0\rangle_{z}}^{0} \left[  \right. \nonumber \\
   && \left. \langle 0^{(i_1,z)} \rvert f_m^{(i_1,z)} ic_{A,i_1}c_{B,j_0} (f_n^{(i_0,z)})^{\dagger}\rvert 0^{(i_0,z)} \rangle \right.  \nonumber \\
    && \left. + u_{\langle i_1 j_0\rangle_{x}}^{0}\langle 0^{(i_1,z)} \rvert f_m^{(i_1,z)} (f_n^{(i_0,z)})^{\dagger} \rvert 0^{(i_0,z)} \rangle\right],
\end{eqnarray}
\begin{eqnarray}\label{eq:hop2_even}
 \left(T_{i_1,i_0}^{yz}\right)_{mn}  &&= \Gamma u_{\langle i_0 j_0 \rangle _{z} }^{0}  \left[ \right. \nonumber \\
 && \left. \langle 0^{i_1,y} \rvert f_m^{(i_1,y)} ic_{A,i_1}c_{B,j_0} (f_n^{i_0,z})^{\dagger} \rvert 0^{i_0,z} \rangle \right. \nonumber \\
&& \left. -  u_{\langle i_1 j_0\rangle _{x} }^{0}\langle 0^{(i_1,y)} \rvert f_m^{(i_1,y)} (f_n^{i_0,z})^{\dagger}  \rvert 0^{(i_0,z)} \rangle \right],
\end{eqnarray}
\begin{eqnarray}\label{eq:hop3_even}
 \left(T_{k,i_0}^{yz}\right)_{mn}  &&=  \Gamma u_{\langle k j_1 \rangle _{z} }^{0}  \left[ \right. \nonumber \\ 
 && \left. \langle 0^{(k,y)} \rvert f_m^{(k,y)} ic_{A,i_0}c_{B,j_1} (f_n^{(i_0,z)})^{\dagger} \rvert 0^{(i_0,z)} \rangle  \right. \nonumber \\
 &&  \left. -  u_{\langle i_0 j_1 \rangle_{x} }^{0}\langle 0^{(k,y)} \rvert f_m^{(k,y)} (f_n^{(i_0,z)})^{\dagger}  \rvert 0^{(i_0,z)} \rangle \right].
\end{eqnarray} 
The hopping matrix associated with the other symmetry related hopping processes  can be  derived in a similar way.
The translational invariance of the above Hamiltonian ${\cal H}_{even}$ implies that the center of mass momentum ${\bm Q}$ is a good quantum number. Correspondingly, we introduce the Fourier transform of the basis $\rvert j,\gamma,n\rangle$ in Eq.~(\ref{eq:Wannier-2}):
\begin{eqnarray}
\rvert \bm{Q}, \gamma, n \rangle & \equiv & \sqrt{{2\over N}} \sum_{j \in A} e^{i\bm{Q}\cdot\bm{r}_j} \rvert j, \gamma, n \rangle,
\label{eq:Wannier-3}
\end{eqnarray}
where ${\bm r}_j$ refers to the coordinate of lattice site $j$. For a given center of mass momentum ${\bm Q}$, the diagonalization of ${\cal H}_{even}$ produces the eigenvectors $\sum_{n,\gamma} c_{{\bm Q},p}(n,\gamma)\rvert \bm{Q}, \gamma, n \rangle$ with energy eigenvalues  $E_p({\bm Q})$, $p=1,2,...,3N/2$.

\subsection{Formation of bound states}
\label{sec:bound_state_B}

\begin{figure*}[t!]
\centering
\includegraphics[scale=0.5]{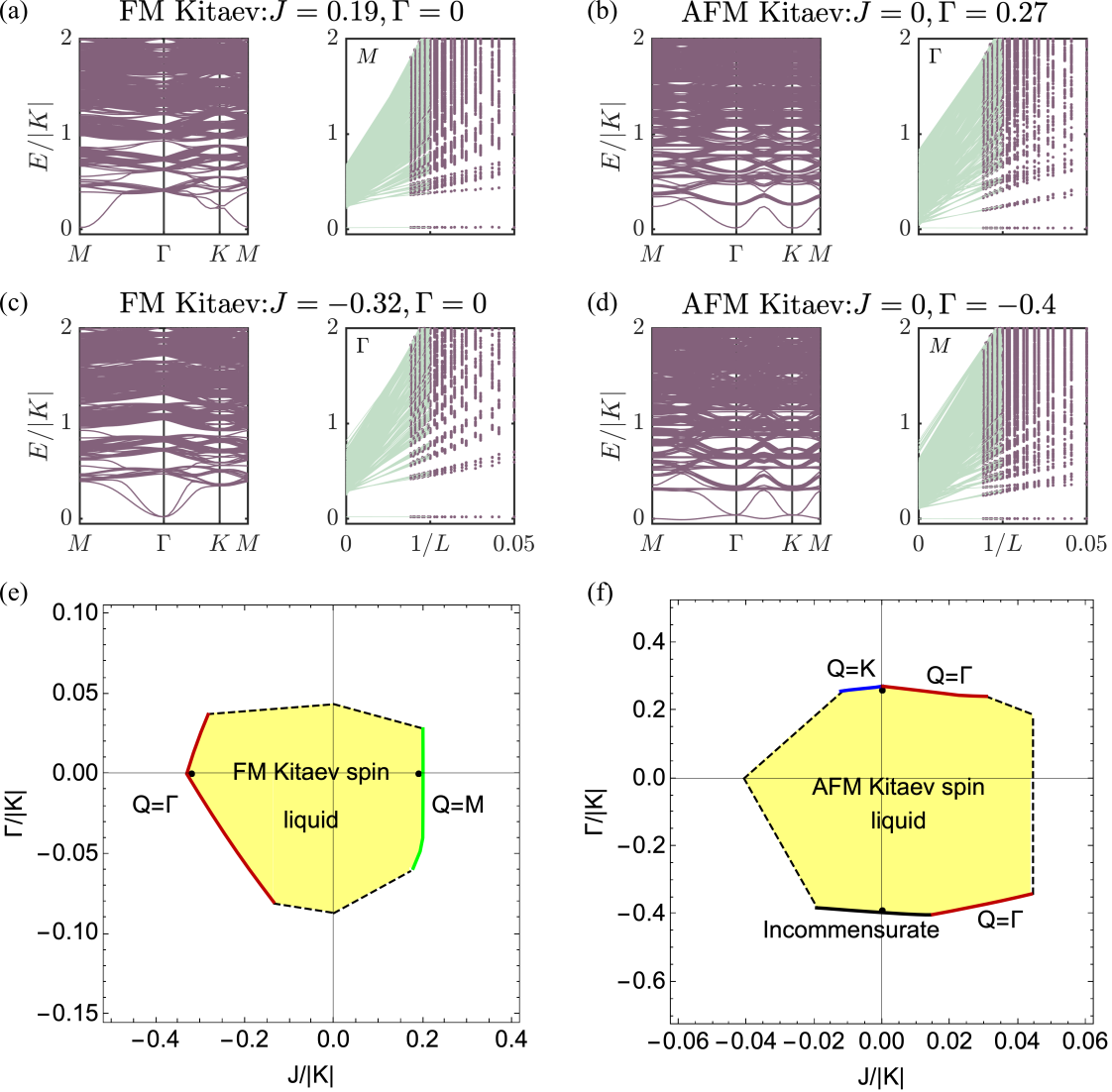}
\caption{(a)-(d) Two-body energy spectrum of a flux pair and a matter
fermion (corresponding to the even-fermion-parity sector) for the FM (a),(c) and AFM (b),(d) Kitaev
models with four representative perturbations indicated by black dots in panels (e) and (f). The right panels
of (a)-(d) show the finite size scaling of the spectrum for the wave
vector that minimizes the excitation energy (the maximum linear size
of the finite lattices is  $52\times 52$ unit cells). (e)-(f) Stability ranges of the FM (e) and AFM (f) Kitaev spin liquids. The dashed lines indicate phase boundaries where bare flux pairs become gapless [see also Figs.~\ref{pd_odd} (c) and (d)], while the solid lines denote phase boundaries induced by a softening of a magnon-like bound state between a flux pair and a matter fermion. The color of each line segment shows the center-of-mass momentum ${\bm Q}$ of the bound state.} \label{pd_even}
\end{figure*}

The spectrum of ${\cal H}_{even}$ forms a flat continuum (localized flux pair plus gapless matter fermions) in  the pure Kitaev limit because of the static nature of the fluxes. However, it becomes dispersive upon  addition of
the non-Kitaev terms because  the flux pair can propagate coherently through the honeycomb lattice.
As we discussed before, the Heisenberg perturbation leads to a spectrum that is much more dispersive for the AFM Kitaev
model than for the FM Kitaev model.
For the FM case, i.e., for the Kitaev liquid state that is more robust against the perturbation,
a discrete level appears below the continuum
when the strength of the Heisenberg interaction exceeds a critical value $J=J_c$  [see Fig.~\ref{pd_even}~(a)].
For $J<0$, we obtain $J_c =-0.25\rvert K \rvert$ and the minimum of the bound state dispersion is located at the $\Gamma$ point. For $J>0$, the critical Heisenberg exchange is $J_c = 0.16\rvert K \rvert$ and the minimum of the bound state dispersion is located at the $M$ point.

We note that extended states  have a rather strong finite size dependence.
Fig.~\ref{pd_even}~(a) shows the finite size scaling of the spectrum at the
$\Gamma$  point, which is the wave vector that minimizes the bound state dispersion.
In the continuum limit, the size dependence of the energy levels follows from the linear dispersion $E_p({\bm Q};L) = E_p({\bm Q}) + A_p L^{-1}$. By contrast, the energy of the bound state has negligible dependence on $L$ because the linear size of the
bound state  is much smaller than the $L$ values used in our  calculations.
The  $L\rightarrow \infty$ extrapolation of the finite size results  indicates  that the binding energy
(gap between the bound state and the continuum) remains finite in the thermodynamic limit.

The positive and negative signs of the Heisenberg interaction are related by a four sublattice spin rotation~\cite{chaloupka2010kitaev,khaliullin2005orbital,schaffer2012quantum} that
produces the following transformations:  ${\cal H}_H \rightarrow -{\cal H}_H + 2{\cal H}_K$ and ${\cal H}_K \rightarrow {\cal H}_K$.
This unitary operation then establishes a correspondence between points of the quantum phase diagram with exactly the same energy spectrum. In terms of the dimensionless ratio $x=J/\rvert K \rvert$~\footnote{$x_+=0.5$ with FM Kitaev interaction is mapped into the pure FM Heisenberg model, implying that the local magnetic moment is fully saturated.}, this transformation maps $1/2>x_{+}>0$ to $x _{-}< 0$,
 \begin{equation}
 x_{-} = -  x_{+}/(1 - 2 x_{+}).
 \label{unitcorr}
 \end{equation}
This property of the \emph{exact solution of the problem} provides a quantitative test  for our approximation scheme.
In other words, the instability that we are finding for positive $x$ at $x_+= 0.20$ must have its negative $x$ counterpart at $x_-=-0.3333$.
This value agrees very well with the result $x_-=-0.3312$ from our approximation scheme.

The  Gamma term also makes the continuum spectrum dispersive and it  induces a bound state between
the flux pair and a matter fermion. However, in this case the bound state only appears for the AFM Kitaev model because the FM model is much more fragile against the Gamma term (the flux-pair gap closes before the bound state is formed).
There are two interesting features associated with the bound state dispersion. 
First, in the absence of Heisenberg interaction, the spectra are identical for the $\Gamma$ and the $K$ points of the Brillouin zone
because the dual Kagome lattice is subdivided into three disconnected Kagome sublattices [see Fig.~\ref{dual_lattice}~(a)]. Consequently, the bound state dispersion has global minima at both momentum points for \textit{positive} Gamma interaction. 
This degeneracy is lifted by the Heisenberg term: the $\Gamma$ ($K$) point has a lower energy for a small positive (negative)  $J$. As the bound state becomes soft for a particular wave vector ${\bm Q}$, the system develops a divergent magnetic susceptibility,  $\chi({\bm Q}, \omega)$, at $\omega=\omega_{\bm Q} \to 0$.
Consequently, the vertical line $J=0$ can be regarded as a phase boundary between $\Gamma$ and $K$ magnetic orderings.

Secondly, for \textit{negative} Gamma interaction,  multiple bands of  bound states are formed and the lowest energy band has a
quasi-flat dispersion. This is a direct consequence of the flat  lowest energy band of
a bare flux pair for the  AFM Kitaev model with $\Gamma<0$ [see Fig.~\ref{pd_odd}~(b)]. For $J<0.0145\rvert K \rvert$,
the bound state becomes soft at  six incommensurate wave vectors related by the $C_6$ symmetry of ${\cal H}$.
These wave vectors are located on
the paths that connect the $M$ points of the Brillouin zone with the zone center (the $\Gamma$ point).
For instance, one of these wave vectors is ${\bm Q}=(0,{4\pi q/{\sqrt{3}a}})$ with $0<q<0.35$. The six wave vectors converge at the $\Gamma$ point for $J>0.0145\rvert K \rvert$.

In summary, by contrast to the odd-parity sector, the even-parity sector can include  bosonic modes if the specific perturbation  induces a bound state of two elementary fermionic particles. The softening of one of these bosonic modes indicates an instability towards magnetic ordering. The corresponding stability boundaries are shown in Fig.~\ref{pd_even}~(e) and (f). In the next section we analyze the type of magnetic ordering that has dominant divergent susceptibility at each point of these boundaries. We stress that the actual phase transition does not need to be continuous.
Correspondingly, the area enclosed by the stability boundaries is an upper bound for the area occupied by the Kitaev liquid phase.
 It is important to note that  the hybridization with the open flux pair has been ignored when considering the even-parity sector.
As we have shown for the odd sector, this hybridization  reduces the energy of an individual closed flux pair and, consequently, the upper bound for the area occupied by the Kitaev liquid phase.
To account for  this effect in the even-parity sector, one must enlarge the variational space by including states with an open flux pair and two matter fermions.


\section{Magnetic order}

The magnetic ordering with dominant susceptibility
can be obtained by computing the  matrix element of the spin operator between the overall ground state $\rvert \Phi_0;f_0 \rangle\equiv2^{N/2}P\{\rvert u_{0}\rangle \otimes \rvert 0^{u_{0}} \rangle\}$ of the pure Kitaev model and the bound state $\rvert B_{\bm{Q}}\rangle$ with center of mass momentum $\bm{Q}$ that becomes soft,
\begin{equation}
C_{\alpha,\bm{r}}^{\mu}=   \langle \Phi_0;f_0 \rvert\hat{S}_{\alpha,\bm{r}}^{\mu} \rvert B_{\bm{Q}} \rangle,
\label{eq:me}
\end{equation}
where $\Phi_0$ refers to the flux-free sector represented by the gauge field configuration $u_0$, $\alpha=A,B$ and ${\bm r}$ denotes the position of each unit cell of the honeycomb lattice. Throughout this work, the position of the unit cells is labeled by the lattice site on the $A$ sublattice.
The quantity
\begin{equation}
{\hat O} =  \sum_{\mu, \alpha, {\bm r}} \text{Re}[C_{\alpha,\bm{r}}^{\mu}] \hat{S}_{\alpha,\bm{r}}^{\mu},
\end{equation}
is the corresponding order parameter. 
We note that $C_{\alpha,\bm{r}}^{\mu}$ is defined up to a U(1) phase   (relative phase between the states with $n$ and $n+1$ bound states), which   is replaced by a Z$_2$ freedom (${\cal H}$ has only discrete symmetries)  upon including processes that create a pair of bound states.~\footnote{Note that  those processes can only be included by enlarging the variational space to states containing two flux pairs and an even number of matter fermions.}

The consequent local magnetic moment at each lattice site is given by
\begin{equation}
M_{\alpha,\bm{r}}^{\mu} \propto  \text{Re} [C_{\alpha,\bm{r}}^{\mu} ].
\label{eq:moments}
\end{equation}
The spin components  $\mu=x,y,z$ refer to the directions parallel to the cubic axes of the underlying ligand octahedra
of O$^{2-}$ ions that mediate the Kitaev, Heisenberg and Gamma interactions in most realizations of this model.
In this reference frame, the honeycomb layer is perpendicular to the [111] axis.

As discussed in the previous section, the bound state $\rvert B_{\bm{Q}}\rangle$ is obtained as the eigenstate of the variational Hamiltonian ${\cal H}_{even}$ at the wave vector ${\bm Q}$, which is formally written as 
\begin{equation}
\rvert B_{\bm{Q}} \rangle = \sum_{n,\mu} c_{\bm{Q}}(n,\mu) \rvert {\bm Q}, \mu, n \rangle.
\end{equation}
The degeneracy $\nu$ of the lowest energy bound state depends on the perturbations.
For $\nu > 1$, we should in principle evaluate the matrix element \eqref{eq:me} for each degenerate bound state.
However, this is not necessary because the magnetic orderings associated with different matrix elements are related by symmetry operations. In principle, the boson can condense in a linear combination of bound state wave functions $\rvert B_{\bm{Q}}\rangle$ with different symmetry related wave vectors ${\bm Q}$ (multi-${\bm Q}$ magnetic ordering).
For simplicity, we will assume that the condensation occurs at a bound state with well defined ordering wave vector ${\bm Q}$ (single-${\bm Q}$ magnetic ordering). However, we should keep in mind that in these cases the actual linear combination of
bound state wave functions is determined by interaction terms between the bound states, whose derivation is beyond the scope of the present work.

To determine the magnetic moments explicitly, we computed the two matrix elements on the $A$ and $B$ sublattices:
\begin{eqnarray}
C_{A,\bm{r}}^{\mu}&=&\frac{2}{\sqrt{N/2}}  e^{-i\bm{Q}\cdot\bm{r}} \tilde{C}_{A,{\bm Q}}^{\mu}, \\
C_{B,\bm{r}}^{\mu}&=&\frac{2}{\sqrt{N/2}} e^{-i\bm{Q}\cdot\bm{r}_{\mu}} \tilde{C}_{B,{\bm Q}}^{\mu},
\label{eq:mam}
\end{eqnarray}
where ${\bm r}_{\mu} \in A$ is connected to the $B$ site of the unit cell at ${\bm r}$ via the $\mu$-th bond, and
\begin{eqnarray}
\tilde{C}_{A,{\bm Q}}^{\mu} &=& i \sum_n c_{\bm Q}(n,\mu) \langle 0^{({\bm r}_{\mu},\mu)} \rvert f_n^{({\bm r}_{\mu},\mu)} c_{A,\bm{r}} \rvert 0^{u_0} \rangle, \\
\tilde{C}_{B,{\bm Q}}^{\mu} &=& \sum_n c_{\bm Q}(n,\mu) \langle 0^{({\bm r}_{\mu},\mu)} \rvert  f_n^{({\bm r}_{\mu},\mu)} c_{B,\bm{r}} \rvert 0^{u_0} \rangle.
\end{eqnarray}
According to Eq.~(\ref{eq:moments}), the magnetic moment at each lattice site is explicitly given by
\begin{equation}
M_{\alpha,\bm{r}}^{\mu} \propto  \cos(\phi-{\bm Q}\cdot {\bm r}) n_{\alpha,1}^{\mu} - \sin(\phi-{\bm Q}\cdot {\bm r}) n_{\alpha,2}^{\mu} ,
\label{eq:moments_2}
\end{equation}
with
\begin{equation}
n_{\alpha,1}^{\mu} = \text{Re}[\tilde{C}_{\alpha,{\bm Q}}^{\mu}],\\
n_{\alpha,2}^{\mu} = \text{Im}[\tilde{C}_{\alpha,{\bm Q}}^{\mu}].
\end{equation}
In other words, the local magnetic moments on sublattice $\alpha=A,B$ are spanned by the two vectors $\bm{n}_{\alpha,1}$ and $\bm{n}_{\alpha,2}$ (a bold variable ${\bm n}\equiv (n^x,n^y,n^z)$ refers to a three-component vector). 
The U(1) degree of freedom, $\tilde{C}_{\alpha,{\bm Q}}^{\mu} \to \tilde{C}_{\alpha,{\bm Q}}^{\mu} e^{i\phi}$, corresponds to a uniform rotation of the local moments on the 2D plane spanned by $\{\bm{n}_{\alpha,1},\bm{n}_{\alpha,2}\}$.
The U(1) phase does not affect the nature of the magnetic orderings that we discuss below, except for the 120$^\circ$ order. Correspondingly, the analysis of the $\phi$-dependence will be restricted to that case.

\begin{figure*}[!t]
\includegraphics[width=\textwidth]{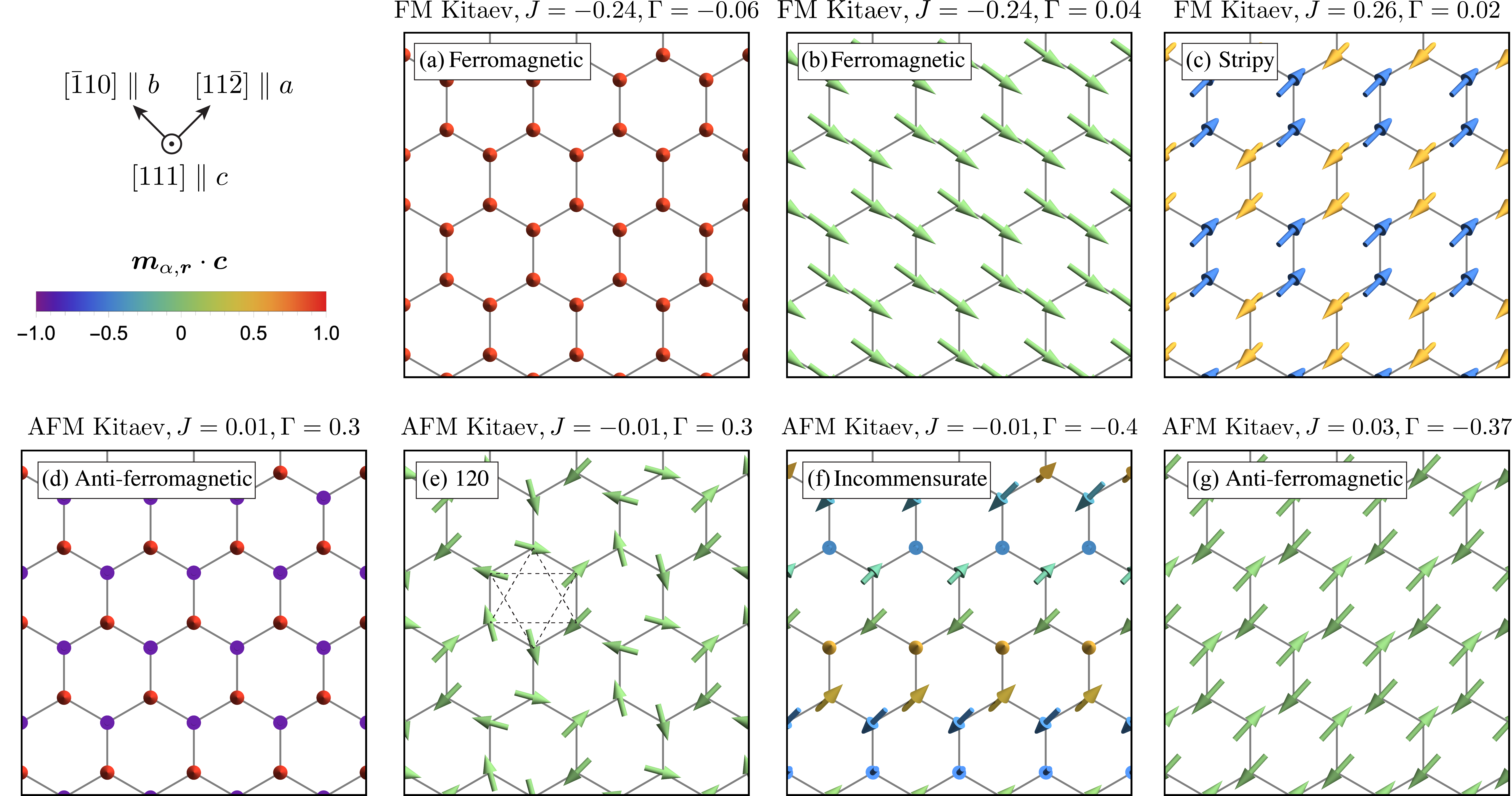}
\caption{ Magnetic order deduced from the condensation of the
lowest-energy bound state for different model parameters.}
\label{fig_order}
\end{figure*}

\subsection{Ferromagnetic Kitaev model}

\subsubsection{Ferromagnetic order ($J <0$)}

For the FM Kitaev model, a strong enough FM Heisenberg term ($J<0$) gives rise to three degenerate bound states
at the $\Gamma$ point.  The bond orientation $\gamma=x,y,z$ of the flux pair is preserved to   first-order  in the Heisenberg
interaction. The resulting three degenerate bound states can then be labeled by the quantum number $\rvert \gamma \rangle$.
The condensation at one of the three bound states gives rise to a collinear magnetic ordering with the moments oriented along
$\gamma$ axis, namely $M^{\mu}_{{\bm r}} = M_{{\bm r}}\delta_{\mu \gamma}$.

The  Gamma term splits the degenerate triplet into a singlet and a doublet.
For $\Gamma<0$, the lowest energy bound state  is the singlet
\begin{equation}
\rvert s \rangle =  (\rvert x \rangle+\rvert y \rangle+\rvert z \rangle)/\sqrt{3},
\end{equation}
that belongs to the trivial representation of the $C_3$ symmetry group. Consequently,
$\tilde{C}_{\alpha,{\bm 0}}^{x}=\tilde{C}_{\alpha,{\bm 0}}^{y}=\tilde{C}_{\alpha,{\bm 0}}^{z} \equiv \tilde{C}_{\alpha}, {\alpha}=A,B$.
Note that the ground state and the bound state are both invariant up to a phase factor under the inversion symmetry,
$c_{A,{\bm r}} \rightarrow c_{B,{\bm r}}$, $c_{B,{\bm r}} \rightarrow -c_{A,{\bm r}}$:
$\rvert 0^{u_0}\rangle \rightarrow \rvert 0^{u_0}\rangle$,
$\rvert b \rangle \equiv \sum_{n\mu} c_{{\bm Q}}(n,\mu) f_n^{({\bm r},\mu)\dagger} \rvert 0^{{\bm r},\mu} \rangle \rightarrow  i  \rvert b \rangle$, implying that
$\tilde{C}_{A}=\tilde{C}_B$.
According to Eq.~(\ref{eq:moments}), this indicates
an instability towards \textit{ferromagnetic} ordering along the [111] direction
[see Fig.~\ref{fig_order} (a)].

For $\Gamma>0$, the lowest energy bound states are spanned by the doublet
\begin{eqnarray}
\rvert d_1 \rangle &=& \frac{1}{\sqrt{3}} (\rvert x \rangle+e^{i2\pi/3}\rvert y \rangle+e^{-i2\pi/3}\rvert z \rangle),
\nonumber \\
\rvert d_2 \rangle &=&  \frac{1}{\sqrt{3}} (\rvert x \rangle+e^{-i2\pi/3}\rvert y \rangle+e^{i2\pi/3}\rvert z \rangle)/\sqrt{3}.
\end{eqnarray}
The bosons can then condense in  any  linear combination of these two states.
If we  assume that the condensation takes place in $\rvert d_1 \rangle$ or $\rvert d_2 \rangle$, we obtain
$\bm{n}_{A2},\bm{n}_{B2}={\bm 0}$ and $\bm{n}_{A1}=\bm{n}_{B1} \bot $[111]. This result implies an instability towards
\textit{ferromagnetic} ordering in the plane perpendicular to the [111] direction [see Fig.~\ref{fig_order} (b)].
Different in-plane moment directions correspond to condensations in different linear combinations of
$\rvert d_1 \rangle$ and $\rvert d_2 \rangle$.
We note that this magnetic order coincides with the one that is obtained in the classical limit of the model~\cite{rau2014generic}. The same is true for the remaining magnetic orders that we discuss below.

\subsubsection{Stripy order ($J > 0$)}

For the FM Kitaev model, a strong enough AFM Heisenberg term ($J>0$) gives rise to three degenerate bound states
at the three $M$ points. For concreteness, we will assume that the bound state condenses at the $M$ point
${\bm Q}=(2\pi/\sqrt{3}a, 0)$ (single-${\bm Q}$ ordering). The magnetic orders associated with condensations at  the other two $M$ points are obtained by applying  a $C_3$ rotation.
In the absence of the $\Gamma$-term, the condensation gives rise to a
collinear magnetic order along the [001] direction. As shown in Fig.~\ref{fig_order}~(c), spins that are connected by the $z$-bonds are oriented in the same direction, forming the so-called stripy magnetic order~\cite{gotfryd2017phase,rau2014generic,rau2014trigonal}. A finite Gamma term induces a global rotation of the magnetic moments about the $b$-axis.

\subsection{Antiferromagnetic Kitaev model}

\subsubsection{Antiferromagnetic order  ($\Gamma > 0, J>0$)}

We will consider now the instabilities of the  AFM Kitaev model. As shown in Fig.~\ref{pd_even} (b), for $\Gamma>0$ and $J=0$,
there are three degenerate bound states. One of them is located at the $\Gamma$ point, while the other two are located  at the  two $K$ points.
This degeneracy is a direct consequence of the restricted motion of the  bound state to one of the three  sublattices
shown in Fig.~\ref{dual_lattice}~(b).
An arbitrarily small Heisenberg interaction lifts the degeneracy between the
$\Gamma$ and $\pm K$ points. The $\Gamma$ state has lower (higher) energy if $J>0$ ($J<0$). 
In this subsection we consider the case $J>0$, while the case  $J<0$ is left for the next subsection. 

The bound state at the $\Gamma$ point picks up a phase $-i$ under inversion. Together with the $C_3$ symmetry of $\mathcal{H}$, 
this property dictates that
$\tilde{C}_{\alpha,{\bm 0}}^x=\tilde{C}_{\alpha,{\bm 0}}^y = \tilde{C}_{\alpha,{\bm 0}}^z \equiv \tilde{C}_{\alpha}, \alpha =A,B$ and $\tilde{C}_{A} = - \tilde{C}_{B}$.
According to Eq.~(\ref{eq:moments}), this indicates an instability towards \textit{antiferromagnetic} ordering along the [111] direction. 
The resulting antiferromagnetic order  is shown in Fig.~\ref{fig_order}~(d)

\subsubsection{$120^{\circ}$ magnetic order  ($\Gamma > 0, J<0$)}

As explained in the last subsection, the bound state dispersion induced by a positive Gamma interaction has a global minimum at the $\pm K$ points for arbitrarily small and negative FM Heisenberg interaction ($J<0$). The magnetic order produced by
the condensation in one of the two $K$ points corresponds to a
six-sublattice structure.
For instance, there are three nonequivalent magnetic moments 
on the A sublattice,
\begin{align*}
\bm{M}_{A,\bm{0}} & \propto \bm{n}_{A1} \cos(\phi) - \bm{n}_{A2} \sin(\phi), \nonumber  \\
\bm{M}_{A,\bm{a}_{1}}  &\propto \bm{n}_{A1} \cos(\phi-{4\pi \over 3}) - \bm{n}_{A2} \sin(\phi-{4\pi \over 3}),\nonumber\\
\bm{M}_{A,2\bm{a}_{1}} & \propto \bm{n}_{A1} \cos(\phi-{2\pi \over 3}) - \bm{n}_{A2} \sin(\phi-{2\pi \over 3}),\nonumber
\end{align*}
and three on the B sublattice,
\begin{align*}
\bm{M}_{B,\bm{0}} & \propto \tilde{\bm{n}}_{B1} \cos(\phi) - \tilde{\bm{n}}_{B2} \sin(\phi), \nonumber \\
\bm{M}_{B,\bm{a}_{1}}  &\propto \tilde{\bm{n}}_{B1} \cos(\phi-{4\pi \over 3}) - \tilde{\bm{n}}_{B2} \sin(\phi-{4\pi \over 3}),  \nonumber\\
\bm{M}_{B,2\bm{a}_{1}} & \propto \tilde{\bm{n}}_{B1} \cos(\phi-{2\pi \over 3}) - \tilde{\bm{n}}_{B2} \sin(\phi-{2\pi \over 3}), \nonumber
\end{align*}
where 
\begin{align*}
\tilde{\bm{n}}_{B1}^{x} &= -\frac{1}{2}\bm{n}_{B1}^{x}+\frac{\sqrt{3}}{2}\bm{n}_{B2}^{x}, &  \tilde{\bm{n}}_{B2}^{x}&=-\frac{\sqrt{3}}{2}\bm{n}_{B1}^{x}-\frac{1}{2}\bm{n}_{B2}^{x}, \\
\tilde{\bm{n}}_{B1}^{y} &= -\frac{1}{2}\bm{n}_{B1}^{y}-\frac{\sqrt{3}}{2}\bm{n}_{B2}^{y}, &   \tilde{\bm{n}}_{B2}^{y}&=\frac{\sqrt{3}}{2}\bm{n}_{B1}^{y}-\frac{1}{2}\bm{n}_{B2}^{y},\\
\tilde{\bm{n}}_{B1}^{z} &= \bm{n}_{B1}^{z},  & \tilde{\bm{n}}_{B2}^{z}&=\bm{n}_{B2}^{z}.
\end{align*}
Here $\phi$ is the angle associated with the above-mentioned U(1) freedom.

The calculation of the matrix element \eqref{eq:me} leads to the following constraints: $\rvert\bm{n}_{A1}\rvert=\rvert\bm{n}_{A2}\rvert=\rvert\tilde{\bm{n}}_{B1}\rvert=\rvert\tilde{\bm{n}}_{B2}\rvert\simeq0.66$
and $\bm{n}_{A1}\bot\bm{n}_{A2}$, $\tilde{\bm{n}}_{B1}\bot\tilde{\bm{n}}_{B2}$.
These constraints imply that the magnitude of the magnetic moments is uniform over the whole lattice.
As a result, the magnetic moments form a $120^{\circ}$ structure on each A and B sublattice.
Another constraint that results  from the calculation is that $\tilde{\bm{n}}_{B1}=-\bm{n}_{A1}$ and $\tilde{\bm{n}}_{B2}=\bm{n}_{A2}$.
This constraint locks the relative vector chirality of the two different sublattices:  for each hexagonal plaquette of the honeycomb lattice, the vector spin chirality along the up-triangles on the A sublattice is the same as the vector chirality along the
down-triangles  on the B sublattice [see Fig.~\ref{fig_order}~(e)].
There is still a remaining degree of freedom in the magnetic pattern  because
a change of $\phi$ corresponds to a uniform rotation of the magnetic moments on the A and B sublattices 
in opposite  directions.
This freedom, that  also exists in the classical limit of the model~\cite{rau2014generic},
is removed by the processes that create and  annihilate pairs of bound states (vacuum fluctuations) with opposite wave vectors $K$ and $-K$.
As it is shown in Fig.~\ref{fig_order}~(e) for a fixed value of $\phi$, the resulting coplanar  order lies in the [111] plane and it agrees with the result that is obtained for the classical limit of the model~\cite{rau2014generic}.

\subsubsection{Spiral magnetic and antiferromagnetic order ($\Gamma < 0$)}

As we found in Sec.~\ref{sec:bound_state_B}, for $\Gamma < 0$ and $J \leq 0.0145 \rvert K\rvert$, the bound state becomes gapless at six incommensurate wave vectors related by $C_6$ symmetry [see Fig.~\ref{pd_even} (f)].
A (single-${\bm Q}$) condensation in one of these wave vectors, such as ${\bm Q}=(0,{4\pi q/{\sqrt{3}a}})$ with $0<q<0.35$,
leads to a coplanar magnetic order that lies in the plane perpendicular to the $[\bar{1}10]$ axis because
${\bm n}_{A1} \times {\bm n}_{A2},{\bm n}_{B1} \times {\bm n}_{B2} \parallel [\bar{1}10]$ [see Fig.~\ref{fig_order} (f)].
By choosing a proper phase of the bound state wave-function such that ${\bm n}_{A1}$ is orthogonal to $ {\bm n}_{A2}$, the two vectors ${\bm n}_{A1}$ and ${\bm n}_{A2}$ have different lengths. 
The same holds true for the B sublattice.
This property leads to a spin density wave, i.e., to a modulation of the \emph{magnitude} of the magnetic moments.

The six incommensurate wave vectors converge to the $\Gamma$ point upon increasing $J$,
 giving rise to a doublet of bound states.
Meanwhile, the magnetic moments rotate about the $[\bar{1}10]$ axis and form an AFM structure parallel to the $[11\bar{2}]$ axis.
 Given the double degeneracy of the lowest energy bound state, the AFM moments can point in any direction of the [111] plane.
We note that \textit{positive} and \textit{negative} signs of the
Gamma interaction added to the AFM Kitaev model both lead to AFM order.
However, the $C_3$ symmetry about the [111] axis is preserved in the former case, while it is  broken in the latter case.
According to numerical simulations~\cite{rau2014generic,rau2014trigonal}, the two AFM orders are adjacent to each other
on the $J-\Gamma$ phase diagram, in agreement with Fig.~\ref{pd_even} (f) where $J>0$. Our analysis indicates that there should be a magnetic moment reorientation phase transition between the two types of AFM order.

\subsection{Comparison with numerical results}

We close this section with  a comparison against the existing numerical simulations.
The ordering wave vectors that we extracted from the analysis presented in this section are consistent with
exact-diagonalization (ED) studies of a $24$-site cluster~\cite{gotfryd2017phase,rau2014generic,rau2014trigonal}.
The orientation of the magnetic moments was determined from studies of the classical limit of the model~\cite{rau2014generic} and it also agrees with our results.  Tables~\ref{table1} and ~\ref{table2} include a comparison between the critical values of the perturbations at which the two-fermion bound state becomes a gapless excitation  and the phase boundaries obtained from ED studies~\cite{gotfryd2017phase,rau2014generic,rau2014trigonal}
and density matrix renormalization group (DMRG)~\cite{gohlke2017dynamics}.
In all cases, the present analysis predicts a larger  stability range of the Kitaev liquid because of two reasons.
The first reason is that in our simple analysis we have neglected processes, such as the hybridization with the open flux pair or the creation/annihilation of two flux pairs [see Fig.~\ref{fig:matrix}~(a)],
that produce a significant  reduction in  the energy of the flux pair (Fig.~\ref{open_FP}).
The second reason is that these phase transitions are expected to be of first order, implying that the transition occurs before the bound state excitation becomes gapless.

\begin{table}[t!]
\centering
\caption{Phase boundaries  of the FM Kitaev spin liquid in the presence of a Heisenberg interaction ($\Gamma=0$).}
\label{table1}
\begin{tabular}{|c|c|c|c|}
\hline
$J_c/\rvert K \rvert $ & this work  & ED~\cite{gotfryd2017phase} & DMRG~\cite{gohlke2017dynamics}   \\
\hline
\hline
$J<0$  & -0.33   & -0.1648 &  -0.1167  \\
\hline
$J>0$ & 0.20  & 0.1231 &  0.0938   \\
\hline
\end{tabular}
\end{table}

\begin{table}[t!]
\centering
\caption{Phase boundaries  of the  AFM Kitaev spin liquid in the presence of a Gamma interaction ($J=0$).}
\label{table2}
\begin{tabular}{|c|c|c|}
\hline
$\Gamma_c/\rvert K \rvert $ & this work  & ED~\cite{rau2014generic,rau2014trigonal}   \\
\hline
\hline
$\Gamma<0$  &  -0.4  & -0.14 \\
\hline
$\Gamma>0$ &  0.27  & 0.095   \\
\hline
\end{tabular}
\end{table}

In the previous section, we have also shown that the FM and AFM Kitaev spin liquids have very  different  stability against the
Heisenberg and the Gamma terms considered in this work. This asymmetry  arises from the constructive/destructive  interference between processes that contribute to the effective hopping of flux pairs and it is also verified by numerical studies of this model.

Finally, we note that some magnetic orders reported in the ED study~\cite{rau2014generic} are not explained by the current approach
because of the absence of preformed bound states in the excitation spectrum.
In this case, the excitations that become soft are the fractional (fermionic) modes.
This result is consistent with a recent DMRG study~\cite{gordon2019theory}, where it is found that  an AFM Gamma interaction induces a transition from the FM Kitaev spin liquid to a different type of spin liquid. The transition is characterized by a discontinuous change in the average flux $\langle W_p \rangle$ per plaquette.

\section{Dynamical spin structure factor}
\label{dynamics}

\begin{figure}[!t]
\includegraphics[width=0.5\textwidth]{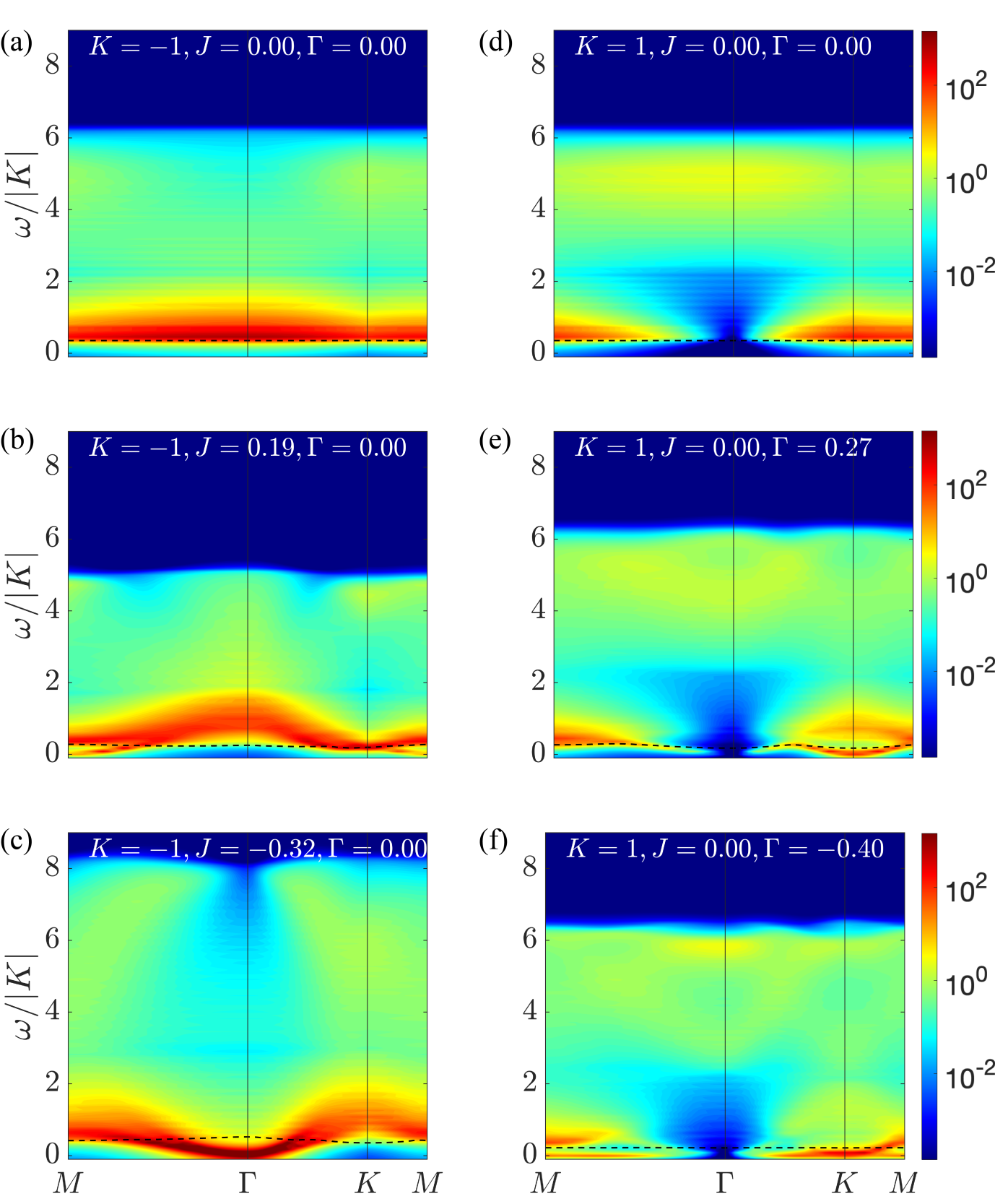}
\caption{(Color online) Dynamical structure factor $S({\bm
q},\omega) \equiv \sum_{\mu} S^{\mu\mu}({\bm q},\omega)$ for
different perturbations of the FM Kitaev model (a-c) and AFM Kitaev
model (d-f). The black dashed line indicates the lower edge of the
continuum. The calculation has been done on a finite lattice of
$71\times 71$ unit cells and the artificial broadening of the delta
functions in Eq.~\eqref{sqw} is $\eta=0.08 \rvert K \rvert$.}
\label{fig:dynamics}
\end{figure}

The above-mentioned two-fermion bound state should appear as a sharp mode in the dynamical spin structure factor that is measured with inelastic neutron scattering. As explained above, the  emergence of these coherent quasi-particle modes signals  the proximity to a  quantum phase transition into a magnetically ordered state. This phenomenon can be exploited to reveal the coherent nature of the continuum of magnetic excitations produced by the unbounded flux pair and matter fermion modes.

The diagonal components of the dynamical structure factor are
\begin{eqnarray}
S^{\mu\mu}({\bm q},\omega) = \sum_n \delta(\omega-E_n +E_G) \rvert \langle G\rvert S^{\mu}_{\bm q} \rvert n\rangle \rvert^2,
\label{sqw}
\end{eqnarray}
where  $\rvert n \rangle$ and $\rvert G \rangle$ denote the exact  eigenstates and the ground state of $\mathcal{H}$  with  eigenvalues $E_n$ and $E_G$, respectively, and $\mu=x,y,z$.
$S^{\mu\mu}({\bm q},\omega)$ can be exactly computed at $T=0$ for the pure Kitaev model.
In this case, the ground state has zero flux and the excited states contributing to the matrix elements in Eq.~\eqref{sqw} have only one flux pair. As we argued in previous sections, the main contribution to these matrix elements arises from states  containing only one  matter fermion (Bogoliubov quasiparticle in the two-flux sector).
Our approach should then provide an accurate estimate of $S^{\mu\mu}({\bm q},\omega)$  for small enough values of $J/K$ and $\Gamma/K$.

The excited eigenstates that contribute to $S^{\mu\mu}({\bm q},\omega)$ are approximated by the eigenstates of the effective Hamiltonian $\mathcal{H}_{even}$ [see Eq.~(\ref{H_even})] that acts on the reduced even-parity  sector.
Fig.~\ref{fig:dynamics} (a) and (d) show the resulting $S({\bm q},\omega) \equiv \sum_{\mu} S^{\mu\mu}({\bm q},\omega)$
for the pure FM and AFM Kitaev model. Panels (b) and (c) show the result for two representative sets of perturbations around the
FM Kitaev model. Similarly, panels (e) and (f) show the result for two representative sets of perturbations around the
AFM Kitaev model.

Since  the mass of the flux pair reduces to a finite value for  $J \neq 0$ or $\Gamma \neq 0$, the spectral weight distribution acquires a finite dispersion in the presence of these perturbations. The magnon-like bound state that was discussed in the previous section
gives rise to sharp and dispersive magnetic modes. These modes acquire a finite lifetime (the line becomes broad) upon entering in the  continuum because they can decay into two independent bond and matter Majorana fermions.
The sharp features below the continuum provide unambiguous evidence of a coherent quantum spin liquid (continuum features
can also originate from disorder and/or thermal fluctuations in classical phases with no long-range quantum entanglement).

\section{Summary and outlook}

In this paper, we have studied the low-energy excitation spectrum of
the Kitaev-Heisenberg-Gamma model through a novel variational
approach that is based on the exact fractionalized excitations of
the pure Kitaev honeycomb model. We have demonstrated that this new
approach reveals the microscopic mechanisms behind many important
aspects of the quantum phase diagram from previous numerical
works.~\cite{chaloupka2010kitaev, chaloupka2013zigzag,
rau2014generic, rau2014trigonal, gotfryd2017phase,
gohlke2017dynamics}

First of all, our results provide a simple explanation for the
asymmetric stability of the Kitaev spin liquid phases around the FM
and AFM Kitaev limits. In our picture, the phase transitions out of
the Kitaev spin liquid phases are related to softening of flux-pair
excitations, which are static for the pure Kitaev model but become
dispersive in presence of Heisenberg and/or Gamma interactions. For
$J \neq 0$ and $\Gamma = 0$, the hopping amplitude of such a flux
pair is large (small) for the AFM (FM) Kitaev spin liquid due to a
constructive (destructive) interference between two contributions
arising from different components of the Heisenberg interaction.
Since a larger hopping amplitude means a stronger dispersion for the
same perturbation, this simple result naturally explains why the AFM
Kitaev spin liquid is more fragile against a Heisenberg perturbation
than the FM Kitaev spin liquid. Conversely, for $\Gamma \neq 0$ and
$J = 0$, the hopping amplitude is small (large) for the AFM (FM)
Kitaev spin liquid due to an analogous interference effect, implying
that the AFM Kitaev spin liquid is more robust against a Gamma
perturbation than the FM Kitaev spin liquid.

Importantly, our variational approach can be straightforwardly
generalized to other perturbations on top of the Kitaev model, including a Zeeman term representing an external magnetic
field.\cite{zhu2018robust, gohlke2018dynamical, hickey2019emergence}
By considering the hopping amplitude of a flux pair, the same kind
of interference effect then explains why the AFM Kitaev spin liquid
is more robust (fragile) against a uniform (staggered) magnetic
field than the FM Kitaev spin liquid.\cite{hickey2019emergence}
Further perturbations of interest include different forms of disorder, such as bond disorder,\cite{knolle2019bond} random vacancies,\cite{kao2021vacancy,kao2021disorder} and topological defects.\cite{petrova2014unpaired} In particular, our varational approach may shed light on the localization of flux excitations via the Anderson  mechanism or the flux-binding effect induced by spin vacancies.\cite{kao2021vacancy, willans2010disorder, willans2011site}

Moreover, our results provide valuable insight about the
magnetically ordered phases beyond the transitions out of the two
Kitaev spin liquid phases. In particular, for each phase transition
induced by the kind of perturbation (Heisenberg or Gamma) against
which the given Kitaev spin liquid (FM or AFM) is more robust, we
find a softening of a bosonic magnon-like excitation at a wave
vector that coincides with the ordering wave vector of the resulting
magnetically ordered phase. Since this bosonic excitation is
topologically trivial (i.e., it does not carry any gauge charge),
its condensation at a continuous transition would lead to a
magnetically ordered quantum spin liquid phase. To reach a
conventional magnetically ordered phase, one would then require two
subsequent phase transitions for generating the magnetic order and
destroying the quantum spin liquid, respectively. However, according
to numerical studies, these two transitions are preempted by a
single discontinuous (first-order) transition which immediately
destroys the quantum spin liquid in favor of a conventional
magnetically ordered phase. It is then an intriguing question
whether including additional perturbations could make this phase
transition continuous, thereby stabilizing a novel quantum phase in
which the fractionalized excitations of the Kitaev spin liquid
coexist with long-range magnetic order.

Finally, our work predicts that the bosonic magnon-like excitation,
which forms further away from the pure Kitaev limit, manifests as a
sharp mode in the dynamical spin structure factor. This exotic bound
state of a flux pair and a Majorana fermion can then be detected in
inelastic neutron scattering and can serve as a direct signature of
the Kitaev spin liquid phase. While we have focused on the dynamical
spin structure factor, straightforward extensions of our approach
can also be used to compute other dynamical responses. In turn,
computing such dynamical responses away from the exactly solvable
limit is crucial for characterizing real candidate materials both
for the actual quantum spin liquid phase as well as the
``proximate'' magnetically ordered phases.


\begin{acknowledgments}

S.-S.~Z.~and C.~D.~B.~are supported by funding from the Lincoln
Chair of Excellence in Physics and from the Los Alamos National
Laboratory Directed Research and Development program.
W.Z. is supported by the Key R\&D Program of Zhejiang Province, China (2021C01002).
G.~B.~H.~was supported by the U.S. Department of Energy,
Office of Science, National Quantum Information Science Research Centers, Quantum Science Center.

\end{acknowledgments}

\bibliographystyle{apsrev4-1}
\bibliography{paper}

\begin{thebibliography}{79}%
\makeatletter
\providecommand \@ifxundefined [1]{%
 \@ifx{#1\undefined}
}%
\providecommand \@ifnum [1]{%
 \ifnum #1\expandafter \@firstoftwo
 \else \expandafter \@secondoftwo
 \fi
}%
\providecommand \@ifx [1]{%
 \ifx #1\expandafter \@firstoftwo
 \else \expandafter \@secondoftwo
 \fi
}%
\providecommand \natexlab [1]{#1}%
\providecommand \enquote  [1]{``#1''}%
\providecommand \bibnamefont  [1]{#1}%
\providecommand \bibfnamefont [1]{#1}%
\providecommand \citenamefont [1]{#1}%
\providecommand \href@noop [0]{\@secondoftwo}%
\providecommand \href [0]{\begingroup \@sanitize@url \@href}%
\providecommand \@href[1]{\@@startlink{#1}\@@href}%
\providecommand \@@href[1]{\endgroup#1\@@endlink}%
\providecommand \@sanitize@url [0]{\catcode `\\12\catcode `\$12\catcode
  `\&12\catcode `\#12\catcode `\^12\catcode `\_12\catcode `\%12\relax}%
\providecommand \@@startlink[1]{}%
\providecommand \@@endlink[0]{}%
\providecommand \url  [0]{\begingroup\@sanitize@url \@url }%
\providecommand \@url [1]{\endgroup\@href {#1}{\urlprefix }}%
\providecommand \urlprefix  [0]{URL }%
\providecommand \Eprint [0]{\href }%
\providecommand \doibase [0]{http://dx.doi.org/}%
\providecommand \selectlanguage [0]{\@gobble}%
\providecommand \bibinfo  [0]{\@secondoftwo}%
\providecommand \bibfield  [0]{\@secondoftwo}%
\providecommand \translation [1]{[#1]}%
\providecommand \BibitemOpen [0]{}%
\providecommand \bibitemStop [0]{}%
\providecommand \bibitemNoStop [0]{.\EOS\space}%
\providecommand \EOS [0]{\spacefactor3000\relax}%
\providecommand \BibitemShut  [1]{\csname bibitem#1\endcsname}%
\let\auto@bib@innerbib\@empty
\bibitem [{\citenamefont {Balents}(2010)}]{balents2010spin}%
  \BibitemOpen
  \bibfield  {author} {\bibinfo {author} {\bibfnamefont {L.}~\bibnamefont
  {Balents}},\ }\href@noop {} {\bibfield  {journal} {\bibinfo  {journal}
  {Nature}\ }\textbf {\bibinfo {volume} {464}},\ \bibinfo {pages} {199}
  (\bibinfo {year} {2010})}\BibitemShut {NoStop}%
\bibitem [{\citenamefont {Savary}\ and\ \citenamefont
  {Balents}(2016)}]{savary2016quantum}%
  \BibitemOpen
  \bibfield  {author} {\bibinfo {author} {\bibfnamefont {L.}~\bibnamefont
  {Savary}}\ and\ \bibinfo {author} {\bibfnamefont {L.}~\bibnamefont
  {Balents}},\ }\href@noop {} {\bibfield  {journal} {\bibinfo  {journal}
  {Reports on Progress in Physics}\ }\textbf {\bibinfo {volume} {80}},\
  \bibinfo {pages} {016502} (\bibinfo {year} {2016})}\BibitemShut {NoStop}%
\bibitem [{\citenamefont {Kitaev}(2003)}]{kitaev2003fault}%
  \BibitemOpen
  \bibfield  {author} {\bibinfo {author} {\bibfnamefont {A.}~\bibnamefont
  {Kitaev}},\ }\href {\doibase https://doi.org/10.1016/S0003-4916(02)00018-0}
  {\bibfield  {journal} {\bibinfo  {journal} {Annals of Physics}\ }\textbf
  {\bibinfo {volume} {303}},\ \bibinfo {pages} {2} (\bibinfo {year}
  {2003})}\BibitemShut {NoStop}%
\bibitem [{\citenamefont {Nayak}\ \emph {et~al.}(2008)\citenamefont {Nayak},
  \citenamefont {Simon}, \citenamefont {Stern}, \citenamefont {Freedman},\ and\
  \citenamefont {Das~Sarma}}]{nayak2008nonabelian}%
  \BibitemOpen
  \bibfield  {author} {\bibinfo {author} {\bibfnamefont {C.}~\bibnamefont
  {Nayak}}, \bibinfo {author} {\bibfnamefont {S.~H.}\ \bibnamefont {Simon}},
  \bibinfo {author} {\bibfnamefont {A.}~\bibnamefont {Stern}}, \bibinfo
  {author} {\bibfnamefont {M.}~\bibnamefont {Freedman}}, \ and\ \bibinfo
  {author} {\bibfnamefont {S.}~\bibnamefont {Das~Sarma}},\ }\href {\doibase
  10.1103/RevModPhys.80.1083} {\bibfield  {journal} {\bibinfo  {journal} {Rev.
  Mod. Phys.}\ }\textbf {\bibinfo {volume} {80}},\ \bibinfo {pages} {1083}
  (\bibinfo {year} {2008})}\BibitemShut {NoStop}%
\bibitem [{\citenamefont {Kitaev}(2006)}]{kitaev2006anyons}%
  \BibitemOpen
  \bibfield  {author} {\bibinfo {author} {\bibfnamefont {A.}~\bibnamefont
  {Kitaev}},\ }\href@noop {} {\bibfield  {journal} {\bibinfo  {journal} {Annals
  of Physics}\ }\textbf {\bibinfo {volume} {321}},\ \bibinfo {pages} {2}
  (\bibinfo {year} {2006})}\BibitemShut {NoStop}%
\bibitem [{\citenamefont {Jackeli}\ and\ \citenamefont
  {Khaliullin}(2009)}]{jackeli2009mott}%
  \BibitemOpen
  \bibfield  {author} {\bibinfo {author} {\bibfnamefont {G.}~\bibnamefont
  {Jackeli}}\ and\ \bibinfo {author} {\bibfnamefont {G.}~\bibnamefont
  {Khaliullin}},\ }\href {\doibase 10.1103/PhysRevLett.102.017205} {\bibfield
  {journal} {\bibinfo  {journal} {Phys. Rev. Lett.}\ }\textbf {\bibinfo
  {volume} {102}},\ \bibinfo {pages} {017205} (\bibinfo {year}
  {2009})}\BibitemShut {NoStop}%
\bibitem [{\citenamefont {Rau}\ \emph {et~al.}(2016)\citenamefont {Rau},
  \citenamefont {Lee},\ and\ \citenamefont {Kee}}]{rau2016spin}%
  \BibitemOpen
  \bibfield  {author} {\bibinfo {author} {\bibfnamefont {J.~G.}\ \bibnamefont
  {Rau}}, \bibinfo {author} {\bibfnamefont {E.~K.-H.}\ \bibnamefont {Lee}}, \
  and\ \bibinfo {author} {\bibfnamefont {H.-Y.}\ \bibnamefont {Kee}},\ }\href
  {\doibase 10.1146/annurev-conmatphys-031115-011319} {\bibfield  {journal}
  {\bibinfo  {journal} {Annual Review of Condensed Matter Physics}\ }\textbf
  {\bibinfo {volume} {7}},\ \bibinfo {pages} {195} (\bibinfo {year} {2016})},\
  \Eprint
  {http://arxiv.org/abs/https://doi.org/10.1146/annurev-conmatphys-031115-011319}
  {https://doi.org/10.1146/annurev-conmatphys-031115-011319} \BibitemShut
  {NoStop}%
\bibitem [{\citenamefont {{Trebst}}(2017)}]{trebst2017kitaev}%
  \BibitemOpen
  \bibfield  {author} {\bibinfo {author} {\bibfnamefont {S.}~\bibnamefont
  {{Trebst}}},\ }\href@noop {} {\bibfield  {journal} {\bibinfo  {journal}
  {ArXiv e-prints}\ } (\bibinfo {year} {2017})},\ \Eprint
  {http://arxiv.org/abs/1701.07056} {arXiv:1701.07056 [cond-mat.str-el]}
  \BibitemShut {NoStop}%
\bibitem [{\citenamefont {Hermanns}\ \emph {et~al.}(2018)\citenamefont
  {Hermanns}, \citenamefont {Kimchi},\ and\ \citenamefont
  {Knolle}}]{hermanns2018physics}%
  \BibitemOpen
  \bibfield  {author} {\bibinfo {author} {\bibfnamefont {M.}~\bibnamefont
  {Hermanns}}, \bibinfo {author} {\bibfnamefont {I.}~\bibnamefont {Kimchi}}, \
  and\ \bibinfo {author} {\bibfnamefont {J.}~\bibnamefont {Knolle}},\ }\href
  {\doibase 10.1146/annurev-conmatphys-033117-053934} {\bibfield  {journal}
  {\bibinfo  {journal} {Annual Review of Condensed Matter Physics}\ }\textbf
  {\bibinfo {volume} {9}},\ \bibinfo {pages} {17} (\bibinfo {year} {2018})},\
  \Eprint
  {http://arxiv.org/abs/https://doi.org/10.1146/annurev-conmatphys-033117-053934}
  {https://doi.org/10.1146/annurev-conmatphys-033117-053934} \BibitemShut
  {NoStop}%
\bibitem [{\citenamefont {Takagi}\ \emph {et~al.}(2019)\citenamefont {Takagi},
  \citenamefont {Takayama}, \citenamefont {Jackeli}, \citenamefont
  {Khaliullin},\ and\ \citenamefont {Nagler}}]{takagi2019concept}%
  \BibitemOpen
  \bibfield  {author} {\bibinfo {author} {\bibfnamefont {H.}~\bibnamefont
  {Takagi}}, \bibinfo {author} {\bibfnamefont {T.}~\bibnamefont {Takayama}},
  \bibinfo {author} {\bibfnamefont {G.}~\bibnamefont {Jackeli}}, \bibinfo
  {author} {\bibfnamefont {G.}~\bibnamefont {Khaliullin}}, \ and\ \bibinfo
  {author} {\bibfnamefont {S.~E.}\ \bibnamefont {Nagler}},\ }\href {\doibase
  10.1038/s42254-019-0038-2} {\bibfield  {journal} {\bibinfo  {journal} {Nature
  Reviews Physics}\ }\textbf {\bibinfo {volume} {1}},\ \bibinfo {pages} {264}
  (\bibinfo {year} {2019})}\BibitemShut {NoStop}%
\bibitem [{\citenamefont {Singh}\ and\ \citenamefont
  {Gegenwart}(2010)}]{singh2010antiferromagnetic}%
  \BibitemOpen
  \bibfield  {author} {\bibinfo {author} {\bibfnamefont {Y.}~\bibnamefont
  {Singh}}\ and\ \bibinfo {author} {\bibfnamefont {P.}~\bibnamefont
  {Gegenwart}},\ }\href {\doibase 10.1103/PhysRevB.82.064412} {\bibfield
  {journal} {\bibinfo  {journal} {Phys. Rev. B}\ }\textbf {\bibinfo {volume}
  {82}},\ \bibinfo {pages} {064412} (\bibinfo {year} {2010})}\BibitemShut
  {NoStop}%
\bibitem [{\citenamefont {Liu}\ \emph {et~al.}(2011)\citenamefont {Liu},
  \citenamefont {Berlijn}, \citenamefont {Yin}, \citenamefont {Ku},
  \citenamefont {Tsvelik}, \citenamefont {Kim}, \citenamefont {Gretarsson},
  \citenamefont {Singh}, \citenamefont {Gegenwart},\ and\ \citenamefont
  {Hill}}]{liu2011longrange}%
  \BibitemOpen
  \bibfield  {author} {\bibinfo {author} {\bibfnamefont {X.}~\bibnamefont
  {Liu}}, \bibinfo {author} {\bibfnamefont {T.}~\bibnamefont {Berlijn}},
  \bibinfo {author} {\bibfnamefont {W.-G.}\ \bibnamefont {Yin}}, \bibinfo
  {author} {\bibfnamefont {W.}~\bibnamefont {Ku}}, \bibinfo {author}
  {\bibfnamefont {A.}~\bibnamefont {Tsvelik}}, \bibinfo {author} {\bibfnamefont
  {Y.-J.}\ \bibnamefont {Kim}}, \bibinfo {author} {\bibfnamefont
  {H.}~\bibnamefont {Gretarsson}}, \bibinfo {author} {\bibfnamefont
  {Y.}~\bibnamefont {Singh}}, \bibinfo {author} {\bibfnamefont
  {P.}~\bibnamefont {Gegenwart}}, \ and\ \bibinfo {author} {\bibfnamefont
  {J.~P.}\ \bibnamefont {Hill}},\ }\href {\doibase 10.1103/PhysRevB.83.220403}
  {\bibfield  {journal} {\bibinfo  {journal} {Phys. Rev. B}\ }\textbf {\bibinfo
  {volume} {83}},\ \bibinfo {pages} {220403} (\bibinfo {year}
  {2011})}\BibitemShut {NoStop}%
\bibitem [{\citenamefont {Choi}\ \emph {et~al.}(2012)\citenamefont {Choi},
  \citenamefont {Coldea}, \citenamefont {Kolmogorov}, \citenamefont
  {Lancaster}, \citenamefont {Mazin}, \citenamefont {Blundell}, \citenamefont
  {Radaelli}, \citenamefont {Singh}, \citenamefont {Gegenwart}, \citenamefont
  {Choi}, \citenamefont {Cheong}, \citenamefont {Baker}, \citenamefont
  {Stock},\ and\ \citenamefont {Taylor}}]{choi2012spin}%
  \BibitemOpen
  \bibfield  {author} {\bibinfo {author} {\bibfnamefont {S.~K.}\ \bibnamefont
  {Choi}}, \bibinfo {author} {\bibfnamefont {R.}~\bibnamefont {Coldea}},
  \bibinfo {author} {\bibfnamefont {A.~N.}\ \bibnamefont {Kolmogorov}},
  \bibinfo {author} {\bibfnamefont {T.}~\bibnamefont {Lancaster}}, \bibinfo
  {author} {\bibfnamefont {I.~I.}\ \bibnamefont {Mazin}}, \bibinfo {author}
  {\bibfnamefont {S.~J.}\ \bibnamefont {Blundell}}, \bibinfo {author}
  {\bibfnamefont {P.~G.}\ \bibnamefont {Radaelli}}, \bibinfo {author}
  {\bibfnamefont {Y.}~\bibnamefont {Singh}}, \bibinfo {author} {\bibfnamefont
  {P.}~\bibnamefont {Gegenwart}}, \bibinfo {author} {\bibfnamefont {K.~R.}\
  \bibnamefont {Choi}}, \bibinfo {author} {\bibfnamefont {S.-W.}\ \bibnamefont
  {Cheong}}, \bibinfo {author} {\bibfnamefont {P.~J.}\ \bibnamefont {Baker}},
  \bibinfo {author} {\bibfnamefont {C.}~\bibnamefont {Stock}}, \ and\ \bibinfo
  {author} {\bibfnamefont {J.}~\bibnamefont {Taylor}},\ }\href {\doibase
  10.1103/PhysRevLett.108.127204} {\bibfield  {journal} {\bibinfo  {journal}
  {Phys. Rev. Lett.}\ }\textbf {\bibinfo {volume} {108}},\ \bibinfo {pages}
  {127204} (\bibinfo {year} {2012})}\BibitemShut {NoStop}%
\bibitem [{\citenamefont {Ye}\ \emph {et~al.}(2012)\citenamefont {Ye},
  \citenamefont {Chi}, \citenamefont {Cao}, \citenamefont {Chakoumakos},
  \citenamefont {Fernandez-Baca}, \citenamefont {Custelcean}, \citenamefont
  {Qi}, \citenamefont {Korneta},\ and\ \citenamefont {Cao}}]{ye2012direct}%
  \BibitemOpen
  \bibfield  {author} {\bibinfo {author} {\bibfnamefont {F.}~\bibnamefont
  {Ye}}, \bibinfo {author} {\bibfnamefont {S.}~\bibnamefont {Chi}}, \bibinfo
  {author} {\bibfnamefont {H.}~\bibnamefont {Cao}}, \bibinfo {author}
  {\bibfnamefont {B.~C.}\ \bibnamefont {Chakoumakos}}, \bibinfo {author}
  {\bibfnamefont {J.~A.}\ \bibnamefont {Fernandez-Baca}}, \bibinfo {author}
  {\bibfnamefont {R.}~\bibnamefont {Custelcean}}, \bibinfo {author}
  {\bibfnamefont {T.~F.}\ \bibnamefont {Qi}}, \bibinfo {author} {\bibfnamefont
  {O.~B.}\ \bibnamefont {Korneta}}, \ and\ \bibinfo {author} {\bibfnamefont
  {G.}~\bibnamefont {Cao}},\ }\href {\doibase 10.1103/PhysRevB.85.180403}
  {\bibfield  {journal} {\bibinfo  {journal} {Phys. Rev. B}\ }\textbf {\bibinfo
  {volume} {85}},\ \bibinfo {pages} {180403} (\bibinfo {year}
  {2012})}\BibitemShut {NoStop}%
\bibitem [{\citenamefont {Comin}\ \emph {et~al.}(2012)\citenamefont {Comin},
  \citenamefont {Levy}, \citenamefont {Ludbrook}, \citenamefont {Zhu},
  \citenamefont {Veenstra}, \citenamefont {Rosen}, \citenamefont {Singh},
  \citenamefont {Gegenwart}, \citenamefont {Stricker}, \citenamefont {Hancock},
  \citenamefont {van~der Marel}, \citenamefont {Elfimov},\ and\ \citenamefont
  {Damascelli}}]{comin2012novel}%
  \BibitemOpen
  \bibfield  {author} {\bibinfo {author} {\bibfnamefont {R.}~\bibnamefont
  {Comin}}, \bibinfo {author} {\bibfnamefont {G.}~\bibnamefont {Levy}},
  \bibinfo {author} {\bibfnamefont {B.}~\bibnamefont {Ludbrook}}, \bibinfo
  {author} {\bibfnamefont {Z.-H.}\ \bibnamefont {Zhu}}, \bibinfo {author}
  {\bibfnamefont {C.~N.}\ \bibnamefont {Veenstra}}, \bibinfo {author}
  {\bibfnamefont {J.~A.}\ \bibnamefont {Rosen}}, \bibinfo {author}
  {\bibfnamefont {Y.}~\bibnamefont {Singh}}, \bibinfo {author} {\bibfnamefont
  {P.}~\bibnamefont {Gegenwart}}, \bibinfo {author} {\bibfnamefont
  {D.}~\bibnamefont {Stricker}}, \bibinfo {author} {\bibfnamefont {J.~N.}\
  \bibnamefont {Hancock}}, \bibinfo {author} {\bibfnamefont {D.}~\bibnamefont
  {van~der Marel}}, \bibinfo {author} {\bibfnamefont {I.~S.}\ \bibnamefont
  {Elfimov}}, \ and\ \bibinfo {author} {\bibfnamefont {A.}~\bibnamefont
  {Damascelli}},\ }\href {\doibase 10.1103/PhysRevLett.109.266406} {\bibfield
  {journal} {\bibinfo  {journal} {Phys. Rev. Lett.}\ }\textbf {\bibinfo
  {volume} {109}},\ \bibinfo {pages} {266406} (\bibinfo {year}
  {2012})}\BibitemShut {NoStop}%
\bibitem [{\citenamefont {Hwan~Chun}\ \emph {et~al.}(2015)\citenamefont
  {Hwan~Chun}, \citenamefont {Kim}, \citenamefont {Kim}, \citenamefont {Zheng},
  \citenamefont {Stoumpos}, \citenamefont {Malliakas}, \citenamefont
  {Mitchell}, \citenamefont {Mehlawat}, \citenamefont {Singh}, \citenamefont
  {Choi}, \citenamefont {Gog}, \citenamefont {Al-Zein}, \citenamefont {Sala},
  \citenamefont {Krisch}, \citenamefont {Chaloupka}, \citenamefont {Jackeli},
  \citenamefont {Khaliullin},\ and\ \citenamefont {Kim}}]{chun2015direct}%
  \BibitemOpen
  \bibfield  {author} {\bibinfo {author} {\bibfnamefont {S.}~\bibnamefont
  {Hwan~Chun}}, \bibinfo {author} {\bibfnamefont {J.-W.}\ \bibnamefont {Kim}},
  \bibinfo {author} {\bibfnamefont {J.}~\bibnamefont {Kim}}, \bibinfo {author}
  {\bibfnamefont {H.}~\bibnamefont {Zheng}}, \bibinfo {author} {\bibfnamefont
  {C.~C.}\ \bibnamefont {Stoumpos}}, \bibinfo {author} {\bibfnamefont {C.~D.}\
  \bibnamefont {Malliakas}}, \bibinfo {author} {\bibfnamefont {J.~F.}\
  \bibnamefont {Mitchell}}, \bibinfo {author} {\bibfnamefont {K.}~\bibnamefont
  {Mehlawat}}, \bibinfo {author} {\bibfnamefont {Y.}~\bibnamefont {Singh}},
  \bibinfo {author} {\bibfnamefont {Y.}~\bibnamefont {Choi}}, \bibinfo {author}
  {\bibfnamefont {T.}~\bibnamefont {Gog}}, \bibinfo {author} {\bibfnamefont
  {A.}~\bibnamefont {Al-Zein}}, \bibinfo {author} {\bibfnamefont {M.~M.}\
  \bibnamefont {Sala}}, \bibinfo {author} {\bibfnamefont {M.}~\bibnamefont
  {Krisch}}, \bibinfo {author} {\bibfnamefont {J.}~\bibnamefont {Chaloupka}},
  \bibinfo {author} {\bibfnamefont {G.}~\bibnamefont {Jackeli}}, \bibinfo
  {author} {\bibfnamefont {G.}~\bibnamefont {Khaliullin}}, \ and\ \bibinfo
  {author} {\bibfnamefont {B.~J.}\ \bibnamefont {Kim}},\ }\href
  {http://dx.doi.org/10.1038/nphys3322} {\bibfield  {journal} {\bibinfo
  {journal} {Nat. Phys.}\ }\textbf {\bibinfo {volume} {11}},\ \bibinfo {pages}
  {462 } (\bibinfo {year} {2015})}\BibitemShut {NoStop}%
\bibitem [{\citenamefont {Singh}\ \emph {et~al.}(2012)\citenamefont {Singh},
  \citenamefont {Manni}, \citenamefont {Reuther}, \citenamefont {Berlijn},
  \citenamefont {Thomale}, \citenamefont {Ku}, \citenamefont {Trebst},\ and\
  \citenamefont {Gegenwart}}]{singh2012relevance}%
  \BibitemOpen
  \bibfield  {author} {\bibinfo {author} {\bibfnamefont {Y.}~\bibnamefont
  {Singh}}, \bibinfo {author} {\bibfnamefont {S.}~\bibnamefont {Manni}},
  \bibinfo {author} {\bibfnamefont {J.}~\bibnamefont {Reuther}}, \bibinfo
  {author} {\bibfnamefont {T.}~\bibnamefont {Berlijn}}, \bibinfo {author}
  {\bibfnamefont {R.}~\bibnamefont {Thomale}}, \bibinfo {author} {\bibfnamefont
  {W.}~\bibnamefont {Ku}}, \bibinfo {author} {\bibfnamefont {S.}~\bibnamefont
  {Trebst}}, \ and\ \bibinfo {author} {\bibfnamefont {P.}~\bibnamefont
  {Gegenwart}},\ }\href {\doibase 10.1103/PhysRevLett.108.127203} {\bibfield
  {journal} {\bibinfo  {journal} {Phys. Rev. Lett.}\ }\textbf {\bibinfo
  {volume} {108}},\ \bibinfo {pages} {127203} (\bibinfo {year}
  {2012})}\BibitemShut {NoStop}%
\bibitem [{\citenamefont {Williams}\ \emph {et~al.}(2016)\citenamefont
  {Williams}, \citenamefont {Johnson}, \citenamefont {Freund}, \citenamefont
  {Choi}, \citenamefont {Jesche}, \citenamefont {Kimchi}, \citenamefont
  {Manni}, \citenamefont {Bombardi}, \citenamefont {Manuel}, \citenamefont
  {Gegenwart},\ and\ \citenamefont {Coldea}}]{williams2016incommensurate}%
  \BibitemOpen
  \bibfield  {author} {\bibinfo {author} {\bibfnamefont {S.~C.}\ \bibnamefont
  {Williams}}, \bibinfo {author} {\bibfnamefont {R.~D.}\ \bibnamefont
  {Johnson}}, \bibinfo {author} {\bibfnamefont {F.}~\bibnamefont {Freund}},
  \bibinfo {author} {\bibfnamefont {S.}~\bibnamefont {Choi}}, \bibinfo {author}
  {\bibfnamefont {A.}~\bibnamefont {Jesche}}, \bibinfo {author} {\bibfnamefont
  {I.}~\bibnamefont {Kimchi}}, \bibinfo {author} {\bibfnamefont
  {S.}~\bibnamefont {Manni}}, \bibinfo {author} {\bibfnamefont
  {A.}~\bibnamefont {Bombardi}}, \bibinfo {author} {\bibfnamefont
  {P.}~\bibnamefont {Manuel}}, \bibinfo {author} {\bibfnamefont
  {P.}~\bibnamefont {Gegenwart}}, \ and\ \bibinfo {author} {\bibfnamefont
  {R.}~\bibnamefont {Coldea}},\ }\href {\doibase 10.1103/PhysRevB.93.195158}
  {\bibfield  {journal} {\bibinfo  {journal} {Phys. Rev. B}\ }\textbf {\bibinfo
  {volume} {93}},\ \bibinfo {pages} {195158} (\bibinfo {year}
  {2016})}\BibitemShut {NoStop}%
\bibitem [{\citenamefont {Kitagawa}\ \emph {et~al.}(2018)\citenamefont
  {Kitagawa}, \citenamefont {Takayama}, \citenamefont {Matsumoto},
  \citenamefont {Kato}, \citenamefont {Takano}, \citenamefont {Kishimoto},
  \citenamefont {Dinnebier}, \citenamefont {Jackeli},\ and\ \citenamefont
  {Takagi}}]{kitagawa2018spin}%
  \BibitemOpen
  \bibfield  {author} {\bibinfo {author} {\bibfnamefont {K.}~\bibnamefont
  {Kitagawa}}, \bibinfo {author} {\bibfnamefont {T.}~\bibnamefont {Takayama}},
  \bibinfo {author} {\bibfnamefont {Y.}~\bibnamefont {Matsumoto}}, \bibinfo
  {author} {\bibfnamefont {A.}~\bibnamefont {Kato}}, \bibinfo {author}
  {\bibfnamefont {R.}~\bibnamefont {Takano}}, \bibinfo {author} {\bibfnamefont
  {Y.}~\bibnamefont {Kishimoto}}, \bibinfo {author} {\bibfnamefont
  {R.}~\bibnamefont {Dinnebier}}, \bibinfo {author} {\bibfnamefont
  {G.}~\bibnamefont {Jackeli}}, \ and\ \bibinfo {author} {\bibfnamefont
  {H.}~\bibnamefont {Takagi}},\ }\href {\doibase
  http://dx.doi.org/10.1038/nature25482} {\bibfield  {journal} {\bibinfo
  {journal} {Nature}\ }\textbf {\bibinfo {volume} {554}},\ \bibinfo {pages}
  {341} (\bibinfo {year} {2018})}\BibitemShut {NoStop}%
\bibitem [{\citenamefont {Bahrami}\ \emph {et~al.}(2019)\citenamefont
  {Bahrami}, \citenamefont {Lafargue-Dit-Hauret}, \citenamefont {Lebedev},
  \citenamefont {Movshovich}, \citenamefont {Yang}, \citenamefont {Broido},
  \citenamefont {Rocquefelte},\ and\ \citenamefont
  {Tafti}}]{bahrami2019thermodynamic}%
  \BibitemOpen
  \bibfield  {author} {\bibinfo {author} {\bibfnamefont {F.}~\bibnamefont
  {Bahrami}}, \bibinfo {author} {\bibfnamefont {W.}~\bibnamefont
  {Lafargue-Dit-Hauret}}, \bibinfo {author} {\bibfnamefont {O.~I.}\
  \bibnamefont {Lebedev}}, \bibinfo {author} {\bibfnamefont {R.}~\bibnamefont
  {Movshovich}}, \bibinfo {author} {\bibfnamefont {H.-Y.}\ \bibnamefont
  {Yang}}, \bibinfo {author} {\bibfnamefont {D.}~\bibnamefont {Broido}},
  \bibinfo {author} {\bibfnamefont {X.}~\bibnamefont {Rocquefelte}}, \ and\
  \bibinfo {author} {\bibfnamefont {F.}~\bibnamefont {Tafti}},\ }\href
  {\doibase 10.1103/PhysRevLett.123.237203} {\bibfield  {journal} {\bibinfo
  {journal} {Phys. Rev. Lett.}\ }\textbf {\bibinfo {volume} {123}},\ \bibinfo
  {pages} {237203} (\bibinfo {year} {2019})}\BibitemShut {NoStop}%
\bibitem [{\citenamefont {Plumb}\ \emph {et~al.}(2014)\citenamefont {Plumb},
  \citenamefont {Clancy}, \citenamefont {Sandilands}, \citenamefont {Shankar},
  \citenamefont {Hu}, \citenamefont {Burch}, \citenamefont {Kee},\ and\
  \citenamefont {Kim}}]{plumb2014spin}%
  \BibitemOpen
  \bibfield  {author} {\bibinfo {author} {\bibfnamefont {K.~W.}\ \bibnamefont
  {Plumb}}, \bibinfo {author} {\bibfnamefont {J.~P.}\ \bibnamefont {Clancy}},
  \bibinfo {author} {\bibfnamefont {L.~J.}\ \bibnamefont {Sandilands}},
  \bibinfo {author} {\bibfnamefont {V.~V.}\ \bibnamefont {Shankar}}, \bibinfo
  {author} {\bibfnamefont {Y.~F.}\ \bibnamefont {Hu}}, \bibinfo {author}
  {\bibfnamefont {K.~S.}\ \bibnamefont {Burch}}, \bibinfo {author}
  {\bibfnamefont {H.-Y.}\ \bibnamefont {Kee}}, \ and\ \bibinfo {author}
  {\bibfnamefont {Y.-J.}\ \bibnamefont {Kim}},\ }\href {\doibase
  10.1103/PhysRevB.90.041112} {\bibfield  {journal} {\bibinfo  {journal} {Phys.
  Rev. B}\ }\textbf {\bibinfo {volume} {90}},\ \bibinfo {pages} {041112}
  (\bibinfo {year} {2014})}\BibitemShut {NoStop}%
\bibitem [{\citenamefont {Sandilands}\ \emph {et~al.}(2015)\citenamefont
  {Sandilands}, \citenamefont {Tian}, \citenamefont {Plumb}, \citenamefont
  {Kim},\ and\ \citenamefont {Burch}}]{sandilands2015scattering}%
  \BibitemOpen
  \bibfield  {author} {\bibinfo {author} {\bibfnamefont {L.~J.}\ \bibnamefont
  {Sandilands}}, \bibinfo {author} {\bibfnamefont {Y.}~\bibnamefont {Tian}},
  \bibinfo {author} {\bibfnamefont {K.~W.}\ \bibnamefont {Plumb}}, \bibinfo
  {author} {\bibfnamefont {Y.-J.}\ \bibnamefont {Kim}}, \ and\ \bibinfo
  {author} {\bibfnamefont {K.~S.}\ \bibnamefont {Burch}},\ }\href {\doibase
  10.1103/PhysRevLett.114.147201} {\bibfield  {journal} {\bibinfo  {journal}
  {Phys. Rev. Lett.}\ }\textbf {\bibinfo {volume} {114}},\ \bibinfo {pages}
  {147201} (\bibinfo {year} {2015})}\BibitemShut {NoStop}%
\bibitem [{\citenamefont {Sears}\ \emph {et~al.}(2015)\citenamefont {Sears},
  \citenamefont {Songvilay}, \citenamefont {Plumb}, \citenamefont {Clancy},
  \citenamefont {Qiu}, \citenamefont {Zhao}, \citenamefont {Parshall},\ and\
  \citenamefont {Kim}}]{sears2015magnetic}%
  \BibitemOpen
  \bibfield  {author} {\bibinfo {author} {\bibfnamefont {J.~A.}\ \bibnamefont
  {Sears}}, \bibinfo {author} {\bibfnamefont {M.}~\bibnamefont {Songvilay}},
  \bibinfo {author} {\bibfnamefont {K.~W.}\ \bibnamefont {Plumb}}, \bibinfo
  {author} {\bibfnamefont {J.~P.}\ \bibnamefont {Clancy}}, \bibinfo {author}
  {\bibfnamefont {Y.}~\bibnamefont {Qiu}}, \bibinfo {author} {\bibfnamefont
  {Y.}~\bibnamefont {Zhao}}, \bibinfo {author} {\bibfnamefont {D.}~\bibnamefont
  {Parshall}}, \ and\ \bibinfo {author} {\bibfnamefont {Y.-J.}\ \bibnamefont
  {Kim}},\ }\href {\doibase 10.1103/PhysRevB.91.144420} {\bibfield  {journal}
  {\bibinfo  {journal} {Phys. Rev. B}\ }\textbf {\bibinfo {volume} {91}},\
  \bibinfo {pages} {144420} (\bibinfo {year} {2015})}\BibitemShut {NoStop}%
\bibitem [{\citenamefont {Majumder}\ \emph {et~al.}(2015)\citenamefont
  {Majumder}, \citenamefont {Schmidt}, \citenamefont {Rosner}, \citenamefont
  {Tsirlin}, \citenamefont {Yasuoka},\ and\ \citenamefont
  {Baenitz}}]{majumder2015anisotropic}%
  \BibitemOpen
  \bibfield  {author} {\bibinfo {author} {\bibfnamefont {M.}~\bibnamefont
  {Majumder}}, \bibinfo {author} {\bibfnamefont {M.}~\bibnamefont {Schmidt}},
  \bibinfo {author} {\bibfnamefont {H.}~\bibnamefont {Rosner}}, \bibinfo
  {author} {\bibfnamefont {A.~A.}\ \bibnamefont {Tsirlin}}, \bibinfo {author}
  {\bibfnamefont {H.}~\bibnamefont {Yasuoka}}, \ and\ \bibinfo {author}
  {\bibfnamefont {M.}~\bibnamefont {Baenitz}},\ }\href {\doibase
  10.1103/PhysRevB.91.180401} {\bibfield  {journal} {\bibinfo  {journal} {Phys.
  Rev. B}\ }\textbf {\bibinfo {volume} {91}},\ \bibinfo {pages} {180401}
  (\bibinfo {year} {2015})}\BibitemShut {NoStop}%
\bibitem [{\citenamefont {Johnson}\ \emph {et~al.}(2015)\citenamefont
  {Johnson}, \citenamefont {Williams}, \citenamefont {Haghighirad},
  \citenamefont {Singleton}, \citenamefont {Zapf}, \citenamefont {Manuel},
  \citenamefont {Mazin}, \citenamefont {Li}, \citenamefont {Jeschke},
  \citenamefont {Valent\'{\i}},\ and\ \citenamefont
  {Coldea}}]{johnson2015monoclinic}%
  \BibitemOpen
  \bibfield  {author} {\bibinfo {author} {\bibfnamefont {R.~D.}\ \bibnamefont
  {Johnson}}, \bibinfo {author} {\bibfnamefont {S.~C.}\ \bibnamefont
  {Williams}}, \bibinfo {author} {\bibfnamefont {A.~A.}\ \bibnamefont
  {Haghighirad}}, \bibinfo {author} {\bibfnamefont {J.}~\bibnamefont
  {Singleton}}, \bibinfo {author} {\bibfnamefont {V.}~\bibnamefont {Zapf}},
  \bibinfo {author} {\bibfnamefont {P.}~\bibnamefont {Manuel}}, \bibinfo
  {author} {\bibfnamefont {I.~I.}\ \bibnamefont {Mazin}}, \bibinfo {author}
  {\bibfnamefont {Y.}~\bibnamefont {Li}}, \bibinfo {author} {\bibfnamefont
  {H.~O.}\ \bibnamefont {Jeschke}}, \bibinfo {author} {\bibfnamefont
  {R.}~\bibnamefont {Valent\'{\i}}}, \ and\ \bibinfo {author} {\bibfnamefont
  {R.}~\bibnamefont {Coldea}},\ }\href {\doibase 10.1103/PhysRevB.92.235119}
  {\bibfield  {journal} {\bibinfo  {journal} {Phys. Rev. B}\ }\textbf {\bibinfo
  {volume} {92}},\ \bibinfo {pages} {235119} (\bibinfo {year}
  {2015})}\BibitemShut {NoStop}%
\bibitem [{\citenamefont {Sandilands}\ \emph {et~al.}(2016)\citenamefont
  {Sandilands}, \citenamefont {Tian}, \citenamefont {Reijnders}, \citenamefont
  {Kim}, \citenamefont {Plumb}, \citenamefont {Kim}, \citenamefont {Kee},\ and\
  \citenamefont {Burch}}]{sandilands2016spin}%
  \BibitemOpen
  \bibfield  {author} {\bibinfo {author} {\bibfnamefont {L.~J.}\ \bibnamefont
  {Sandilands}}, \bibinfo {author} {\bibfnamefont {Y.}~\bibnamefont {Tian}},
  \bibinfo {author} {\bibfnamefont {A.~A.}\ \bibnamefont {Reijnders}}, \bibinfo
  {author} {\bibfnamefont {H.-S.}\ \bibnamefont {Kim}}, \bibinfo {author}
  {\bibfnamefont {K.~W.}\ \bibnamefont {Plumb}}, \bibinfo {author}
  {\bibfnamefont {Y.-J.}\ \bibnamefont {Kim}}, \bibinfo {author} {\bibfnamefont
  {H.-Y.}\ \bibnamefont {Kee}}, \ and\ \bibinfo {author} {\bibfnamefont
  {K.~S.}\ \bibnamefont {Burch}},\ }\href {\doibase 10.1103/PhysRevB.93.075144}
  {\bibfield  {journal} {\bibinfo  {journal} {Phys. Rev. B}\ }\textbf {\bibinfo
  {volume} {93}},\ \bibinfo {pages} {075144} (\bibinfo {year}
  {2016})}\BibitemShut {NoStop}%
\bibitem [{\citenamefont {Banerjee}\ \emph {et~al.}(2016)\citenamefont
  {Banerjee}, \citenamefont {Bridges}, \citenamefont {Yan}, \citenamefont
  {Aczel}, \citenamefont {Li}, \citenamefont {Stone}, \citenamefont {Granroth},
  \citenamefont {Lumsden}, \citenamefont {Yiu}, \citenamefont {Knolle},
  \citenamefont {Bhattacharjee}, \citenamefont {Kovrizhin}, \citenamefont
  {Moessner}, \citenamefont {Tennant}, \citenamefont {G.},\ and\ \citenamefont
  {Nagler}}]{banerjee2016proximate}%
  \BibitemOpen
  \bibfield  {author} {\bibinfo {author} {\bibfnamefont {A.}~\bibnamefont
  {Banerjee}}, \bibinfo {author} {\bibfnamefont {C.~A.}\ \bibnamefont
  {Bridges}}, \bibinfo {author} {\bibfnamefont {J.-Q.}\ \bibnamefont {Yan}},
  \bibinfo {author} {\bibfnamefont {A.~A.}\ \bibnamefont {Aczel}}, \bibinfo
  {author} {\bibfnamefont {L.}~\bibnamefont {Li}}, \bibinfo {author}
  {\bibfnamefont {M.~B.}\ \bibnamefont {Stone}}, \bibinfo {author}
  {\bibfnamefont {G.~E.}\ \bibnamefont {Granroth}}, \bibinfo {author}
  {\bibfnamefont {M.~D.}\ \bibnamefont {Lumsden}}, \bibinfo {author}
  {\bibfnamefont {Y.}~\bibnamefont {Yiu}}, \bibinfo {author} {\bibfnamefont
  {J.}~\bibnamefont {Knolle}}, \bibinfo {author} {\bibfnamefont
  {S.}~\bibnamefont {Bhattacharjee}}, \bibinfo {author} {\bibfnamefont {D.~L.}\
  \bibnamefont {Kovrizhin}}, \bibinfo {author} {\bibfnamefont {R.}~\bibnamefont
  {Moessner}}, \bibinfo {author} {\bibfnamefont {D.~A.}\ \bibnamefont
  {Tennant}}, \bibinfo {author} {\bibfnamefont {M.~D.}\ \bibnamefont {G.}}, \
  and\ \bibinfo {author} {\bibfnamefont {S.~E.}\ \bibnamefont {Nagler}},\
  }\href {\doibase 10.1038/nmat4604} {\bibfield  {journal} {\bibinfo  {journal}
  {Nature materials}\ } (\bibinfo {year} {2016}),\
  10.1038/nmat4604}\BibitemShut {NoStop}%
\bibitem [{\citenamefont {Banerjee}\ \emph {et~al.}(2017)\citenamefont
  {Banerjee}, \citenamefont {Yan}, \citenamefont {Knolle}, \citenamefont
  {Bridges}, \citenamefont {Stone}, \citenamefont {Lumsden}, \citenamefont
  {Mandrus}, \citenamefont {Tennant}, \citenamefont {Moessner},\ and\
  \citenamefont {Nagler}}]{banerjee2017neutron}%
  \BibitemOpen
  \bibfield  {author} {\bibinfo {author} {\bibfnamefont {A.}~\bibnamefont
  {Banerjee}}, \bibinfo {author} {\bibfnamefont {J.}~\bibnamefont {Yan}},
  \bibinfo {author} {\bibfnamefont {J.}~\bibnamefont {Knolle}}, \bibinfo
  {author} {\bibfnamefont {C.~A.}\ \bibnamefont {Bridges}}, \bibinfo {author}
  {\bibfnamefont {M.~B.}\ \bibnamefont {Stone}}, \bibinfo {author}
  {\bibfnamefont {M.~D.}\ \bibnamefont {Lumsden}}, \bibinfo {author}
  {\bibfnamefont {D.~G.}\ \bibnamefont {Mandrus}}, \bibinfo {author}
  {\bibfnamefont {D.~A.}\ \bibnamefont {Tennant}}, \bibinfo {author}
  {\bibfnamefont {R.}~\bibnamefont {Moessner}}, \ and\ \bibinfo {author}
  {\bibfnamefont {S.~E.}\ \bibnamefont {Nagler}},\ }\href {\doibase
  10.1126/science.aah6015} {\bibfield  {journal} {\bibinfo  {journal}
  {Science}\ }\textbf {\bibinfo {volume} {356}},\ \bibinfo {pages} {1055}
  (\bibinfo {year} {2017})}\BibitemShut {NoStop}%
\bibitem [{\citenamefont {Do}\ \emph {et~al.}(2017)\citenamefont {Do},
  \citenamefont {Park}, \citenamefont {Yoshitake}, \citenamefont {Nasu},
  \citenamefont {Motome}, \citenamefont {Kwon}, \citenamefont {Adroja},
  \citenamefont {Voneshen}, \citenamefont {Kim}, \citenamefont {Jang},
  \citenamefont {Park}, \citenamefont {Choi},\ and\ \citenamefont
  {Ji}}]{do2017majorana}%
  \BibitemOpen
  \bibfield  {author} {\bibinfo {author} {\bibfnamefont {S.-H.}\ \bibnamefont
  {Do}}, \bibinfo {author} {\bibfnamefont {S.-Y.}\ \bibnamefont {Park}},
  \bibinfo {author} {\bibfnamefont {J.}~\bibnamefont {Yoshitake}}, \bibinfo
  {author} {\bibfnamefont {J.}~\bibnamefont {Nasu}}, \bibinfo {author}
  {\bibfnamefont {Y.}~\bibnamefont {Motome}}, \bibinfo {author} {\bibfnamefont
  {Y.~S.}\ \bibnamefont {Kwon}}, \bibinfo {author} {\bibfnamefont {D.~T.}\
  \bibnamefont {Adroja}}, \bibinfo {author} {\bibfnamefont {D.~J.}\
  \bibnamefont {Voneshen}}, \bibinfo {author} {\bibfnamefont {K.}~\bibnamefont
  {Kim}}, \bibinfo {author} {\bibfnamefont {T.-H.}\ \bibnamefont {Jang}},
  \bibinfo {author} {\bibfnamefont {J.-H.}\ \bibnamefont {Park}}, \bibinfo
  {author} {\bibfnamefont {K.-Y.}\ \bibnamefont {Choi}}, \ and\ \bibinfo
  {author} {\bibfnamefont {S.}~\bibnamefont {Ji}},\ }\href {\doibase
  10.1038/nphys4264} {\bibfield  {journal} {\bibinfo  {journal} {Nature
  Physics}\ }\textbf {\bibinfo {volume} {13}},\ \bibinfo {pages} {1079}
  (\bibinfo {year} {2017})}\BibitemShut {NoStop}%
\bibitem [{\citenamefont {Yamaji}\ \emph {et~al.}(2014)\citenamefont {Yamaji},
  \citenamefont {Nomura}, \citenamefont {Kurita}, \citenamefont {Arita},\ and\
  \citenamefont {Imada}}]{yamaji2014first}%
  \BibitemOpen
  \bibfield  {author} {\bibinfo {author} {\bibfnamefont {Y.}~\bibnamefont
  {Yamaji}}, \bibinfo {author} {\bibfnamefont {Y.}~\bibnamefont {Nomura}},
  \bibinfo {author} {\bibfnamefont {M.}~\bibnamefont {Kurita}}, \bibinfo
  {author} {\bibfnamefont {R.}~\bibnamefont {Arita}}, \ and\ \bibinfo {author}
  {\bibfnamefont {M.}~\bibnamefont {Imada}},\ }\href {\doibase
  10.1103/PhysRevLett.113.107201} {\bibfield  {journal} {\bibinfo  {journal}
  {Phys. Rev. Lett.}\ }\textbf {\bibinfo {volume} {113}},\ \bibinfo {pages}
  {107201} (\bibinfo {year} {2014})}\BibitemShut {NoStop}%
\bibitem [{\citenamefont {Sizyuk}\ \emph {et~al.}(2014)\citenamefont {Sizyuk},
  \citenamefont {Price}, \citenamefont {W\"olfle},\ and\ \citenamefont
  {Perkins}}]{sizyuk2014importance}%
  \BibitemOpen
  \bibfield  {author} {\bibinfo {author} {\bibfnamefont {Y.}~\bibnamefont
  {Sizyuk}}, \bibinfo {author} {\bibfnamefont {C.}~\bibnamefont {Price}},
  \bibinfo {author} {\bibfnamefont {P.}~\bibnamefont {W\"olfle}}, \ and\
  \bibinfo {author} {\bibfnamefont {N.~B.}\ \bibnamefont {Perkins}},\ }\href
  {\doibase 10.1103/PhysRevB.90.155126} {\bibfield  {journal} {\bibinfo
  {journal} {Phys. Rev. B}\ }\textbf {\bibinfo {volume} {90}},\ \bibinfo
  {pages} {155126} (\bibinfo {year} {2014})}\BibitemShut {NoStop}%
\bibitem [{\citenamefont {Kim}\ \emph {et~al.}(2015)\citenamefont {Kim},
  \citenamefont {V.}, \citenamefont {Catuneanu},\ and\ \citenamefont
  {Kee}}]{kim2015kitaev}%
  \BibitemOpen
  \bibfield  {author} {\bibinfo {author} {\bibfnamefont {H.-S.}\ \bibnamefont
  {Kim}}, \bibinfo {author} {\bibfnamefont {V.~S.}\ \bibnamefont {V.}},
  \bibinfo {author} {\bibfnamefont {A.}~\bibnamefont {Catuneanu}}, \ and\
  \bibinfo {author} {\bibfnamefont {H.-Y.}\ \bibnamefont {Kee}},\ }\href
  {\doibase 10.1103/PhysRevB.91.241110} {\bibfield  {journal} {\bibinfo
  {journal} {Phys. Rev. B}\ }\textbf {\bibinfo {volume} {91}},\ \bibinfo
  {pages} {241110} (\bibinfo {year} {2015})}\BibitemShut {NoStop}%
\bibitem [{\citenamefont {Chaloupka}\ and\ \citenamefont
  {Khaliullin}(2015)}]{chaloupka2015hidden}%
  \BibitemOpen
  \bibfield  {author} {\bibinfo {author} {\bibfnamefont {J.~c.~v.}\
  \bibnamefont {Chaloupka}}\ and\ \bibinfo {author} {\bibfnamefont
  {G.}~\bibnamefont {Khaliullin}},\ }\href {\doibase
  10.1103/PhysRevB.92.024413} {\bibfield  {journal} {\bibinfo  {journal} {Phys.
  Rev. B}\ }\textbf {\bibinfo {volume} {92}},\ \bibinfo {pages} {024413}
  (\bibinfo {year} {2015})}\BibitemShut {NoStop}%
\bibitem [{\citenamefont {Rousochatzakis}\ \emph {et~al.}(2015)\citenamefont
  {Rousochatzakis}, \citenamefont {Reuther}, \citenamefont {Thomale},
  \citenamefont {Rachel},\ and\ \citenamefont
  {Perkins}}]{rousochatzakis2015phase}%
  \BibitemOpen
  \bibfield  {author} {\bibinfo {author} {\bibfnamefont {I.}~\bibnamefont
  {Rousochatzakis}}, \bibinfo {author} {\bibfnamefont {J.}~\bibnamefont
  {Reuther}}, \bibinfo {author} {\bibfnamefont {R.}~\bibnamefont {Thomale}},
  \bibinfo {author} {\bibfnamefont {S.}~\bibnamefont {Rachel}}, \ and\ \bibinfo
  {author} {\bibfnamefont {N.~B.}\ \bibnamefont {Perkins}},\ }\href {\doibase
  10.1103/PhysRevX.5.041035} {\bibfield  {journal} {\bibinfo  {journal} {Phys.
  Rev. X}\ }\textbf {\bibinfo {volume} {5}},\ \bibinfo {pages} {041035}
  (\bibinfo {year} {2015})}\BibitemShut {NoStop}%
\bibitem [{\citenamefont {Nishimoto}\ \emph {et~al.}(2016)\citenamefont
  {Nishimoto}, \citenamefont {Katukuri}, \citenamefont {Yushankhai},
  \citenamefont {Stoll}, \citenamefont {R{\"o}{\ss}ler}, \citenamefont {Hozoi},
  \citenamefont {Rousochatzakis},\ and\ \citenamefont {van~den
  Brink}}]{nishimoto2016strongly}%
  \BibitemOpen
  \bibfield  {author} {\bibinfo {author} {\bibfnamefont {S.}~\bibnamefont
  {Nishimoto}}, \bibinfo {author} {\bibfnamefont {V.~M.}\ \bibnamefont
  {Katukuri}}, \bibinfo {author} {\bibfnamefont {V.}~\bibnamefont
  {Yushankhai}}, \bibinfo {author} {\bibfnamefont {H.}~\bibnamefont {Stoll}},
  \bibinfo {author} {\bibfnamefont {U.~K.}\ \bibnamefont {R{\"o}{\ss}ler}},
  \bibinfo {author} {\bibfnamefont {L.}~\bibnamefont {Hozoi}}, \bibinfo
  {author} {\bibfnamefont {I.}~\bibnamefont {Rousochatzakis}}, \ and\ \bibinfo
  {author} {\bibfnamefont {J.}~\bibnamefont {van~den Brink}},\ }\href {\doibase
  10.1038/ncomms10273} {\bibfield  {journal} {\bibinfo  {journal} {Nature
  Communications}\ }\textbf {\bibinfo {volume} {7}},\ \bibinfo {pages} {10273}
  (\bibinfo {year} {2016})}\BibitemShut {NoStop}%
\bibitem [{\citenamefont {Kim}\ and\ \citenamefont
  {Kee}(2016)}]{kim2016crystal}%
  \BibitemOpen
  \bibfield  {author} {\bibinfo {author} {\bibfnamefont {H.-S.}\ \bibnamefont
  {Kim}}\ and\ \bibinfo {author} {\bibfnamefont {H.-Y.}\ \bibnamefont {Kee}},\
  }\href {\doibase 10.1103/PhysRevB.93.155143} {\bibfield  {journal} {\bibinfo
  {journal} {Phys. Rev. B}\ }\textbf {\bibinfo {volume} {93}},\ \bibinfo
  {pages} {155143} (\bibinfo {year} {2016})}\BibitemShut {NoStop}%
\bibitem [{\citenamefont {Winter}\ \emph {et~al.}(2016)\citenamefont {Winter},
  \citenamefont {Li}, \citenamefont {Jeschke},\ and\ \citenamefont
  {Valent\'{\i}}}]{winter2016challenges}%
  \BibitemOpen
  \bibfield  {author} {\bibinfo {author} {\bibfnamefont {S.~M.}\ \bibnamefont
  {Winter}}, \bibinfo {author} {\bibfnamefont {Y.}~\bibnamefont {Li}}, \bibinfo
  {author} {\bibfnamefont {H.~O.}\ \bibnamefont {Jeschke}}, \ and\ \bibinfo
  {author} {\bibfnamefont {R.}~\bibnamefont {Valent\'{\i}}},\ }\href {\doibase
  10.1103/PhysRevB.93.214431} {\bibfield  {journal} {\bibinfo  {journal} {Phys.
  Rev. B}\ }\textbf {\bibinfo {volume} {93}},\ \bibinfo {pages} {214431}
  (\bibinfo {year} {2016})}\BibitemShut {NoStop}%
\bibitem [{\citenamefont {Chaloupka}\ and\ \citenamefont
  {Khaliullin}(2016)}]{chaloupka2016magnetic}%
  \BibitemOpen
  \bibfield  {author} {\bibinfo {author} {\bibfnamefont {J.~c.~v.}\
  \bibnamefont {Chaloupka}}\ and\ \bibinfo {author} {\bibfnamefont
  {G.}~\bibnamefont {Khaliullin}},\ }\href {\doibase
  10.1103/PhysRevB.94.064435} {\bibfield  {journal} {\bibinfo  {journal} {Phys.
  Rev. B}\ }\textbf {\bibinfo {volume} {94}},\ \bibinfo {pages} {064435}
  (\bibinfo {year} {2016})}\BibitemShut {NoStop}%
\bibitem [{\citenamefont {Winter}\ \emph
  {et~al.}(2017{\natexlab{a}})\citenamefont {Winter}, \citenamefont {Riedl},
  \citenamefont {Maksimov}, \citenamefont {Chernyshev}, \citenamefont
  {Honecker},\ and\ \citenamefont {Valent\'i­}}]{winter2017breakdown}%
  \BibitemOpen
  \bibfield  {author} {\bibinfo {author} {\bibfnamefont {S.~M.}\ \bibnamefont
  {Winter}}, \bibinfo {author} {\bibfnamefont {K.}~\bibnamefont {Riedl}},
  \bibinfo {author} {\bibfnamefont {P.~A.}\ \bibnamefont {Maksimov}}, \bibinfo
  {author} {\bibfnamefont {A.~L.}\ \bibnamefont {Chernyshev}}, \bibinfo
  {author} {\bibfnamefont {A.}~\bibnamefont {Honecker}}, \ and\ \bibinfo
  {author} {\bibfnamefont {R.}~\bibnamefont {Valent\'i­}},\ }\href {\doibase
  10.1038/s41467-017-01177-0} {\bibfield  {journal} {\bibinfo  {journal}
  {Nature Communications}\ }\textbf {\bibinfo {volume} {8}},\ \bibinfo {pages}
  {1152} (\bibinfo {year} {2017}{\natexlab{a}})}\BibitemShut {NoStop}%
\bibitem [{\citenamefont {Winter}\ \emph
  {et~al.}(2017{\natexlab{b}})\citenamefont {Winter}, \citenamefont {Tsirlin},
  \citenamefont {Daghofer}, \citenamefont {van~den Brink}, \citenamefont
  {Singh}, \citenamefont {Gegenwart},\ and\ \citenamefont
  {Valent\'i}}]{winter2017models}%
  \BibitemOpen
  \bibfield  {author} {\bibinfo {author} {\bibfnamefont {S.~M.}\ \bibnamefont
  {Winter}}, \bibinfo {author} {\bibfnamefont {A.~A.}\ \bibnamefont {Tsirlin}},
  \bibinfo {author} {\bibfnamefont {M.}~\bibnamefont {Daghofer}}, \bibinfo
  {author} {\bibfnamefont {J.}~\bibnamefont {van~den Brink}}, \bibinfo {author}
  {\bibfnamefont {Y.}~\bibnamefont {Singh}}, \bibinfo {author} {\bibfnamefont
  {P.}~\bibnamefont {Gegenwart}}, \ and\ \bibinfo {author} {\bibfnamefont
  {R.}~\bibnamefont {Valent\'i}},\ }\href
  {http://stacks.iop.org/0953-8984/29/i=49/a=493002} {\bibfield  {journal}
  {\bibinfo  {journal} {Journal of Physics: Condensed Matter}\ }\textbf
  {\bibinfo {volume} {29}},\ \bibinfo {pages} {493002} (\bibinfo {year}
  {2017}{\natexlab{b}})}\BibitemShut {NoStop}%
\bibitem [{\citenamefont {Eichstaedt}\ \emph {et~al.}(2019)\citenamefont
  {Eichstaedt}, \citenamefont {Zhang}, \citenamefont {Laurell}, \citenamefont
  {Okamoto}, \citenamefont {Eguiluz},\ and\ \citenamefont
  {Berlijn}}]{eichstaedt2019deriving}%
  \BibitemOpen
  \bibfield  {author} {\bibinfo {author} {\bibfnamefont {C.}~\bibnamefont
  {Eichstaedt}}, \bibinfo {author} {\bibfnamefont {Y.}~\bibnamefont {Zhang}},
  \bibinfo {author} {\bibfnamefont {P.}~\bibnamefont {Laurell}}, \bibinfo
  {author} {\bibfnamefont {S.}~\bibnamefont {Okamoto}}, \bibinfo {author}
  {\bibfnamefont {A.~G.}\ \bibnamefont {Eguiluz}}, \ and\ \bibinfo {author}
  {\bibfnamefont {T.}~\bibnamefont {Berlijn}},\ }\href {\doibase
  10.1103/PhysRevB.100.075110} {\bibfield  {journal} {\bibinfo  {journal}
  {Phys. Rev. B}\ }\textbf {\bibinfo {volume} {100}},\ \bibinfo {pages}
  {075110} (\bibinfo {year} {2019})}\BibitemShut {NoStop}%
\bibitem [{\citenamefont {Chaloupka}\ \emph {et~al.}(2010)\citenamefont
  {Chaloupka}, \citenamefont {Jackeli},\ and\ \citenamefont
  {Khaliullin}}]{chaloupka2010kitaev}%
  \BibitemOpen
  \bibfield  {author} {\bibinfo {author} {\bibfnamefont {J.}~\bibnamefont
  {Chaloupka}}, \bibinfo {author} {\bibfnamefont {G.}~\bibnamefont {Jackeli}},
  \ and\ \bibinfo {author} {\bibfnamefont {G.}~\bibnamefont {Khaliullin}},\
  }\href {\doibase 10.1103/PhysRevLett.105.027204} {\bibfield  {journal}
  {\bibinfo  {journal} {Phys. Rev. Lett.}\ }\textbf {\bibinfo {volume} {105}},\
  \bibinfo {pages} {027204} (\bibinfo {year} {2010})}\BibitemShut {NoStop}%
\bibitem [{\citenamefont {Chaloupka}\ \emph {et~al.}(2013)\citenamefont
  {Chaloupka}, \citenamefont {Jackeli},\ and\ \citenamefont
  {Khaliullin}}]{chaloupka2013zigzag}%
  \BibitemOpen
  \bibfield  {author} {\bibinfo {author} {\bibfnamefont {J.}~\bibnamefont
  {Chaloupka}}, \bibinfo {author} {\bibfnamefont {G.}~\bibnamefont {Jackeli}},
  \ and\ \bibinfo {author} {\bibfnamefont {G.}~\bibnamefont {Khaliullin}},\
  }\href {\doibase 10.1103/PhysRevLett.110.097204} {\bibfield  {journal}
  {\bibinfo  {journal} {Phys. Rev. Lett.}\ }\textbf {\bibinfo {volume} {110}},\
  \bibinfo {pages} {097204} (\bibinfo {year} {2013})}\BibitemShut {NoStop}%
\bibitem [{\citenamefont {Rau}\ \emph {et~al.}(2014)\citenamefont {Rau},
  \citenamefont {Lee},\ and\ \citenamefont {Kee}}]{rau2014generic}%
  \BibitemOpen
  \bibfield  {author} {\bibinfo {author} {\bibfnamefont {J.~G.}\ \bibnamefont
  {Rau}}, \bibinfo {author} {\bibfnamefont {E.~K.-H.}\ \bibnamefont {Lee}}, \
  and\ \bibinfo {author} {\bibfnamefont {H.-Y.}\ \bibnamefont {Kee}},\ }\href
  {\doibase 10.1103/PhysRevLett.112.077204} {\bibfield  {journal} {\bibinfo
  {journal} {Phys. Rev. Lett.}\ }\textbf {\bibinfo {volume} {112}},\ \bibinfo
  {pages} {077204} (\bibinfo {year} {2014})}\BibitemShut {NoStop}%
\bibitem [{\citenamefont {Rau}\ and\ \citenamefont
  {Kee}(2014)}]{rau2014trigonal}%
  \BibitemOpen
  \bibfield  {author} {\bibinfo {author} {\bibfnamefont {J.~G.}\ \bibnamefont
  {Rau}}\ and\ \bibinfo {author} {\bibfnamefont {H.-Y.}\ \bibnamefont {Kee}},\
  }\href@noop {} {\bibfield  {journal} {\bibinfo  {journal} {arXiv preprint
  arXiv:1408.4811}\ } (\bibinfo {year} {2014})}\BibitemShut {NoStop}%
\bibitem [{\citenamefont {Gotfryd}\ \emph {et~al.}(2017)\citenamefont
  {Gotfryd}, \citenamefont {Rusna\ifmmode~\check{c}\else \v{c}\fi{}ko},
  \citenamefont {Wohlfeld}, \citenamefont {Jackeli}, \citenamefont
  {Chaloupka},\ and\ \citenamefont {Ole\ifmmode~\acute{s}\else
  \'{s}\fi{}}}]{gotfryd2017phase}%
  \BibitemOpen
  \bibfield  {author} {\bibinfo {author} {\bibfnamefont {D.}~\bibnamefont
  {Gotfryd}}, \bibinfo {author} {\bibfnamefont {J.}~\bibnamefont
  {Rusna\ifmmode~\check{c}\else \v{c}\fi{}ko}}, \bibinfo {author}
  {\bibfnamefont {K.}~\bibnamefont {Wohlfeld}}, \bibinfo {author}
  {\bibfnamefont {G.}~\bibnamefont {Jackeli}}, \bibinfo {author} {\bibfnamefont
  {J.~c.~v.}\ \bibnamefont {Chaloupka}}, \ and\ \bibinfo {author}
  {\bibfnamefont {A.~M.}\ \bibnamefont {Ole\ifmmode~\acute{s}\else
  \'{s}\fi{}}},\ }\href {\doibase 10.1103/PhysRevB.95.024426} {\bibfield
  {journal} {\bibinfo  {journal} {Phys. Rev. B}\ }\textbf {\bibinfo {volume}
  {95}},\ \bibinfo {pages} {024426} (\bibinfo {year} {2017})}\BibitemShut
  {NoStop}%
\bibitem [{\citenamefont {Gohlke}\ \emph {et~al.}(2017)\citenamefont {Gohlke},
  \citenamefont {Verresen}, \citenamefont {Moessner},\ and\ \citenamefont
  {Pollmann}}]{gohlke2017dynamics}%
  \BibitemOpen
  \bibfield  {author} {\bibinfo {author} {\bibfnamefont {M.}~\bibnamefont
  {Gohlke}}, \bibinfo {author} {\bibfnamefont {R.}~\bibnamefont {Verresen}},
  \bibinfo {author} {\bibfnamefont {R.}~\bibnamefont {Moessner}}, \ and\
  \bibinfo {author} {\bibfnamefont {F.}~\bibnamefont {Pollmann}},\ }\href
  {\doibase 10.1103/PhysRevLett.119.157203} {\bibfield  {journal} {\bibinfo
  {journal} {Phys. Rev. Lett.}\ }\textbf {\bibinfo {volume} {119}},\ \bibinfo
  {pages} {157203} (\bibinfo {year} {2017})}\BibitemShut {NoStop}%
\bibitem [{\citenamefont {Gohlke}\ \emph
  {et~al.}(2018{\natexlab{a}})\citenamefont {Gohlke}, \citenamefont {Wachtel},
  \citenamefont {Yamaji}, \citenamefont {Pollmann},\ and\ \citenamefont
  {Kim}}]{gohlke2017quantum}%
  \BibitemOpen
  \bibfield  {author} {\bibinfo {author} {\bibfnamefont {M.}~\bibnamefont
  {Gohlke}}, \bibinfo {author} {\bibfnamefont {G.}~\bibnamefont {Wachtel}},
  \bibinfo {author} {\bibfnamefont {Y.}~\bibnamefont {Yamaji}}, \bibinfo
  {author} {\bibfnamefont {F.}~\bibnamefont {Pollmann}}, \ and\ \bibinfo
  {author} {\bibfnamefont {Y.~B.}\ \bibnamefont {Kim}},\ }\href {\doibase
  10.1103/PhysRevB.97.075126} {\bibfield  {journal} {\bibinfo  {journal} {Phys.
  Rev. B}\ }\textbf {\bibinfo {volume} {97}},\ \bibinfo {pages} {075126}
  (\bibinfo {year} {2018}{\natexlab{a}})}\BibitemShut {NoStop}%
\bibitem [{\citenamefont {Gordon}\ \emph {et~al.}(2019)\citenamefont {Gordon},
  \citenamefont {Catuneanu}, \citenamefont {S{\o}rensen},\ and\ \citenamefont
  {Kee}}]{gordon2019theory}%
  \BibitemOpen
  \bibfield  {author} {\bibinfo {author} {\bibfnamefont {J.~S.}\ \bibnamefont
  {Gordon}}, \bibinfo {author} {\bibfnamefont {A.}~\bibnamefont {Catuneanu}},
  \bibinfo {author} {\bibfnamefont {E.~S.}\ \bibnamefont {S{\o}rensen}}, \ and\
  \bibinfo {author} {\bibfnamefont {H.-Y.}\ \bibnamefont {Kee}},\ }\href@noop
  {} {\bibfield  {journal} {\bibinfo  {journal} {Nature communications}\
  }\textbf {\bibinfo {volume} {10}},\ \bibinfo {pages} {2470} (\bibinfo {year}
  {2019})}\BibitemShut {NoStop}%
\bibitem [{\citenamefont {Osorio~Iregui}\ \emph {et~al.}(2014)\citenamefont
  {Osorio~Iregui}, \citenamefont {Corboz},\ and\ \citenamefont
  {Troyer}}]{iregui2014probing}%
  \BibitemOpen
  \bibfield  {author} {\bibinfo {author} {\bibfnamefont {J.}~\bibnamefont
  {Osorio~Iregui}}, \bibinfo {author} {\bibfnamefont {P.}~\bibnamefont
  {Corboz}}, \ and\ \bibinfo {author} {\bibfnamefont {M.}~\bibnamefont
  {Troyer}},\ }\href {\doibase 10.1103/PhysRevB.90.195102} {\bibfield
  {journal} {\bibinfo  {journal} {Phys. Rev. B}\ }\textbf {\bibinfo {volume}
  {90}},\ \bibinfo {pages} {195102} (\bibinfo {year} {2014})}\BibitemShut
  {NoStop}%
\bibitem [{\citenamefont {Lee}\ \emph {et~al.}(2020)\citenamefont {Lee},
  \citenamefont {Kaneko}, \citenamefont {Chern}, \citenamefont {Okubo},
  \citenamefont {Yamaji}, \citenamefont {Kawashima},\ and\ \citenamefont
  {Kim}}]{lee2020magnetic}%
  \BibitemOpen
  \bibfield  {author} {\bibinfo {author} {\bibfnamefont {H.-Y.}\ \bibnamefont
  {Lee}}, \bibinfo {author} {\bibfnamefont {R.}~\bibnamefont {Kaneko}},
  \bibinfo {author} {\bibfnamefont {L.~E.}\ \bibnamefont {Chern}}, \bibinfo
  {author} {\bibfnamefont {T.}~\bibnamefont {Okubo}}, \bibinfo {author}
  {\bibfnamefont {Y.}~\bibnamefont {Yamaji}}, \bibinfo {author} {\bibfnamefont
  {N.}~\bibnamefont {Kawashima}}, \ and\ \bibinfo {author} {\bibfnamefont
  {Y.~B.}\ \bibnamefont {Kim}},\ }\href {\doibase 10.1038/s41467-020-15320-x}
  {\bibfield  {journal} {\bibinfo  {journal} {Nature Communications}\ }\textbf
  {\bibinfo {volume} {11}},\ \bibinfo {pages} {1639} (\bibinfo {year}
  {2020})}\BibitemShut {NoStop}%
\bibitem [{\citenamefont {Schaffer}\ \emph
  {et~al.}(2012{\natexlab{a}})\citenamefont {Schaffer}, \citenamefont
  {Bhattacharjee},\ and\ \citenamefont {Kim}}]{schaffer2012probing}%
  \BibitemOpen
  \bibfield  {author} {\bibinfo {author} {\bibfnamefont {R.}~\bibnamefont
  {Schaffer}}, \bibinfo {author} {\bibfnamefont {S.}~\bibnamefont
  {Bhattacharjee}}, \ and\ \bibinfo {author} {\bibfnamefont {Y.~B.}\
  \bibnamefont {Kim}},\ }\href {\doibase 10.1103/PhysRevB.86.224417} {\bibfield
   {journal} {\bibinfo  {journal} {Phys. Rev. B}\ }\textbf {\bibinfo {volume}
  {86}},\ \bibinfo {pages} {224417} (\bibinfo {year}
  {2012}{\natexlab{a}})}\BibitemShut {NoStop}%
\bibitem [{\citenamefont {Wang}\ \emph {et~al.}(2019)\citenamefont {Wang},
  \citenamefont {Normand},\ and\ \citenamefont {Liu}}]{wang2019one}%
  \BibitemOpen
  \bibfield  {author} {\bibinfo {author} {\bibfnamefont {J.}~\bibnamefont
  {Wang}}, \bibinfo {author} {\bibfnamefont {B.}~\bibnamefont {Normand}}, \
  and\ \bibinfo {author} {\bibfnamefont {Z.-X.}\ \bibnamefont {Liu}},\ }\href
  {\doibase 10.1103/PhysRevLett.123.197201} {\bibfield  {journal} {\bibinfo
  {journal} {Phys. Rev. Lett.}\ }\textbf {\bibinfo {volume} {123}},\ \bibinfo
  {pages} {197201} (\bibinfo {year} {2019})}\BibitemShut {NoStop}%
\bibitem [{\citenamefont {Baskaran}\ \emph {et~al.}(2007)\citenamefont
  {Baskaran}, \citenamefont {Mandal},\ and\ \citenamefont
  {Shankar}}]{baskaran2007exact}%
  \BibitemOpen
  \bibfield  {author} {\bibinfo {author} {\bibfnamefont {G.}~\bibnamefont
  {Baskaran}}, \bibinfo {author} {\bibfnamefont {S.}~\bibnamefont {Mandal}}, \
  and\ \bibinfo {author} {\bibfnamefont {R.}~\bibnamefont {Shankar}},\
  }\href@noop {} {\bibfield  {journal} {\bibinfo  {journal} {Physical review
  letters}\ }\textbf {\bibinfo {volume} {98}},\ \bibinfo {pages} {247201}
  (\bibinfo {year} {2007})}\BibitemShut {NoStop}%
\bibitem [{\citenamefont {Knolle}\ \emph
  {et~al.}(2014{\natexlab{a}})\citenamefont {Knolle}, \citenamefont
  {Kovrizhin}, \citenamefont {Chalker},\ and\ \citenamefont
  {Moessner}}]{knolle2014dynamics}%
  \BibitemOpen
  \bibfield  {author} {\bibinfo {author} {\bibfnamefont {J.}~\bibnamefont
  {Knolle}}, \bibinfo {author} {\bibfnamefont {D.~L.}\ \bibnamefont
  {Kovrizhin}}, \bibinfo {author} {\bibfnamefont {J.~T.}\ \bibnamefont
  {Chalker}}, \ and\ \bibinfo {author} {\bibfnamefont {R.}~\bibnamefont
  {Moessner}},\ }\href {\doibase 10.1103/PhysRevLett.112.207203} {\bibfield
  {journal} {\bibinfo  {journal} {Phys. Rev. Lett.}\ }\textbf {\bibinfo
  {volume} {112}},\ \bibinfo {pages} {207203} (\bibinfo {year}
  {2014}{\natexlab{a}})}\BibitemShut {NoStop}%
\bibitem [{\citenamefont {Knolle}\ \emph {et~al.}(2015)\citenamefont {Knolle},
  \citenamefont {Kovrizhin}, \citenamefont {Chalker},\ and\ \citenamefont
  {Moessner}}]{knolle2015dynamics}%
  \BibitemOpen
  \bibfield  {author} {\bibinfo {author} {\bibfnamefont {J.}~\bibnamefont
  {Knolle}}, \bibinfo {author} {\bibfnamefont {D.~L.}\ \bibnamefont
  {Kovrizhin}}, \bibinfo {author} {\bibfnamefont {J.~T.}\ \bibnamefont
  {Chalker}}, \ and\ \bibinfo {author} {\bibfnamefont {R.}~\bibnamefont
  {Moessner}},\ }\href {\doibase 10.1103/PhysRevB.92.115127} {\bibfield
  {journal} {\bibinfo  {journal} {Phys. Rev. B}\ }\textbf {\bibinfo {volume}
  {92}},\ \bibinfo {pages} {115127} (\bibinfo {year} {2015})}\BibitemShut
  {NoStop}%
\bibitem [{\citenamefont {Song}\ \emph {et~al.}(2016)\citenamefont {Song},
  \citenamefont {You},\ and\ \citenamefont {Balents}}]{song2016low}%
  \BibitemOpen
  \bibfield  {author} {\bibinfo {author} {\bibfnamefont {X.-Y.}\ \bibnamefont
  {Song}}, \bibinfo {author} {\bibfnamefont {Y.-Z.}\ \bibnamefont {You}}, \
  and\ \bibinfo {author} {\bibfnamefont {L.}~\bibnamefont {Balents}},\ }\href
  {\doibase 10.1103/PhysRevLett.117.037209} {\bibfield  {journal} {\bibinfo
  {journal} {Phys. Rev. Lett.}\ }\textbf {\bibinfo {volume} {117}},\ \bibinfo
  {pages} {037209} (\bibinfo {year} {2016})}\BibitemShut {NoStop}%
\bibitem [{\citenamefont {Knolle}\ \emph
  {et~al.}(2014{\natexlab{b}})\citenamefont {Knolle}, \citenamefont {Chern},
  \citenamefont {Kovrizhin}, \citenamefont {Moessner},\ and\ \citenamefont
  {Perkins}}]{knolle2014raman}%
  \BibitemOpen
  \bibfield  {author} {\bibinfo {author} {\bibfnamefont {J.}~\bibnamefont
  {Knolle}}, \bibinfo {author} {\bibfnamefont {G.-W.}\ \bibnamefont {Chern}},
  \bibinfo {author} {\bibfnamefont {D.~L.}\ \bibnamefont {Kovrizhin}}, \bibinfo
  {author} {\bibfnamefont {R.}~\bibnamefont {Moessner}}, \ and\ \bibinfo
  {author} {\bibfnamefont {N.~B.}\ \bibnamefont {Perkins}},\ }\href {\doibase
  10.1103/PhysRevLett.113.187201} {\bibfield  {journal} {\bibinfo  {journal}
  {Phys. Rev. Lett.}\ }\textbf {\bibinfo {volume} {113}},\ \bibinfo {pages}
  {187201} (\bibinfo {year} {2014}{\natexlab{b}})}\BibitemShut {NoStop}%
\bibitem [{\citenamefont {Hal\'asz}\ \emph {et~al.}(2016)\citenamefont
  {Hal\'asz}, \citenamefont {Perkins},\ and\ \citenamefont {van~den
  Brink}}]{halasz2016resonant}%
  \BibitemOpen
  \bibfield  {author} {\bibinfo {author} {\bibfnamefont {G.~B.}\ \bibnamefont
  {Hal\'asz}}, \bibinfo {author} {\bibfnamefont {N.~B.}\ \bibnamefont
  {Perkins}}, \ and\ \bibinfo {author} {\bibfnamefont {J.}~\bibnamefont
  {van~den Brink}},\ }\href {\doibase 10.1103/PhysRevLett.117.127203}
  {\bibfield  {journal} {\bibinfo  {journal} {Phys. Rev. Lett.}\ }\textbf
  {\bibinfo {volume} {117}},\ \bibinfo {pages} {127203} (\bibinfo {year}
  {2016})}\BibitemShut {NoStop}%
\bibitem [{\citenamefont {Hal\'asz}\ \emph {et~al.}(2019)\citenamefont
  {Hal\'asz}, \citenamefont {Kourtis}, \citenamefont {Knolle},\ and\
  \citenamefont {Perkins}}]{halasz2019observing}%
  \BibitemOpen
  \bibfield  {author} {\bibinfo {author} {\bibfnamefont {G.~B.}\ \bibnamefont
  {Hal\'asz}}, \bibinfo {author} {\bibfnamefont {S.}~\bibnamefont {Kourtis}},
  \bibinfo {author} {\bibfnamefont {J.}~\bibnamefont {Knolle}}, \ and\ \bibinfo
  {author} {\bibfnamefont {N.~B.}\ \bibnamefont {Perkins}},\ }\href {\doibase
  10.1103/PhysRevB.99.184417} {\bibfield  {journal} {\bibinfo  {journal} {Phys.
  Rev. B}\ }\textbf {\bibinfo {volume} {99}},\ \bibinfo {pages} {184417}
  (\bibinfo {year} {2019})}\BibitemShut {NoStop}%
\bibitem [{\citenamefont {Samarakoon}\ \emph {et~al.}(2017)\citenamefont
  {Samarakoon}, \citenamefont {Banerjee}, \citenamefont {Zhang}, \citenamefont
  {Kamiya}, \citenamefont {Nagler}, \citenamefont {Tennant}, \citenamefont
  {Lee},\ and\ \citenamefont {Batista}}]{samarakoon2017comprehensive}%
  \BibitemOpen
  \bibfield  {author} {\bibinfo {author} {\bibfnamefont {A.~M.}\ \bibnamefont
  {Samarakoon}}, \bibinfo {author} {\bibfnamefont {A.}~\bibnamefont
  {Banerjee}}, \bibinfo {author} {\bibfnamefont {S.-S.}\ \bibnamefont {Zhang}},
  \bibinfo {author} {\bibfnamefont {Y.}~\bibnamefont {Kamiya}}, \bibinfo
  {author} {\bibfnamefont {S.~E.}\ \bibnamefont {Nagler}}, \bibinfo {author}
  {\bibfnamefont {D.~A.}\ \bibnamefont {Tennant}}, \bibinfo {author}
  {\bibfnamefont {S.-H.}\ \bibnamefont {Lee}}, \ and\ \bibinfo {author}
  {\bibfnamefont {C.~D.}\ \bibnamefont {Batista}},\ }\href {\doibase
  10.1103/PhysRevB.96.134408} {\bibfield  {journal} {\bibinfo  {journal} {Phys.
  Rev. B}\ }\textbf {\bibinfo {volume} {96}},\ \bibinfo {pages} {134408}
  (\bibinfo {year} {2017})}\BibitemShut {NoStop}%
\bibitem [{\citenamefont {Zhu}\ \emph {et~al.}(2017)\citenamefont {Zhu},
  \citenamefont {Maksimov}, \citenamefont {White},\ and\ \citenamefont
  {Chernyshev}}]{zhu2017disorder}%
  \BibitemOpen
  \bibfield  {author} {\bibinfo {author} {\bibfnamefont {Z.}~\bibnamefont
  {Zhu}}, \bibinfo {author} {\bibfnamefont {P.~A.}\ \bibnamefont {Maksimov}},
  \bibinfo {author} {\bibfnamefont {S.~R.}\ \bibnamefont {White}}, \ and\
  \bibinfo {author} {\bibfnamefont {A.~L.}\ \bibnamefont {Chernyshev}},\ }\href
  {\doibase 10.1103/PhysRevLett.119.157201} {\bibfield  {journal} {\bibinfo
  {journal} {Phys. Rev. Lett.}\ }\textbf {\bibinfo {volume} {119}},\ \bibinfo
  {pages} {157201} (\bibinfo {year} {2017})}\BibitemShut {NoStop}%
\bibitem [{\citenamefont {Samarakoon}\ \emph {et~al.}(2018)\citenamefont
  {Samarakoon}, \citenamefont {Wachtel}, \citenamefont {Yamaji}, \citenamefont
  {Tennant}, \citenamefont {Batista},\ and\ \citenamefont
  {Kim}}]{samarakoon2018classical}%
  \BibitemOpen
  \bibfield  {author} {\bibinfo {author} {\bibfnamefont {A.~M.}\ \bibnamefont
  {Samarakoon}}, \bibinfo {author} {\bibfnamefont {G.}~\bibnamefont {Wachtel}},
  \bibinfo {author} {\bibfnamefont {Y.}~\bibnamefont {Yamaji}}, \bibinfo
  {author} {\bibfnamefont {D.~A.}\ \bibnamefont {Tennant}}, \bibinfo {author}
  {\bibfnamefont {C.~D.}\ \bibnamefont {Batista}}, \ and\ \bibinfo {author}
  {\bibfnamefont {Y.~B.}\ \bibnamefont {Kim}},\ }\href {\doibase
  10.1103/PhysRevB.98.045121} {\bibfield  {journal} {\bibinfo  {journal} {Phys.
  Rev. B}\ }\textbf {\bibinfo {volume} {98}},\ \bibinfo {pages} {045121}
  (\bibinfo {year} {2018})}\BibitemShut {NoStop}%
\bibitem [{\citenamefont {Hentschel}\ and\ \citenamefont
  {Guinea}(2007)}]{hentschel2007orthogonality}%
  \BibitemOpen
  \bibfield  {author} {\bibinfo {author} {\bibfnamefont {M.}~\bibnamefont
  {Hentschel}}\ and\ \bibinfo {author} {\bibfnamefont {F.}~\bibnamefont
  {Guinea}},\ }\href {\doibase 10.1103/PhysRevB.76.115407} {\bibfield
  {journal} {\bibinfo  {journal} {Phys. Rev. B}\ }\textbf {\bibinfo {volume}
  {76}},\ \bibinfo {pages} {115407} (\bibinfo {year} {2007})}\BibitemShut
  {NoStop}%
\bibitem [{Note1()}]{Note1}%
  \BibitemOpen
  \bibinfo {note} {Note that the fermion parity is defined with respect to the
  fermion vacuum of the flux-free sector and accounts for both bond and matter
  fermions.}\BibitemShut {Stop}%
\bibitem [{\citenamefont {B\'acsi}\ and\ \citenamefont
  {Virosztek}(2010)}]{Basci10}%
  \BibitemOpen
  \bibfield  {author} {\bibinfo {author} {\bibfnamefont {A.}~\bibnamefont
  {B\'acsi}}\ and\ \bibinfo {author} {\bibfnamefont {A.}~\bibnamefont
  {Virosztek}},\ }\href {\doibase 10.1103/PhysRevB.82.193405} {\bibfield
  {journal} {\bibinfo  {journal} {Phys. Rev. B}\ }\textbf {\bibinfo {volume}
  {82}},\ \bibinfo {pages} {193405} (\bibinfo {year} {2010})}\BibitemShut
  {NoStop}%
\bibitem [{\citenamefont {Khaliullin}(2005)}]{khaliullin2005orbital}%
  \BibitemOpen
  \bibfield  {author} {\bibinfo {author} {\bibfnamefont {G.}~\bibnamefont
  {Khaliullin}},\ }\href@noop {} {\bibfield  {journal} {\bibinfo  {journal}
  {Progress of Theoretical Physics Supplement}\ }\textbf {\bibinfo {volume}
  {160}},\ \bibinfo {pages} {155} (\bibinfo {year} {2005})}\BibitemShut
  {NoStop}%
\bibitem [{\citenamefont {Schaffer}\ \emph
  {et~al.}(2012{\natexlab{b}})\citenamefont {Schaffer}, \citenamefont
  {Bhattacharjee},\ and\ \citenamefont {Kim}}]{schaffer2012quantum}%
  \BibitemOpen
  \bibfield  {author} {\bibinfo {author} {\bibfnamefont {R.}~\bibnamefont
  {Schaffer}}, \bibinfo {author} {\bibfnamefont {S.}~\bibnamefont
  {Bhattacharjee}}, \ and\ \bibinfo {author} {\bibfnamefont {Y.~B.}\
  \bibnamefont {Kim}},\ }\href@noop {} {\bibfield  {journal} {\bibinfo
  {journal} {Physical Review B}\ }\textbf {\bibinfo {volume} {86}},\ \bibinfo
  {pages} {224417} (\bibinfo {year} {2012}{\natexlab{b}})}\BibitemShut
  {NoStop}%
\bibitem [{Note2()}]{Note2}%
  \BibitemOpen
  \bibinfo {note} {$x_+=0.5$ with FM Kitaev interaction is mapped into the pure
  FM Heisenberg model, implying that the local magnetic moment is fully
  saturated.}\BibitemShut {Stop}%
\bibitem [{Note3()}]{Note3}%
  \BibitemOpen
  \bibinfo {note} {Note that those processes can only be included by enlarging
  the variational space to states containing two flux pairs and an even number
  of matter fermions.}\BibitemShut {Stop}%
\bibitem [{\citenamefont {Zhu}\ \emph {et~al.}(2018)\citenamefont {Zhu},
  \citenamefont {Kimchi}, \citenamefont {Sheng},\ and\ \citenamefont
  {Fu}}]{zhu2018robust}%
  \BibitemOpen
  \bibfield  {author} {\bibinfo {author} {\bibfnamefont {Z.}~\bibnamefont
  {Zhu}}, \bibinfo {author} {\bibfnamefont {I.}~\bibnamefont {Kimchi}},
  \bibinfo {author} {\bibfnamefont {D.~N.}\ \bibnamefont {Sheng}}, \ and\
  \bibinfo {author} {\bibfnamefont {L.}~\bibnamefont {Fu}},\ }\href {\doibase
  10.1103/PhysRevB.97.241110} {\bibfield  {journal} {\bibinfo  {journal} {Phys.
  Rev. B}\ }\textbf {\bibinfo {volume} {97}},\ \bibinfo {pages} {241110}
  (\bibinfo {year} {2018})}\BibitemShut {NoStop}%
\bibitem [{\citenamefont {Gohlke}\ \emph
  {et~al.}(2018{\natexlab{b}})\citenamefont {Gohlke}, \citenamefont
  {Moessner},\ and\ \citenamefont {Pollmann}}]{gohlke2018dynamical}%
  \BibitemOpen
  \bibfield  {author} {\bibinfo {author} {\bibfnamefont {M.}~\bibnamefont
  {Gohlke}}, \bibinfo {author} {\bibfnamefont {R.}~\bibnamefont {Moessner}}, \
  and\ \bibinfo {author} {\bibfnamefont {F.}~\bibnamefont {Pollmann}},\ }\href
  {\doibase 10.1103/PhysRevB.98.014418} {\bibfield  {journal} {\bibinfo
  {journal} {Phys. Rev. B}\ }\textbf {\bibinfo {volume} {98}},\ \bibinfo
  {pages} {014418} (\bibinfo {year} {2018}{\natexlab{b}})}\BibitemShut
  {NoStop}%
\bibitem [{\citenamefont {Hickey}\ and\ \citenamefont
  {Trebst}(2019)}]{hickey2019emergence}%
  \BibitemOpen
  \bibfield  {author} {\bibinfo {author} {\bibfnamefont {C.}~\bibnamefont
  {Hickey}}\ and\ \bibinfo {author} {\bibfnamefont {S.}~\bibnamefont
  {Trebst}},\ }\href {\doibase 10.1038/s41467-019-08459-9} {\bibfield
  {journal} {\bibinfo  {journal} {Nature Communications}\ }\textbf {\bibinfo
  {volume} {10}},\ \bibinfo {pages} {530} (\bibinfo {year} {2019})}\BibitemShut
  {NoStop}%
\bibitem [{\citenamefont {Knolle}\ \emph {et~al.}(2019)\citenamefont {Knolle},
  \citenamefont {Moessner},\ and\ \citenamefont {Perkins}}]{knolle2019bond}%
  \BibitemOpen
  \bibfield  {author} {\bibinfo {author} {\bibfnamefont {J.}~\bibnamefont
  {Knolle}}, \bibinfo {author} {\bibfnamefont {R.}~\bibnamefont {Moessner}}, \
  and\ \bibinfo {author} {\bibfnamefont {N.~B.}\ \bibnamefont {Perkins}},\
  }\href@noop {} {\bibfield  {journal} {\bibinfo  {journal} {Physical review
  letters}\ }\textbf {\bibinfo {volume} {122}},\ \bibinfo {pages} {047202}
  (\bibinfo {year} {2019})}\BibitemShut {NoStop}%
\bibitem [{\citenamefont {Kao}\ \emph {et~al.}(2021)\citenamefont {Kao},
  \citenamefont {Knolle}, \citenamefont {Hal{\'a}sz}, \citenamefont
  {Moessner},\ and\ \citenamefont {Perkins}}]{kao2021vacancy}%
  \BibitemOpen
  \bibfield  {author} {\bibinfo {author} {\bibfnamefont {W.-H.}\ \bibnamefont
  {Kao}}, \bibinfo {author} {\bibfnamefont {J.}~\bibnamefont {Knolle}},
  \bibinfo {author} {\bibfnamefont {G.~B.}\ \bibnamefont {Hal{\'a}sz}},
  \bibinfo {author} {\bibfnamefont {R.}~\bibnamefont {Moessner}}, \ and\
  \bibinfo {author} {\bibfnamefont {N.~B.}\ \bibnamefont {Perkins}},\
  }\href@noop {} {\bibfield  {journal} {\bibinfo  {journal} {Physical Review
  X}\ }\textbf {\bibinfo {volume} {11}},\ \bibinfo {pages} {011034} (\bibinfo
  {year} {2021})}\BibitemShut {NoStop}%
\bibitem [{\citenamefont {Kao}\ and\ \citenamefont
  {Perkins}(2021)}]{kao2021disorder}%
  \BibitemOpen
  \bibfield  {author} {\bibinfo {author} {\bibfnamefont {W.-H.}\ \bibnamefont
  {Kao}}\ and\ \bibinfo {author} {\bibfnamefont {N.~B.}\ \bibnamefont
  {Perkins}},\ }\href@noop {} {\bibfield  {journal} {\bibinfo  {journal}
  {Annals of Physics}\ ,\ \bibinfo {pages} {168506}} (\bibinfo {year}
  {2021})}\BibitemShut {NoStop}%
\bibitem [{\citenamefont {Petrova}\ \emph {et~al.}(2014)\citenamefont
  {Petrova}, \citenamefont {Mellado},\ and\ \citenamefont
  {Tchernyshyov}}]{petrova2014unpaired}%
  \BibitemOpen
  \bibfield  {author} {\bibinfo {author} {\bibfnamefont {O.}~\bibnamefont
  {Petrova}}, \bibinfo {author} {\bibfnamefont {P.}~\bibnamefont {Mellado}}, \
  and\ \bibinfo {author} {\bibfnamefont {O.}~\bibnamefont {Tchernyshyov}},\
  }\href@noop {} {\bibfield  {journal} {\bibinfo  {journal} {Physical Review
  B}\ }\textbf {\bibinfo {volume} {90}},\ \bibinfo {pages} {134404} (\bibinfo
  {year} {2014})}\BibitemShut {NoStop}%
\bibitem [{\citenamefont {Willans}\ \emph {et~al.}(2010)\citenamefont
  {Willans}, \citenamefont {Chalker},\ and\ \citenamefont
  {Moessner}}]{willans2010disorder}%
  \BibitemOpen
  \bibfield  {author} {\bibinfo {author} {\bibfnamefont {A.~J.}\ \bibnamefont
  {Willans}}, \bibinfo {author} {\bibfnamefont {J.~T.}\ \bibnamefont
  {Chalker}}, \ and\ \bibinfo {author} {\bibfnamefont {R.}~\bibnamefont
  {Moessner}},\ }\href {\doibase 10.1103/PhysRevLett.104.237203} {\bibfield
  {journal} {\bibinfo  {journal} {Phys. Rev. Lett.}\ }\textbf {\bibinfo
  {volume} {104}},\ \bibinfo {pages} {237203} (\bibinfo {year}
  {2010})}\BibitemShut {NoStop}%
\bibitem [{\citenamefont {Willans}\ \emph {et~al.}(2011)\citenamefont
  {Willans}, \citenamefont {Chalker},\ and\ \citenamefont
  {Moessner}}]{willans2011site}%
  \BibitemOpen
  \bibfield  {author} {\bibinfo {author} {\bibfnamefont {A.~J.}\ \bibnamefont
  {Willans}}, \bibinfo {author} {\bibfnamefont {J.~T.}\ \bibnamefont
  {Chalker}}, \ and\ \bibinfo {author} {\bibfnamefont {R.}~\bibnamefont
  {Moessner}},\ }\href {\doibase 10.1103/PhysRevB.84.115146} {\bibfield
  {journal} {\bibinfo  {journal} {Phys. Rev. B}\ }\textbf {\bibinfo {volume}
  {84}},\ \bibinfo {pages} {115146} (\bibinfo {year} {2011})}\BibitemShut
  {NoStop}%
\end{thebibliography}%

\end{document}